\documentclass[aps,twocolumn,reprint,letterpaper,showkeys]{revtex4-2}

\usepackage{amsmath,amssymb,amsfonts} %
\usepackage{graphicx}

\newcommand{\Eq}[1]{Eq.~(\ref{#1})}

\newcommand{\hh}{h}  

\newcommand{\revision}[1]{#1}

\begin{document}

\title{Intermolecular forces at ice and water interfaces:\\ premelting, surface freezing and regelation}

\author{Juan Luengo-M\'arquez}
\affiliation{%
Department of Theoretical Condensed Matter Physics and Instituto Nicol\'as Cabrera, \\
Universidad Autónoma de Madrid, 28049 Madrid (Spain)
}

\author{Fernando Izquierdo-Ruiz}
\affiliation{Departamento.\ de Qu\'{i}mica-F\'{i}sica y Anal\'{i}tica, \\Facultad de Ciencias
Qu\'{i}micas, Universidad de Oviedo, 33006 Oviedo, Spain.}

\author{Luis G.\ MacDowell}
\affiliation{Departamento.\ de Qu\'{i}mica-F\'{i}sica, \\Facultad de Ciencias 
Qu\'{i}micas, Universidad Complutense de Madrid, 28040 Madrid, Spain.}
\email{Corresponding author: lgmac@quim.ucm.es}

\date{\today}%


\begin{abstract}
	Using Lifshitz theory we assess the role of van der Waals forces at
interfaces of ice and water. The results are combined with measured structural forces from computer simulations to develop a quantitative model of the surface free energy of premelting films. This input is employed within the framework of wetting theory and allows us to predict qualitatively the behavior of quasi-liquid layer thickness as a function of ambient conditions. Our results \revision{emphasize} the significance of vapor pressure. The ice vapor interface is shown to exhibit only incomplete premelting, but the situation can shift to a state of complete surface melting above water saturation. The results obtained serve also to assess the role of subsurface freezing at the water-vapor interface, and we show that intermolecular forces favor subsurface ice nucleation only in conditions of water undersaturation. We show ice regelation at ambient pressure may be explained as a process of capillary freezing, without the need to invoke the action of bulk pressure melting. Our results for van der Waals forces are exploited in order to gauge dispersion interactions in empirical point charge models of water.
\end{abstract}


\maketitle

\section{Introduction}

\label{Intro}

The interface of liquid and solid phases of water exposed to air hosts a large
number of complex phenomena of very important practical and theoretical
significance.\cite{bjorneholm16,bartels12,libbrecht22}  
Multiple different compounds, such as
atmospheric gases, ions or surfactants can easily adsorb and \revision{significantly} change the
interfacial properties of water.\cite{pruppacher10} However, without the
need of any additional species, interfaces of ice and \revision{liquid} water in contact with pure
water vapor already exhibit a fascinating and complex physics that has attracted
the attention of researchers for many years.\cite{weyl51} 

One particularly interesting issue is the possibility of condensed phases of 
water to \revision{self-adsorb} one on to the other as the triple point is approached. 
In this situation, ice, \revision{liquid} water and water vapor have a similar chemical potential
and feed one from the other. Of course, macroscopic  samples
of the three bulk phases can only be found simultaneously exactly at the
triple point, but it is not unexpected to find microscopic amounts of a third
phase adsorbing at the interface of the two other  at
coexistence.\cite{lipowsky82,dietrich88,schick90} 

A well known example is that of ice 
premelting.\cite{jellinek67,nenow84,petrenko94,dash95,rosenberg05,dash06,slater19,nagata19}
Here, ice in coexistence with the vapor phase close to the triple point
is said to {\em premelt} thin amounts of ice at the surface, 
forming a so called quasi-liquid layer. 
The cost of forming a nanoscopic amount of premelted ice can be balanced by 
a delicate interplay of surface intermolecular
forces.\cite{elbaum91,wettlaufer99,dash06} 

As the triple point is approached, the free energy penalty of the bulk
liquid phase vanishes, and the question is then whether the premelting
film remains finite or diverges at the triple point. Unfortunately, the answer  has remained elusive and 
controversial, due mainly to a large body of conflicting experimental
results.\cite{petrenko94,li07b,slater19}

In the theory of wetting, the question \revision{of} the size of adsorbed liquid
layers on a solid is discussed in terms of 
the {\em interface potential}, $g(\hh)$, which accounts
for the free energy of a uniform wetting film of thickness $\hh$ adsorbed 
at the interface between two coexisting bulk phases. The experimental question
as to how $\hh$ evolves with temperature is then mapped into the theoretical
question of how does the interface potential depend on film
thickness.\cite{dietrich88,schick90,dash06} As a very important bonus of the
emphasis on wetting, the formalism allows to assess the evolution of film
thickness not only due to changes in temperature, but also due to changes in the
vapor pressure, which we claim is essential for an understanding of atmospheric
ice.\cite{llombart20,sibley21}

Simplified models of condensed matter physics,  where interactions are
assumed \revision{short-range}, and no packing effects are included, predict a 
logarithmic divergence of the film thickness as the triple point is
approached.\cite{lipowsky82,lipowsky89,limmer14,li19} A more complex
scenario follows by taking into account explicitly packing correlations
in the scale of the molecular diameter.\cite{chernov88,henderson94} 
These correlations can lead in principle to oscillations of the interface 
potential \revision{with respect to the spatial coordinate}, which bind the premelting film to local minima of finite thickness.
However, as the triple point is approached solid-vapor interfaces undergo
a roughening transition, and the oscillatory behavior is 
washed out by thermal 
fluctuations.\cite{chernov88,henderson94,henderson05,llombart20} 
The ultimate fate of the premelting film thickness will depend in
such cases on the behavior of the interface potential at long range,
which is dominated by van der Waals forces with algebraic decay.
This point was emphasized long ago by Elbaum and Schick, who
estimated the long range interactions of premelting films
using Dzyaloshinskii-Lifshitz-Pitaevski theory (DLP) of van der Waals forces.
Their results suggested that the interface potential exhibits an absolute
minimum at a film thickness of about 3~nm.\cite{elbaum91b} 

Unfortunately, the predictions of DLP theory heavily rely on the parametrization
of optical properties of ice and water over the full electromagnetic spectrum,
from the static response to well beyond the ultra-violet \revision{(UV)}.\cite{parsegian05}
This is a demanding requirement, because experimental measurements in
the high energy region are far from trivial, while the modeling of the optical
properties over such a large frequency domain is also
difficult and
controversial.\cite{parsegian81,elbaum91b,roth96,dagastine00,fernandez00,wang17,luengo19,luengo20,fiedler20,luengo21,gudarzi21}

In view of these difficulties, here we revisit the role of van der Waals
forces in ice premelting, using a combination of experimental
dielectric 
properties,\cite{zelsmann95,segelstein81,wieliczka89,bertie96,heller74,hayashi15,wang17,buckley58, warren08,seki81,auty52} 
DLP theory,\cite{dzyaloshinskii61,ninham70b,parsegian05} 
and Quantum Mechanical Density Functional Theory calculations
(DFT).\cite{vasp1,vasp2,vasp3,perdew96,shishkin06,fuchs07,gajdos06,nunes01}

\revision{These} results are revised \revision{in light} of our recent work on the 
structure, kinetics and thermodynamics of the 
ice-vapor interface, 
\cite{benet16,benet19,llombart19,llombart20,llombart20b,sibley21,luengo21}
providing a consistent and comprehensive framework for the
description of premelting films as a function of temperature and
pressure. 

Not unexpectedly, we find that the understanding gained in the
problem of surface premelting gives us also new insight \revision{into} a
number of related problems. Firstly, we digress on the phenomenon of regelation, which refers to the adhesion between thawing ice parcels, which has interest in ice sintering. Secondly, we discuss  surface freezing, i.e., the possible formation of
an adsorbed ice layer at the water-vapor interface, a problem that has
received great attention recently in view of its atmospheric
implications.\cite{tabazadeh02,shaw05,li13,hussain21}
As an additional \revision{result}, we show how the understanding 
of van der Waals forces at water interfaces achieved in this work 
can also serve to gauge the choice of Lennard-Jones parameters
in well known point charge models of water interactions.\cite{jorgensen83,berendsen87,abascal05,abascal05b,henriques16}

In section II we summarize DLP theory for later use in the manuscript. Section
III is devoted to the modeling of optical properties and \revision{also presents an improved oscillator model to characterize the 
dielectric responses of water and ice close to the triple point. 
This representation will be used later
as input in DLP theory}. Readers not interested in the
details can skip
section III and move on to section IV, where we present the
results of van der Waals forces at interfaces involving combinations of bulk
ice, water and vapor phases. Section V is devoted to a discussion on the
implications of our results to the understanding of ice premelting, 
regelation and surface freezing.  Finally, section VI summarizes our main 
findings.

\section{Lifshitz Theory of surface van der Waals forces}

Van der Waals forces result from correlated dipole fluctuations over the full frequency domain. \revision{For molecules a distance apart large enough
to not allow overlapping of the electronic wave functions, the strongest dipole correlations 
are athermal high frequency fluctuations \revision{that stem from the electronic
polarizability of the material}. To leading order, this produces
dispersion interactions, which can be described by an an effective pairwise 
potential with the familiar $r^{-6}$ power law dependence. It must be noted that just as the chemical bond, these dipole interactions are quantum-mechanical in nature,
and emerge from the same electrostatic Hamiltonian\cite{cohen05b}.}
The combined
effect of these pairwise forces results in the interaction between
macroscopic bodies. Particularly, for the free energy
of a wetting film of thickness $\hh$ intervening between two macroscopic 
planar bodies \revision{- i.e. with zero curvature, such as 
the walls in a slit pore -}
the pair-wise summation of additive forces yields an effective interaction between the surfaces of the wetting film
which decays as $\hh^{-2}$.\cite{hamaker37}

At larger distances, however, the effect of high frequency
fluctuations becomes suppressed due to retardation and intermolecular
forces become dominated by low frequency dipole correlations in
the infrared \revision{(IR)} and microwave \revision{(MW)} region.\cite{parsegian05} 
Additionally, the effective pair interactions start decaying at a faster 
rate of order $r^{-7}$, which results in an effective interaction between 
surfaces of order $\hh^{-3}$.\cite{gregory81}

The van der Waals free energy of interactions between two semi-infinite bodies, 
$1$ and $2$, across a thick layer of a third medium, $m$, may be described 
in terms of the interface potential, $g_{1m2}(\hh)$, which measures
the surface free energy of the system as a function of the 
thickness of the intervening medium. 

Without loss of generality, $g_{1m2}(\hh)$ is conveniently expressed as:
\begin{equation}
             g_{1m2}(\hh) = -\frac{A_{1m2}(\hh)}{12\pi \hh^2}
		 \label{hamaker}
\end{equation} 
where $A_{1m2}(\hh)$ is the Hamaker
function.\revision{\cite{parsegian05}} At short distances
it has an asymptotic finite value which is 
known as the Hamaker
constant.\revision{\cite{hamaker37,israelachvili11,parsegian05}} At larger distances, however, retardation
becomes significant and $A_{1m2}(\hh)$ becomes $\hh$
dependent.\revision{\cite{dzyaloshinskii61,parsegian05}}

\revision{
A general solution for the difficult problem of calculating Hamaker functions
was provided by the Dzyaloshinskii-Lifshitz-Pitaaevsky theory of van der Waals
forces (DLP)  \cite{dzyaloshinskii61,parsegian05}. 
The main idea of this approach is to calculate the 
exact partition function of the electromagnetic standing waves of the
involved media. These are estimated approximately, by  solving
the equations of continuum electrodynamics and imposing the
continuity of the electric fields at the interfaces. It follows that
this theory assumes structureless interfaces, and neglects
the continuous change of dielectric properties accross an interfacial
region of finite width, $\xi$.  Fortunately, at the low temperatures we
consider the interfacial region is very sharp, of the order of 1~nm at most,
and the corrections to the macroscopic approximation are of order $\xi/\hh$
\cite{parsegian05}. Whence, the macroscopic theory becomes exact in
the limit of large $\hh$. This is all we need here, since in order to
decide whether a film wets or not wets a substrate, it is the long
range behavior of the interface potential which needs to be addressed.
Notice further that an exact solution of the problem is beyond state
of the art, since one neats to deal simultaneously with the finite
interfacial width, the polarizability of the media, the infinite range
of dispersion interactions and the quantization of the electromagnetic modes.
DLP deals with all but the first issue.
In view of these difficulties, we restore here to DLP theory, 
which provides the following
expression for the Hamaker constant in terms of
bulk optical properties of the involved 
materials:\cite{dzyaloshinskii61,parsegian05}
}
\begin{equation}\label{dlp}
   \begin{array}{ccc}
   A_{1m2}(\hh)  & = & - \frac{3}{2} k_B T {\sum_{n=0}^{\infty}}'
     \int_{r_n}^{\infty} x \left [
	  \ln(1 - R^M_{1m2} e^{-x}) \right . + \\ & &  \\ & &
     \left . \ln(1 - R^E_{1m2} e^{-x})\, \right ]
	       dx
    \end{array}
\end{equation}
where $R^{M}_{1m2}(x,n)=\Delta^M_{1m}(x,n)\Delta^M_{2m}(x,n)$,
$R^{E}_{1m2}(x,n)=\Delta^E_{1m}(x,n)\Delta^E_{2m}(x,n)$, and
\begin{equation}
 \begin{array}{lcl}
		     \Delta^M_{ij} = \frac{x_i \epsilon_j - x_j\epsilon_i}{x_i
			  \epsilon_j +
		     x_j\epsilon_i}
		      & &
			 \Delta^E_{ij} = \frac{x_i \mu_j - x_j\mu_i}{x_i \mu_j +
			 x_j\mu_i}
		    \end{array}
\end{equation}
while
\begin{equation}
   x_i^2 = x^2 + (\epsilon_i\mu_i - \epsilon_m\mu_m) (2  \omega_n \hh/c)^2
\end{equation}
Here, it is understood that $\epsilon_j=\epsilon_j(i\omega_n)$ is the complex
dielectric constant of material $j$, which is evaluated at the imaginary
Matsubara frequency  $i\omega_n= i\frac{2\pi k_BT}{\hbar} n$, while frequency
dependent magnetic permitivities $\mu_i=\mu_i(i\omega)$ are approximated to 
unity in all calculations.

Furthermore, the prime next to the sum over $n$ recalls that the $n=0$ term has
an extra factor of $1/2$, while the lower limit of integration is given
by $r_n=2\epsilon_m^{1/2} \omega_n \hh/c$. As usual, $\hbar$ is Planck's constant
in units of angular frequency, $c$ is the speed of light and $k_B$ is
Boltzmann's constant.

The DLP result \revision{as displayed in Eq. \ref{dlp} is very complicated to interpret physically}. However, to a very
good approximation one can write instead:\cite{parsegian05,macdowell19}
\begin{equation}\label{aproxsum}
     A_{1m2}(\hh) = \frac{3}{2} k_B T \sum_{n=0}^{\infty} 
      R_{1m2}(\omega_n) [1 + \nu_n \hh] e^{-\nu_n \hh} 
\end{equation}
where $\nu_n=2\epsilon_m^{1/2}\omega_n/c$ and 
\begin{equation}
   R_{1m2}(\omega_n) = 
     \frac{(\epsilon_1 -\epsilon_m)(\epsilon_2 - \epsilon_m)}
	  {(\epsilon_1 +\epsilon_m)(\epsilon_2 + \epsilon_m) } 
\end{equation} 
From the above result we see that $A_{1m2}(\hh)$ is given by a weighted sum
of the frequency contributions, $R_{1m2}(\omega_n)$. In the limit $\hh\to 0$, each
contribution has equal weight, and the Hamaker constant is obtained
as the unweighted sum of all frequency contributions. However, as $\hh$
increases, high energy contributions are \revision{suppressed} exponentially, 
and the Hamaker 
constant then is dictated by the behavior of $R_{1m2}(\omega_n)$ at
small frequencies.  For two identical materials
interacting across a medium,  the frequency dependent function 
$R_{1m1}(\omega_n)$ is always positive, so that $A_{1m1}(\hh)$ is positive at all
distances.  If the materials are different, however,
$R_{1m2}(\omega_n)$ can become negative and exhibit a complicated 
frequency dependence.  It follows that the
differences $\epsilon_i(i\omega) - \epsilon_j(i\omega)$ not only set
the scale of the van der Waals interaction, but can also determine
their sign. This is a particularly delicate problem for interfaces of
similar materials, such as that of ice and water, since
a very good accuracy is required in the determination of
the dielectric functions in order to ensure the correct sign and, accordingly,
the correct qualitative behavior of the interactions.

\section{Modeling the dielectric functions}

\subsection{Water}

A large number of parametrizations of the dielectric function of water
from the \revision{microwave (MW)} to the Extreme \revision{ultraviolet (UV)}
region may be found in the literature.
\cite{parsegian81,elbaum91b,roth96,dagastine00,fernandez00,wang17,fiedler20,gudarzi21}
However, most of the parametrizations up to
date,\cite{parsegian81,elbaum91b,roth96,dagastine00,fernandez00} rely on
the high energy band measured by Heller,\cite{heller74} which has now been
revised by  Hayashi  over considerably higher energy values.\cite{hayashi15}
Improved parametrizations using the modern data for the electronic transitions
have been proposed.\cite{wang17,fiedler20,gudarzi21} Unfortunately, the model by
Wang and Dagastine is presented only in tabulated form. A fully parametrized
form with detailed description of the MW region was presented recently,
but exhibits refractive indexes in the near \revision{IR} which are somewhat too 
large. Since the dielectric functions of water and ice are rather similar,
a precise evaluation of the refractive indexes is very important to
capture the sign of the Hamaker constant.  For this reason, we have carried
out a new parametrization for liquid water that is very similar to
that of Ref.\cite{fiedler20}, but captures the refractive indexes more
accurately.

\revision{In the supplementary material section we discuss a large number
of literature sources for optical properties of water. }
Based on that discussion, our set of absorption coefficients
for the parametrization of water at the freezing point comprises
the data of Zelsmann et al.\cite{zelsmann95} for the far-IR 
(2.4 meV to 70 meV), Wieliczka et al. (0.066 to 1.01 eV),
and the synchrotron high energy band measurements of Hayashi and Hiraoke
recommended in Ref.\cite{hayashi15,wang17,fiedler20}  
We call this the `Hayashi set'. To \revision{account} 
for the uncertainty in the temperature effect of the
high energy band, we also consider an alternative  set with the same data for 
the far-IR  to the near-UV but with Heller's high energy band instead, which
we will denote as the `Heller set'.  
In both cases we use $\epsilon(0)=88.2$ for the static dielectric constant 
at 0~C.\cite{buckley58}

\subsection{Ice}

\revision{A discussion of experimental optical properties of ice may be found in
the supplementary section.} Based on our literature survey, 
our data set for the parametrization of ice uses the compilation of
optical data by Warren, with includes the important high energy band as
measured by Seki.\cite{warren08,seki81}  For the static
dielectric constant we use the value of $\epsilon(0)=91.5$ reported by 
Auty and Cole.\cite{auty52}

\subsection{Fit to experimental data}

The selected optical data for water and ice are first modeled using
the conventional prescription due to to Parsegian and
Ninham\revision{\cite{ninham70,parsegian05}, based on
a sum of Lorentz oscillators}:
\begin{equation}
   \epsilon(\omega) = 1 +
   \sum_{k=1}^{N_{osc}}\frac{A_{k}}{1-iB_{k}\omega-C_{k}\omega^{2}}
\end{equation}
This form shows readily that  evaluation of $\epsilon(\omega)$ at
a purely imaginary frequency, say $\omega=i\xi$, provides
a well behaved real valued function,
\begin{equation}
    \epsilon(i\xi) = 1 + \sum_{k=1}^{N_{osc}}\frac{A_{k}}{1 + B_{k}\xi +
    C_{k}\xi^{2}}
\end{equation}
which can be  used for the calculation of the Hamaker function,
\Eq{dlp}.

Using this model, we performed fits for both water and ice using a total of 11
Lorentz oscillators. Six were used to fit the MW and IR regions down to
approximately 1~eV,
and five to model the high energy band in the extreme UV region and beyond.
Further details of the fit and the resulting parameters may be found in
the supplementary material.

The results of the fit to the high energy band are shown in Fig.\ref{fig:band}.
 \revision{Further details and results for a fit of the full spectrum may be found in the supplementary materials section.}

\subsection{Improved oscillator model}

Whereas the agreement of the plain Lorentzian parametrization with experimental data is fair,  we find 
for either water or ice, that the decay of the high energy band 
towards the UV region is much slower than is found experimentally. This is not 
a problem of the fitting procedure, but rather, of the Lorentzian model,
as noted already in previous 
work.\cite{dingfelder98,dagastine00,emfietzoglou07,wang17} 
We find that the only way to 
describe the slow decay of the electronic band at high energies is by use of a 
very broad Lorentzian, which then falls slowly also in the low energy region.  
\revision{The failure to reproduce this tail results in refractive
indexes in the visible (VIS) that are too high.} Since the refractive index is an
important target property, the only way that one can
remedy the problem is by deteriorating significantly the agreement with
the high energy band, either by decreasing the intensity at the maximum
or truncating the tail towards high energies. 

It \revision{appears} that the only way to remedy this problem is by introducing
oscillators that are asymmetrical. In this way one
can reproduce the slow
decay towards high energies while having a sharp decay at the beginning
of the high energy band. This strategy has been used occasionally,
by merely truncating the Lorentzian oscillators below some energy
threshold.\cite{dingfelder98,emfietzoglou07} The resulting model
for the extinction coefficient can the be integrated by use of the
\revision{proper} Kramers-Kronig relation \revision{(see the supplementary material)}, but unfortunately, the truncated
Lorentzians  can no longer
be exploited to reproduce in an easy manner the dielectric function at
imaginary frequencies.

In order to improve this situation, we seek for a modified Lorentzian
model for the extinction coefficients which can be made asymmetrical 
by use of a suitable sharply decaying `Heaviside' function, while remaining 
continuous and useful also to model the dielectric response at imaginary
frequencies.

First notice that the extinction coefficient, 
$\kappa(\omega)$ is strongly related to
$\epsilon_2(\omega)$, which reads:
\begin{equation}
   \epsilon_{2}(\omega) = \sum_{k=1}^{N_{osc}}\frac{A_{k}B_{k}\omega}{1 +
   \omega^{2}(B_{k}^{2}-2C_{k}) + C_{k}^{2}\omega^{4}}
   \label{drude_epsiiiiim}
\end{equation}
A sharp decay of $\epsilon_2(\omega)$ as observed for $\kappa(\omega)$ \revision{(see the supplementary material)} may be 
achieved by merely truncating this function beyond some threshold frequency, 
say, $\omega_0$.\cite{dingfelder98,emfietzoglou07} The truncation corresponds in practice to the
introduction of 
a modified Lorentz oscillator with a frequency dependent $B$ parameter which
remains constant for frequencies larger than $\omega_0$ and vanishes otherwise.
However, this transformation needs to be implemented in such a way that the
dielectric function at complex frequencies remains continuous.

Heuristically, we find these two conditions - vanishing of the extinction
   coefficient and continuity of $\epsilon(i\omega)$ -  may be accomplished by
substitution of the coefficient $B$ in the Lorentz oscillator by
a modified  coefficient $B(\omega) = B H(\omega)$, where
$H(\omega)$ is a 'complex Heaviside function', given by
$H = H_1(\omega) + i H_2(\omega)$. Here,  $H_1(\omega)$ is a sharply increasing
real function, which vanishes for $\omega < \omega_0$; and likewise,
$H_2(\omega)$, is a sharply decaying real function which  vanishes in the complementary region
 $\omega > \omega_0$.

With this device, the real and imaginary parts of the complex dielectric
function now read:
\begin{equation}
	\varepsilon_{1}(\omega) = 1 + \sum_{k=1}^{N_{osc}} \frac{A_{k}(1 - \omega^{2}C_{k} + \omega
	H_{2}(\omega)B_{k})}{[1 - \omega^{2}C_{k} + \omega H_{2}(\omega)B_{k}]^2 + [\omega
	H_{1}(\omega)B_{k}]^2}
	\notag
\end{equation}
\begin{equation}
	\varepsilon_{2}(\omega) = \sum_{k=1}^{N_{osc}} \frac{A_{k}\omega H_{1}(\omega)B_{k}}{[1 -
	\omega^{2}C_{k} + \omega H_{2}(\omega)B_{k}]^2 + [\omega H_{1}(\omega)B_{k}]^2}
\label{dielec_w_heavi}
\end{equation}
Where we have replaced $B_{k}$ by $B_{k}(\omega) = B_{k}H(\omega)$.
The dielectric function at complex frequency becomes then:
\begin{equation}
   \epsilon(i\omega) = 1 + \sum_{k=1}^{N_{osc}}\frac{A_{k}}{1 + B_{k}\omega
	H_1(i\omega)
	+ B_{k} \omega iH_{2}(i\omega) + C_{k}\omega^{2}}
    \label{dielec_iw_heavi}
\end{equation}

With the properties we have discussed so far, we see that if $\omega >\omega_0$,
$H_{2}(\omega)$ vanishes and both $\epsilon_1(\omega)$ and $\epsilon_2(\omega)$
recover the usual form of the Lorentz oscillator provided $H_{1}(\omega) \rightarrow 1$. On the other hand, if 
$\omega<\omega_0$, $\epsilon_2(\omega)$ vanishes altogether, as required.
Additionally, for $\epsilon(i\omega)$ to remain continuous at $\omega =
\omega_{0}$, we require $\lim_{\omega\to\omega_{0}}^{(-)}[\omega
iH_{2}(i\omega)] = \omega_{0}$, which is most easily imposed by assuming
$\omega iH_{2}(i\omega) = \omega_{0}$.

As regards the functions $H_1(\omega)$ and $H_2(\omega)$, we need them to remain real for imaginary frequencies. Furthermore, 
we require $H_{1}(\omega)$ to be symmetrical with respect to the transformation $\omega\to-\omega$, and conversely, we need $H_{2}(\omega)$ to be an odd function, so that the whole satisfies $H(-\omega) = \overline{H}(\omega)$. These set of conditions may be satisfied by the choice:
\begin{equation}
   H_{1}(\omega) =
   \frac{1}{2}\left(\tanh{\frac{\omega^{4}-\omega_{0}^{4}}{\Delta\omega}} +
	\tanh{\frac{\omega^{4}+\omega_{0}^{4}}{\Delta\omega}}\right) 
   \label{real_ASF}
\end{equation}
\begin{equation}
   H_{2}(\omega) =
   \frac{\omega_{0}}{2\omega}\left(\tanh{\frac{\omega_{0}^{4}-\omega^{4}}{\Delta\omega}}
	+ \tanh{\frac{\omega^{4}+\omega_{0}^{4}}{\Delta\omega}}\right)
   \label{imaginary_ASF}
\end{equation}
where $\Delta\omega$ is a suitable parameter that ensures a sufficiently fast decay of the extinction coefficient. Henceforth, we call $H(\omega) = H_{1}(\omega) + iH_{2}(\omega)$, with its real and imaginary parts given by Eq. \ref{real_ASF} and Eq. \ref{imaginary_ASF}, the Complex Step Function (CSF).

\begin{figure}[htb!]
\includegraphics[width=0.48\textwidth,height=0.35\textwidth,keepaspectratio]{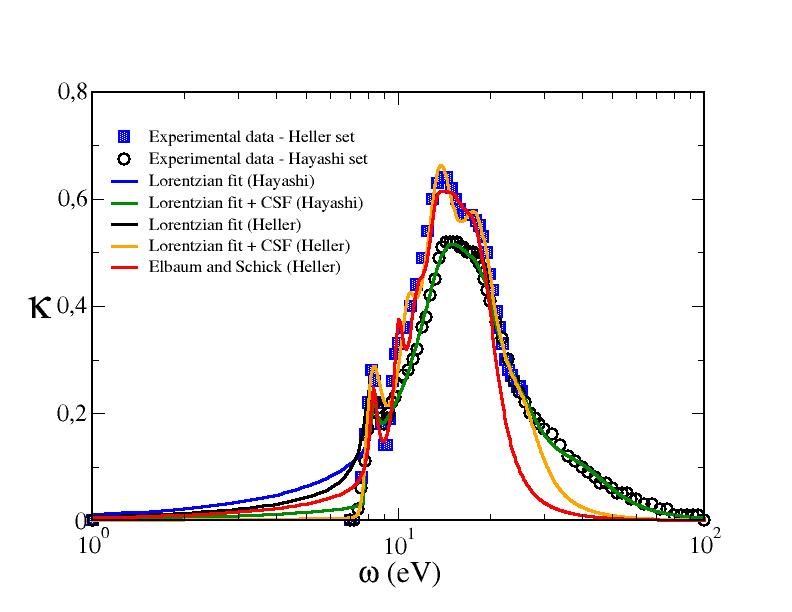}
\\
\includegraphics[width=0.48\textwidth,height=0.35\textwidth,keepaspectratio]{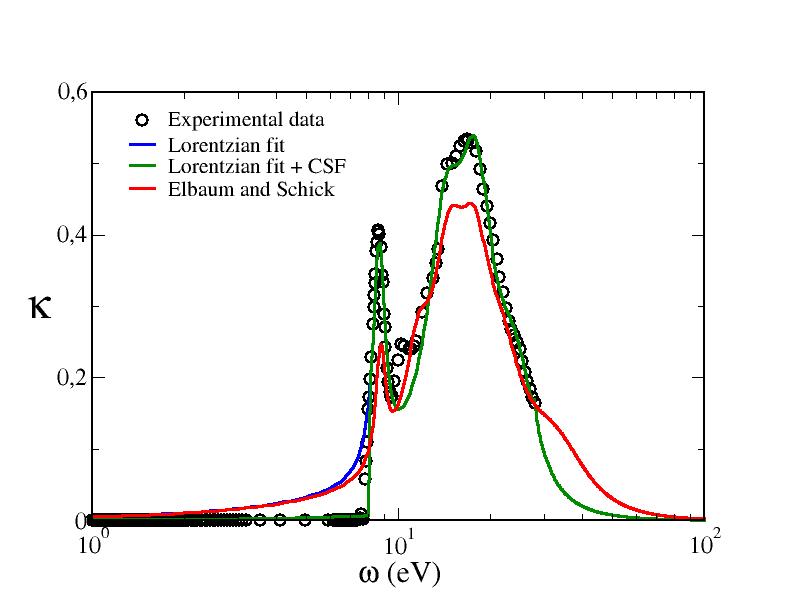}
        \caption{\revision{High energy band of the extinction index ($\kappa$).
The figure compares experimental data (black dots) and parametric
representations with Lorentzian oscillators (lines), as indicated in the legends}. Experimental results for both water (top) and ice (bottom)  are compared to the Lorentz model and the Lorentz-CSF model from this work. Fits of experimental data by \revision{ES} are also displayed.}
\label{fig:band}
\end{figure}

The improvement in the description of extinction coefficients by the modified
oscillator is illustrated in Fig. \ref{fig:band}. Clearly, the fit to the electronic band
remains as good towards high energies, but the model now provides a sharp decay of the band towards small energies as observed in experiment. This is very convenient, because one can now improve any fit of Lorentz oscillators
merely by transforming \revision{the constant coefficient $B$ in the Lorentzian oscillator (see the supplementary material)}, into the modified coefficient $B(\omega) = B H(\omega)$. Therefore, the original fitting parameters remain unchanged, and only the cutoff frequency $\omega_0$ and the decay parameter $\Delta\omega$ need to be added. 
Thanks to this device, we can use an accurate parametrization of the extinction coefficients based on the Lorentz oscillator
to obtain in a simple manner and with real algebra the sought dielectric
function at imaginary frequency $\epsilon(i\omega)$. Unlike the Brendel-Borman oscillator,\cite{brendel92} the nature of our 
model guarantees that optical properties remain meaningful at $\omega=0$, and the correct symmetry of $\epsilon_1(\omega)$ and $\epsilon_2(\omega)$ is preserved.  

\begin{figure}[htb!]
        \includegraphics[width=0.48\textwidth,height=0.35\textwidth,keepaspectratio]{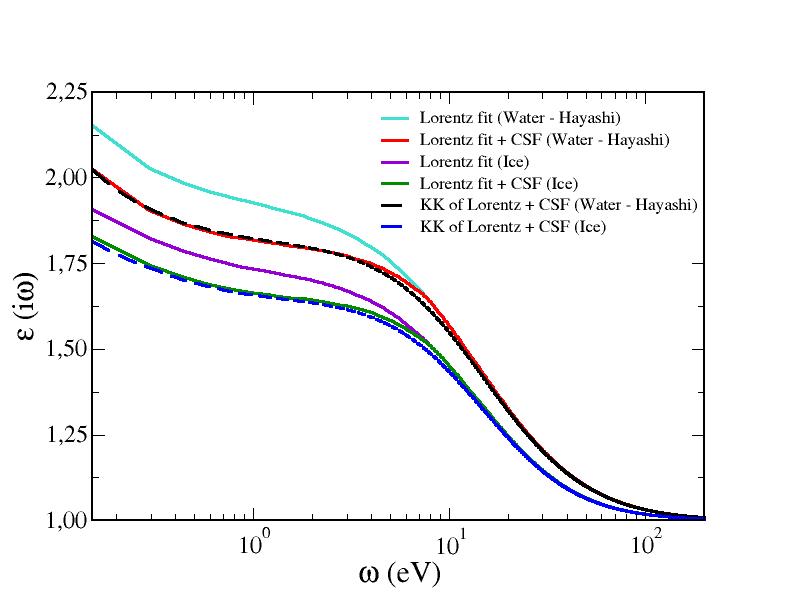}
\\
	\includegraphics[width=0.48\textwidth,height=0.35\textwidth,keepaspectratio]{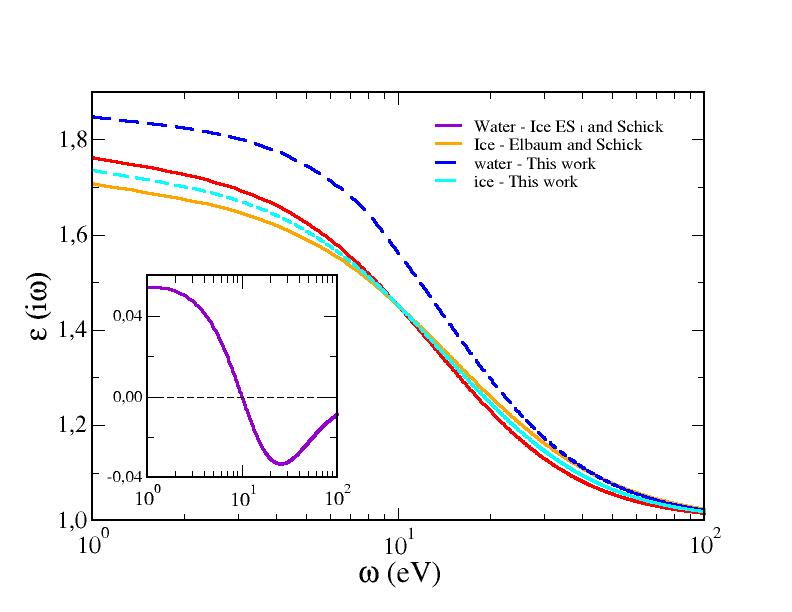}
	\caption{Dielectric response at imaginary frequencies of liquid water
	   (Hayashi set) and
	ice. (Top) Black and blue dashed lines represent $\epsilon(i\omega)$
   obtained from the KK relations over the parametric representation with CSF of
$\kappa$ in this work. The accordance with their corresponding parametric
representation of $\epsilon(i\omega)$ illustrates that although the model with
CSF \revision{does not rigorously obey the Kramers-Kronig relation}, 
it provides an accurate description of the
dielectric response. (Bottom)  A similar representation using the Heller set for
	water. The parametrization from \revision{ES} predicts $\epsilon(i\omega)$
that is smaller for water than for ice at high energies. \revision{This is illustrated in the inset, where we plot the difference between the dielectric response of water and ice.}
In contrast, our revised parametrization of
similar experimental data does not support that prediction and is
positive all the way from MW to higher energies.
\label{fig:dielectric}
}
\end{figure}

Unfortunately, despite these advantages, the model 
\revision{does not strictly obey the Kramers-Kronig relations, which
is a physical constraint that dielectric functions must obey}.
It appears that 
to have a sharply decaying model that is accurate \revision{and obeys
Kramers-Kronig} one cannot
avoid the use of special functions with no simple analytical form for
$\epsilon(i\omega)$.\cite{orosco18b} 
In practice, \revision{the deviations from Kramer-Kronig are very small}.
Fig. \ref{fig:dielectric}-top compares $\epsilon(i\omega)$  obtained
in analytical form from \Eq{dielec_iw_heavi}, 
with the Kramers-Kronig transformation
of the parametric representation of $\kappa(\omega)$ computed through its relation with $\epsilon_{1}(\omega)$ and $\epsilon_{2}(\omega)$. The two curves are
clearly very similar on the scale of the figure and  differ at most by
3\%.  In the same figure we also show the results
obtained from the plain Lorentz model. The curves are almost
identical for energies above the first electronic excitation, but 
differ significantly for lower energies. Thanks to the truncation
of the  Lorentz oscillators, the refractive indexes are now significantly lower
and much closer to experimental results. For water, the Lorentz model at
$\lambda=1000~nm$ provides a refractive index of 1.38, while the Lorentz+CSF
model yields 1.34, far closer to the experimental value of 1.33 
at 0 degrees.\cite{lide94}

For ice, the Lorentz model yields
1.31, while the Lorentz+CSF model yields 1.29, 
to be compared with the experimental value of 1.30 at T=266~K.\cite{warren08}
These considerations provide confidence on our parametrization, particularly for
the important region between the near \revision{IR} and the soft \revision{x-rays (XR) regions}. \revision{Innaccuracies} could occur for
the high energy tails beyond ca.40~eV, particularly for ice, because of lack of
data, but these tails contribute little to the overall result of the Hamaker
constant.

The results for the parametrization of dielectric properties of ice and water 
described here have been  employed prior to publication in 
Ref.\cite{luengo21,li22}


\section{Results}

We now use the fits for the complex dielectric function based 
on the Lorentz+CSF in order to calculate the Hamaker functions for a
number of relevant interfaces involving water and ice. Unless otherwise
stated, we describe the dielectric properties of water as obtained from
fits to the Hayashi set.

\subsection{Interaction of water and ice across air}

\begin{figure}[htb!]
        \includegraphics[width=0.48\textwidth,height=0.35\textwidth,keepaspectratio]{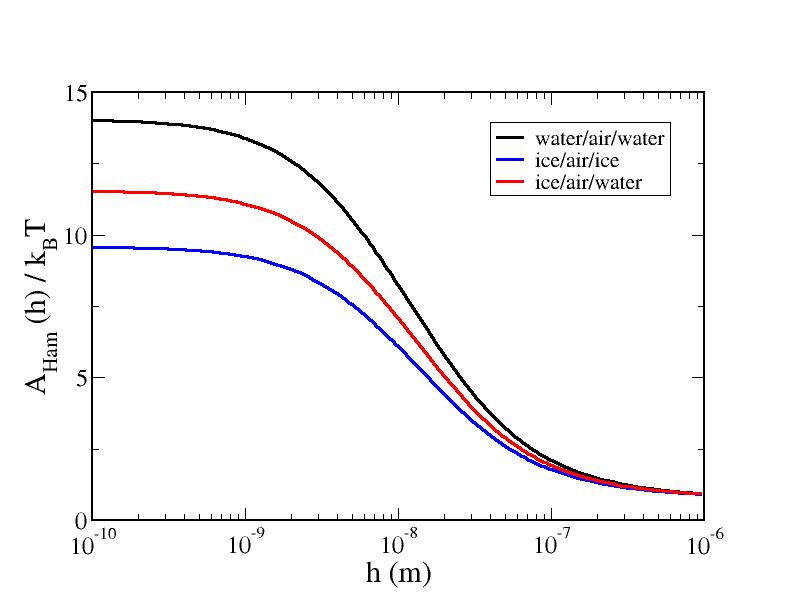}
	\caption{Hamaker coefficients of condensed water phases separated by air.
Results are shown for water/air/water (black), ice/air/ice (blue) and
ice/air/water (red). In all three cases, the Lorentz model + CSF of this work has been used, with the parameterization based on the Hayashi set for liquid water.}
	\label{fig:hamaker_air}
\end{figure}

Whereas our main goal is the study of ice/water interfaces, we first consider
the simpler systems that result from the interaction of either ice or water 
across air, i.e., water/air/water, ice/air/ice and ice/air/water. These
cases pose less problems than systems where ice and water are in contact.
According to DLP theory, Hamaker functions are given by differences of
the form $(\epsilon_i - \epsilon_m)$, with the index $i$ corresponding to
media 1 or 2 interacting  across medium $m$. Here, $m$ simply corresponds to
air (or water vapor), and we can safely assume $\epsilon_m=1$. Accordingly,
the Hamaker function is given by factors of the form $(\epsilon_i-1)$,
which are positive at all frequencies. This implies that
there could be some discrepancies on the actual value of the Hamaker
function depending on the parametrization of dielectric properties, but
there can be no controversy as regards its sign, which must
always be positive (i.e. corresponding to attraction between two identical bodies
across air).

Fig. \ref{fig:hamaker_air} displays our results for the interaction of water-water, ice-ice and
ice-water slabs. Of course, 
 we find that the Hamaker functions are positive irrespective of the
distance of separation, $\hh$, between the condensed media.

At small distances, $A(\hh)$  remains flat up to
about 1~nm. The  extrapolation of this function towards zero distance
 provides what is conventionally known as the Hamaker constant in
the chemical physics literature.\cite{israelachvili11} A fact that is less
often appreciated is that the Hamaker function gradually decreases
as $\hh$ increases, and eventually adopts a much smaller constant value 
corresponding 
exactly to the $n=0$ contribution of the sum in \Eq{hamaker} (c.f.
\Eq{aproxsum}). 
i.e., whereas both in the limit $\hh\to 0$ and $\hh \to \infty$ the
free energy follows the power law $g(\hh)\propto h^{-2}$, the corresponding
proportionality coefficients are completely different. The former is set
by the scale of the first electronic excitation, which usually falls in
the ultraviolet region or beyond; the second one is of
order $k_BT$, whence, usually two orders of magnitude smaller at ambient
temperature. 

This behavior may be rationalized by noting that the Hamaker
function may be split into a static ($n=0$) and a frequency dependent
contribution as $A(\hh) =  A_{\omega=0} +  A_{\omega>0}(\hh)$, with
$A_{\omega=0}$, a constant of order $k_BT$:\cite{israelachvili11}
\begin{equation}
	A_{\omega=0} = \frac{3}{4}\left[  
   \frac{(\epsilon_1(0) -\epsilon_m(0))}{(\epsilon_1(0)+\epsilon_m(0))}
	\frac{(\epsilon_2(0) -\epsilon_m(0))}{(\epsilon_2(0)+\epsilon_m(0))}\right] k_B T 
\end{equation} 
and $A_{\omega>0}(\hh)$, a
function of $\hh$ which is finite at $\hh\to 0$ and vanishes at $\hh\to\infty$.
Whereas this trend may not be obvious from direct
inspection of either \Eq{hamaker}, or its simplified form, \Eq{aproxsum}, we note that
the limiting asymptotic behavior and the crossover from the $\hh\to 0$ and
$\hh\to\infty$ regimes are described qualitatively by the approximate
relation:\cite{macdowell19}
\begin{equation}\label{simple-1}
\begin{array}{ccc}
	 A_{\omega > 0}(\hh) & = & \displaystyle{ \frac{3\hbar
	    c}{32\sqrt{2}\,n_m\hh}
           \left ( \frac{n_1^2-n_m^2}{n_1^2+n_m^2}
		  \frac{n_2^2-n_m^2}{n_2^2+n_m^2}
	  \right ) } \\ & & \\ & &
        \left [
		 (2+\frac{3}{2}\nu_T \hh) e^{-\nu_T \hh} - (2 +
	 \nu_{\infty} \hh ) e^{-\nu_{\infty} \hh}
				        \right ]
     \end{array}
\end{equation}
where $\nu_T$ and $\nu_{\infty}$ are wave numbers that set the relevant 
length scales governing the behavior of $A(\hh)$. 
\revision{ 
The first wave number,
$\nu_T=4\epsilon_{m}^{1/2}\pi k_{B}T/\hbar c$ is known exactly.
At water's triple point,
it falls in the near IR, and sets the length-scale where retardation effects
become exponentially suppressed, $A_{\omega>0}\to 0$ and so $A(\hh)\to
A_{\omega=0}$ for $\hh$ larger than the micrometer. 
}

\revision{The second length-scale,
$\nu_{\infty}$ is an empirical parameter that falls in the UV region
and sets the crossover from the non-retarded regime, with interactions 
falling as $1/\hh^2$, to the retarded regime, where $A_{\omega>0}(\hh)$ becomes dominated by a decay of order $1/\hh^3$ (the Casimir regime). 
Generally, $\nu_{\infty}$ can be calculated numerically \cite{luengo21}, 
but  a good approximation is given by:\cite{macdowell19,luengo22}
\begin{equation}
   \nu_{\infty} = 4 n_m\frac{(n_1^2+n_m^2)^{1/2}(n_2^2+n_m^2)^{1/2}}
             {(n_1^2+n_m^2)^{1/2}+(n_2^2+n_m^2)^{1/2}} \frac{\omega_e}{c}
 \label{nuinf}
\end{equation} 
where $\omega_e$ is of the order of the principal electronic excitation,
and $n_i$ is the refractive index of medium $i$.
}

For most practical matters, in the range of distances smaller than 
1~nm, the Hamaker function might be approximated to a constant
$A(0)$, which, for the interaction between either bulk water or ice
slabs is of the order of a few tens of zJ \revision{(1~zJ$=10^{-21}\ J$)}. 


\subsection{Interactions of the Ice/water/air system -- Premelting}

\begin{figure}[htb!]
        \includegraphics[width=0.48\textwidth,height=0.35\textwidth,keepaspectratio]{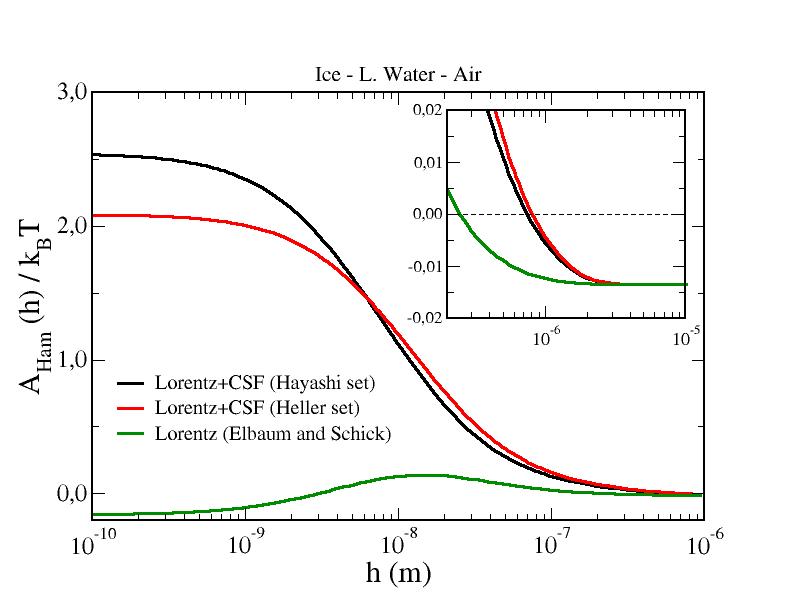}
        \caption{Hamaker coefficients for a premelting liquid film between ice
and water. The figure shows result of the present work
employing the dielectric function of liquid water from 
Hayashi (black) and Heller (red) sets. The green line displays the result 
	of the \revision{ES} parametrization for the Heller set.
	\revision{The inset shows the Hamaker function
		for premelting thickness in the scale of the micrometer.}
	}
	\label{fig:hamaker_ice-water-air}
\end{figure}

We now turn to the more subtle problem of van der Waals interactions relevant
to ice premelting.  Here, we study how the surface free energy
of a liquid film of water intruding between bulk ice and air  depends
on the liquid film thickness, $\hh$. In this case, the Hamaker function is
dictated mainly by the difference between the dielectric functions of ice
and water, which are very similar. Accordingly, not only the scale of the
Hamaker function, but even its sign, depends crucially on an accurate estimation
of the dielectric properties.

A look at Fig. \ref{fig:dielectric} shows that in all
the relevant range of the electromagnetic spectrum above the microwave region,
the complex dielectric function at imaginary frequencies, $\epsilon(i\omega)$ 
is larger for water than it is for ice in our parametric representation, so that we can expect \revision{right} away that the Hamaker
function will  be positive.

Fig. \ref{fig:hamaker_ice-water-air} displays the Hamaker function of ice/water/air system versus
water layer thickness, and confirms this expectation.  Results are shown for the water
dielectric function as obtained from both the Hayashi and the Heller set, in
order to account for possible uncertainty due to the choice of experimental
 dielectric properties.
Although some differences are observed, we see that in both cases the
Hamaker function is positive and presents a \revision{monotonic} decay all the way from
vanishing layer thickness to the micrometer range. 

Again, the extrapolation of the Hamaker function to zero separation provides
the Hamaker constant, which, in this case, is much smaller than that found
for condensed water phases interacting across air, since the
difference $\epsilon_i - \epsilon_m$, with $m$ corresponding to water, is
now very small. Importantly, we also note that the Hamaker function starts to
decrease significantly for thicknesses barely beyond the nanometer range.
This behavior results from the large polarizability of the intervening phase
between ice and air, i.e., water.  Indeed, the value of $\nu_{\infty}$ in
\Eq{simple-1}
may be estimated approximately from \Eq{nuinf},
whence, compared to the interaction between condensed water phases across
air, we see that now $\nu_{\infty}$ increases by a factor of about $n_w$,
implying a faster decay of the Hamaker function (c.f. \Eq{simple-1}).

\revision{For water films thicker than 1 micrometer}, Fig. \ref{fig:hamaker_ice-water-air}  shows that the Hamaker
function intersects the zero axis and becomes negative. This interesting
behavior may be understood from  \Eq{simple-1}, which shows that for 
distances larger than $\nu_T^{-1}$, $A_{\omega>0}(\hh)$ vanishes altogether,
 and only the $n=0$ term of the Hamaker function remains. Compared to
contributions for $n>0$, this term has opposite sign, since the static
dielectric function of ice is larger than that of water. As a result, 
van der Waals interactions oppose the growth of wetting films of small thickness,
but favor growth of thick wetting films beyond the micrometer ($\hh > 10~\mu~m$). It must be
understood, however, that the intensity of van der Waals forces at such
distances is extremely small, and whatever small perturbation, such as
dissolved gases,  \revision{electrolytes} or minute changes away from the triple point
could easily change the overall free energy balance.\cite{elbaum93,wettlaufer99}

Our predictions differ very much from the influential work of
Elbaum and Schick, who first called the attention on the significance of van der
Waals interactions in the study of ice premelting.\cite{elbaum91b} The
Hamaker function predicted by these authors -- Fig.\ref{fig:hamaker_ice-water-air} (green lines) --
is negative in the sub-nanometer range and positive in
the nanometer range, then negative again at distances beyond the decade
of micrometer. 

The \revision{non-monotonic} behavior predicted in Ref.\cite{elbaum91b} 
can be traced to the  
parametrization of the  complex dielectric functions in that work,
which predict $\epsilon(i\omega)$ that is larger for ice than for water
at high energy, as seen in Fig. \ref{fig:dielectric}-bottom. However, we can see clearly in Fig. \ref{fig:band}-top 
that the parametrization performed by \revision{ES} fails to
describe correctly the target high energy band of the Heller set. On the one
hand, it appears to truncate the high energy tail in water's extinction
coefficient, and in the other hand, it exhibits too slow a decay for the
same portion of the extinction coefficient of ice (i.e..  
the high frequency range of $\epsilon(i\omega)$ in water is underestimated,
while the corresponding portion in ice is overestimated),
resulting in the complex dielectric function at imaginary frequencies which
is larger for ice than it is for water. Such behavior does not appear to
be supported by the current experimental data. Our revised parametrizations
for ice and water (Heller set) using similar data as \revision{ES} provide
dielectric functions at imaginary frequencies that are always larger for water
than for ice. 

This result is in agreement with expectations from the f-sum rule
and the Lorentz model of dielectric response.\cite{tanner13}
At very high energies one expects a decay of the complex dielectric
function of the form $\epsilon(i\omega) \propto \omega_p^2 / \omega^2$,
where $\omega_p \propto \rho_e$ is the plasma frequency and $\rho_e$ is the electron density in the material. Since
the density of water is larger than that of ice, one expects that the plasma
frequency should be larger for the former than it is for the latter, and
accordingly, that $\epsilon(i\omega)$ should remain larger in water than in 
ice also at high energies.

In order to further clarify this problem, we calculated the optical
properties of ice and water using Density Functional Theory. 
Previously, we showed that this theory does a  good job
at qualitatively describing the main differences in the absorption spectrum
of ice and water in the region of electronic excitations. We use the
Kramers-Kronig relations and the synthetic spectrum obtained
from DFT in order to assess  $\epsilon(i\omega)$ in an independent 
manner.  

As a check of the theoretical calculations, we note that
the predicted indexes of refraction are in fairly good agreement with
experiment. For water, we obtain $n_w=1.31$, compared with the experimental
value of $n_w=1.33$. For ice, DFT yields $n_i=1.28$, compared with the
experimental value $n_i=1.30$. Whence, the refractive indexes are somewhat
too low, but in the correct order.  

\begin{figure}
  \includegraphics[width=0.48\textwidth,height=0.35\textwidth,keepaspectratio]{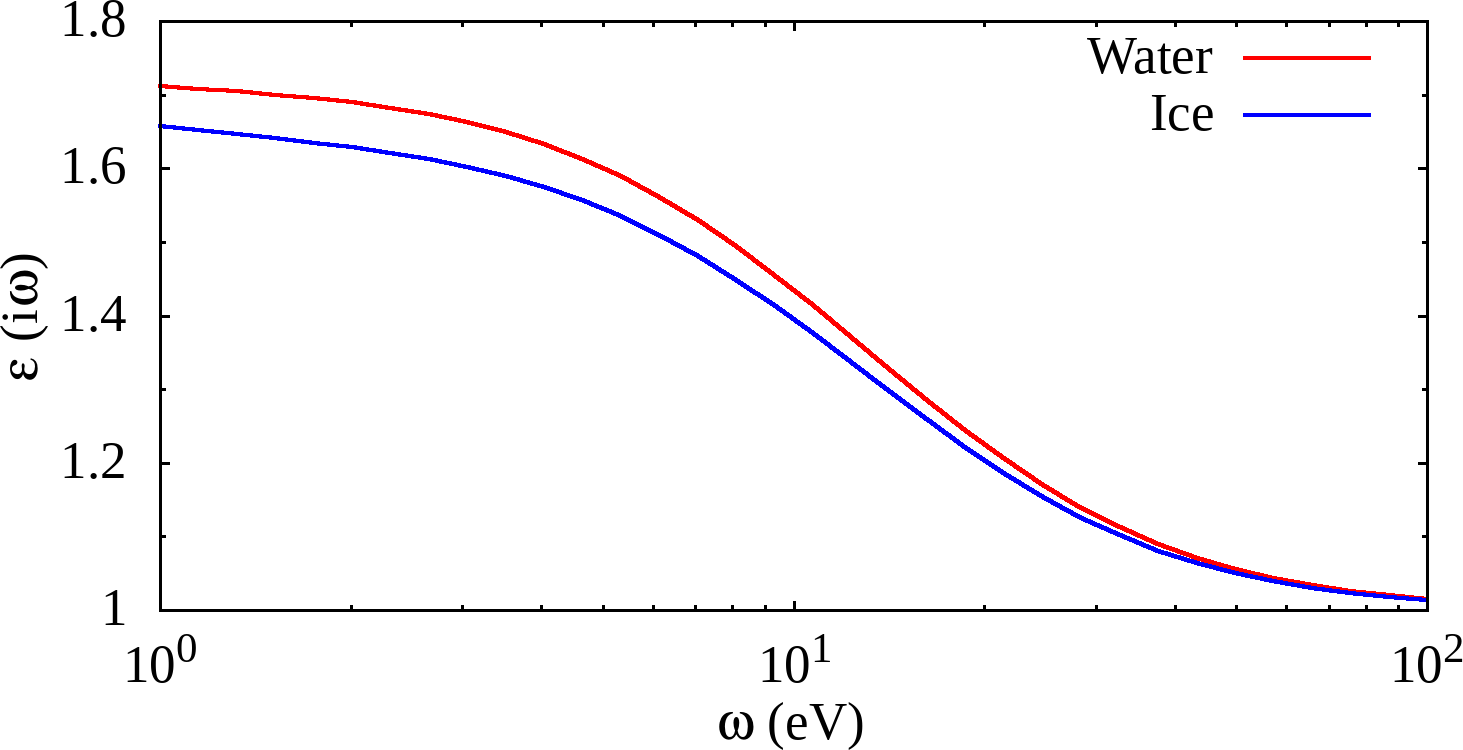}
\quad \\
\quad \\
  \includegraphics[width=0.48\textwidth,height=0.35\textwidth,keepaspectratio]{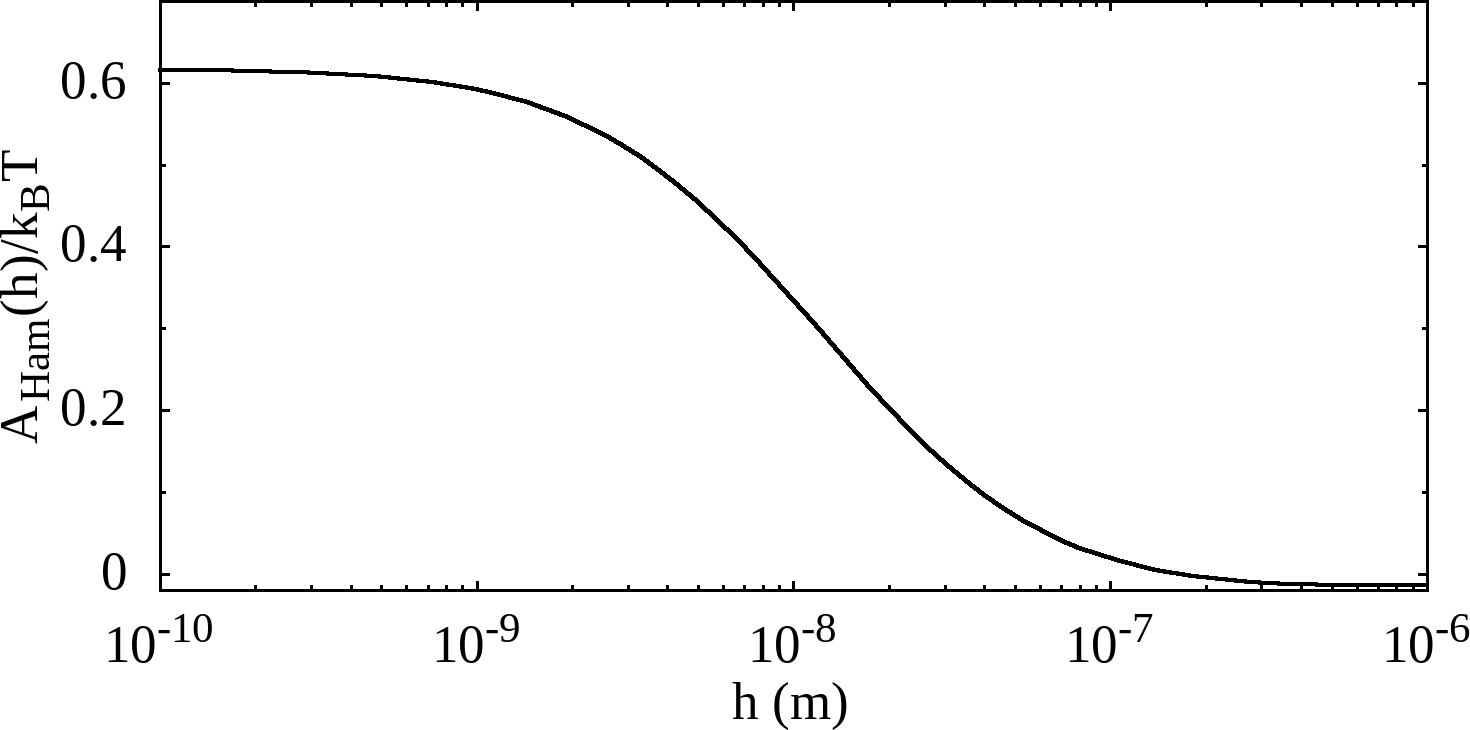}
  \caption{Dielectric response at imaginary frequencies of liquid
water and ice computed in the DFT approximation(top). We can observe that the
dielectric function of water remains always at bigger values than ice in the
high energy region. This is in agreement with our new fits to the experimental
data. 
Hamaker coefficients for a premelting liquid film between ice and
water(bottom). The figure shows results of the calculations using the dielectric 
functions calculated in the DFT approximation.  }
	\label{fig:epsimagdft}
\end{figure}

The complex dielectric function at imaginary frequencies in the UV is displayed
in Fig. \ref{fig:epsimagdft}, and it is seen that it remains
higher for water than for ice in all the UV region and beyond,
in agreement with expectations from the Lorentz model, and fits
to the experimental results. In line with predictions of the 
refractive indexes, the $\epsilon(i\omega)$ differ very little, and
provides a Hamaker functions that is about   an order of magnitude
smaller than predicted by  the Lorentz-CSF fit. However, on qualitative grounds we see that the Hamaker function remains positive
everywhere in the nanometer range.

In summary, we find that fits with a Lorentz model of two different
experimental sets for the dielectric response of water, as well as 
theoretical DFT calculations predict
an optical response in the ultraviolet region that is always higher
for water than for ice, resulting in a positive Hamaker constant for
the adsorption of a liquid water layer intervening between bulk ice
and air. 

\subsection{Implications \revision{for} past work}

From the discussion above, we see that the current improved understanding
of the role of van der Waals forces  on ice premelting differs qualitatively
from the early predictions of Elbaum and Schick.\cite{elbaum91b} This
work, henceforth referred as ES, has been very influential and the results used regularly on a number
of studies,\cite{wilen95,dash06,french10,limmer14,bostrom17,thiyam18} including work
from some of us,\cite{benet19,esteso20,sibley21} so we devote here a few lines 
to discuss how this could affect currently published results.

As far as physical implications are concerned, the ES model of van der Waals
interaction predicts an interface potential with a minimum, implying incomplete
surface melting. On the other hand, our work shows a negative \revision{monotonic}
contribution of van der Waals forces that completely inhibits surface
premelting. Fortunately, the situation is not as bad as it appears. At short
distances, the surface interactions are no longer dominated by van der Waals
forces. Instead, they are governed by short range structural forces related
to the packing of water molecules on the solid
substrate.\cite{derjaguin87,henderson94,henderson05} Using a square gradient model together with
molecular simulations of the mW model, Limmer and Chandler 
showed that at short range, packing effects promote surface 
melting.\cite{limmer14}  The use of the mW model here is very convenient,
because  dispersion contributions are truncated at very short distances, so the results from this
model can be used as a proxy of the effect of short range forces without
any possible entanglement of long van der Waals tails. Similarly, in our recent
work, we calculated the interface potential from simulations for the TIP4P/Ice
model, and found clear evidence of a structural contribution promoting surface
melting.\cite{llombart20,sibley21} Accordingly, as we will discuss in detail
later, the addition of short range contributions to the van der Waals tail
produces an interface potential with a minimum, in qualitative agreement with
the findings of ES. It must be made clear, however, that the origin of this
minimum does not stem from van der Waals forces alone, as implied by
ES. It is  a compromise between \revision{opposing} short range structural forces and 
long range van der Waals forces.

\subsection{Interactions of the water/ice/air system -- Surface freezing}

\begin{figure}[htb!]
        \includegraphics[width=0.48\textwidth,height=0.35\textwidth,keepaspectratio]{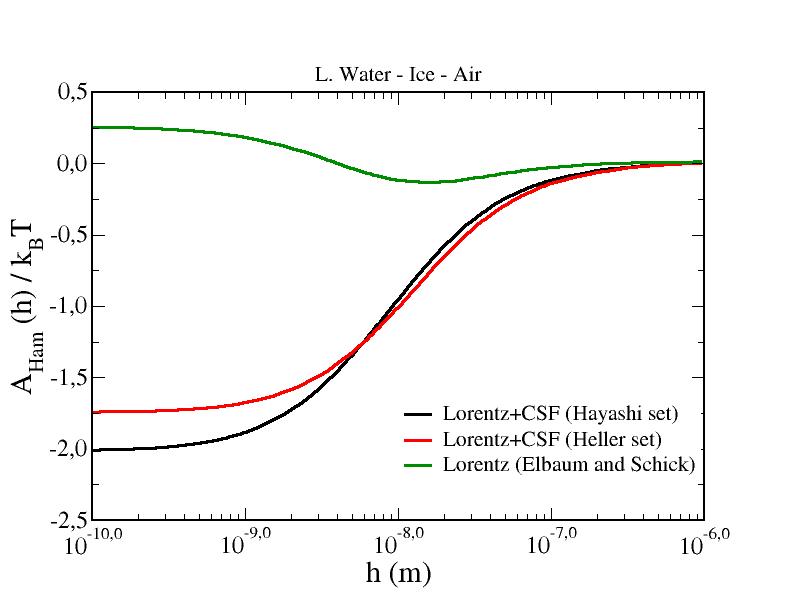}
	\caption{Hamaker coefficients from nanometers to microns compared to the thermal energy for water/ice/air. Black line and red line show the result of the present work employing, respectively, the dielectric function of liquid water with Hayashi and Heller sets. Green line displays the result of \revision{ES}.}
        \label{fig:hamaker_water-ice-air}
\end{figure}

The parametrization used above for the dielectric functions of ice and water
also allows us to study the free energy of ice films formed between bulk
phases of water and air. The results obtained using either the Hayashi and
the Heller sets are shown in Fig. \ref{fig:hamaker_water-ice-air}. We see that the Hamaker function that
results is negative in the relevant range between vanishing ice thickness down
to
the micrometer range, whereupon, it changes sign and becomes positive. In fact,
the behavior observed in this case is nearly a mirror image of that observed
for the ice/water/air system, since the interactions again are mainly governed
by the  difference $\epsilon_w - \epsilon_i$. i.e.,  the same factor
governing the water/ice/air system, but with opposite sign.

This expectation is confirmed in Table \ref{table:hamaker_constants}, where we see that the Hamaker
constant for the system water/ice/air is very similar, but of opposite sign than
that of ice/water/air. As noted before, both are significantly smaller than
the Hamaker constants for condensed phases of water across air. It is worth to
remark that, although DFT computed values of the Hamaker constant result in
sensibly smaller values in all cases, the relative order of the results and the
order of magnitude is conserved.

Our results differ dramatically from expectations based on the \revision{ES}
parametrization, which yields instead a positive Hamaker constant, and thus,
the prediction of  complete suppression of surface freezing.\cite{elbaum91c}

\begin{table}
   \begin{tabular}{c|c|c|c|c|c|c}
  system  & w/a/w     &  i/a/i    & i/a/w     & i/w/a     & w/i/a   & i/w/i and w/i/w   \\
  \hline
  A/zJ    &  52.77    &   36.07   &  43.46    &   9.54    &  -7.58  & 2.02              \\
  \hline
  DFT     & 34.93     &   30.54   &   32.66   &   2.30    & -2.12   & 0.19              \\
  \hline
\end{tabular}
\caption{Table of Hamaker constants involving ice and liquid water as 
obtained from DLP theory with the parametrization of dielectric properties
obtained in this work (using the Hayashi set to model water). Results
are given in units of \revision{zJ=$10^{-21}$J}}
	\label{table:hamaker_constants}
\end{table}

\subsection{Interactions across a condensed phase}

\begin{figure}[htb!]
        \includegraphics[width=0.48\textwidth,height=0.35\textwidth,keepaspectratio]{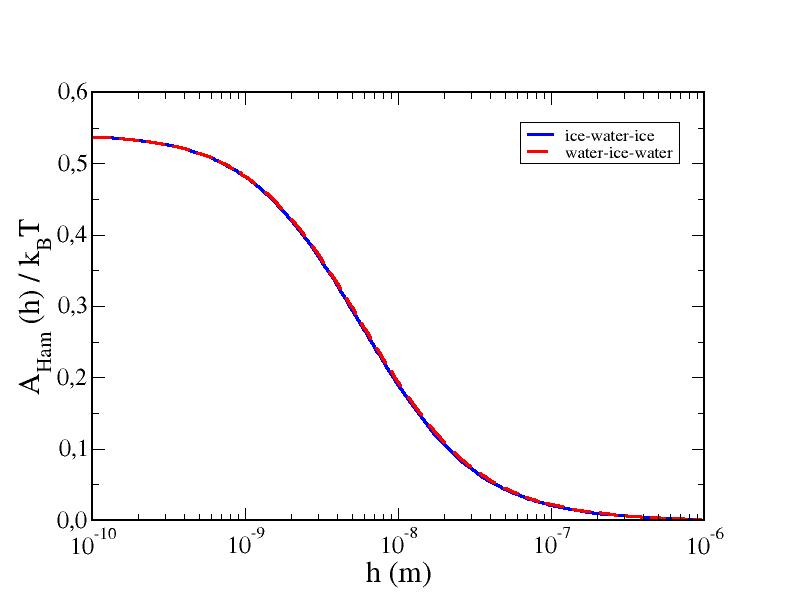}
        \caption{Hamaker function for ice/water/ice and water/ice/water as
	     obtained from the Hayashi set. The continuous blue line corresponds 
	     to ice/water/ice and the dashed red line to water/ice/water. 
        \label{fig:hamaker_iwi}
     }
\end{figure}

As a final result, we now discuss the van der Waals forces when all the bodies
involved are condensed phases, i.e., the interaction of two ice slabs with water
in between (ice/water/ice), and the complementary case of two water slabs 
interacting across ice (water/ice/water). Since both of these settings refer to two identical interacting bodies
across a third medium, van der Waals forces always conspire in favor of the 
two bulk materials to adhere, or, alternatively, the layer in between to vanish.

Figure \ref{fig:hamaker_iwi} shows the Hamaker functions for both of these
cases. As expected, the results are almost identical 
and the Hamaker function is always positive, implying attraction of the two bulk
bodies.
Because the intervening layer is a condensed phase, note that now the decay of 
the Hamaker function starts much sooner than
in the case of two solids interacting across air. Particularly, we see that the
Hamaker function has decayed by 10\% already at a distance of 1 nm.
Additionally, since all three bodies involved are condensed phases of water with
similar refractive indexes, the Hamaker constant is now much smaller than in
previous cases, $A_{ham}=2.02$~zJ. However, as long as the Hamaker function
is positive, the two bulk materials will decrease their free energy by 
decreasing the thickness of the intermediate layer.

\subsection{Comparison with empirical force fields}

Admittedly, the results for the Hamaker constants described above are obtained
after rather involved numerical calculations within the framework of DLP theory,
which accounts explicitly for polarization effects. An interesting question then
is whether simple point charge molecular models that are widely used can
possibly describe the apparently complex behavior embodied in DLP. 

In practice, for non polarizable potentials interacting with the usual
dispersion tail, $u_{ij}(r)\propto -C_{ij}/r^{-6}$, the Hamaker 
constant for the adsorption of phase $m$ at the interface between
phase, $1$, and phase $3$  may be estimated
accurately from a plain sharp-kink approximation of the density profiles
as:\cite{dietrich86,dietrich91,macdowell17}
\begin{equation}
 \label{ska0}
   A_{1m3} =  4\pi^2(C^{1/2}_{11}\rho_{1} - C^{1/2}_{mm}\rho_{m})
   (C^{1/2}_{33}\rho_{3} - C^{1/2}_{mm}\rho_{m} )
\end{equation} 
where $\rho_i$ are bulk number densities of the phases involved, and we have assumed $C_{ij}=C_{ii}^{1/2}C_{jj}^{1/2}$.

In the case of interest here, a single component system at the triple point, all phases are formed from water, so that all $C_{ii}=C$.
In most point charge models, the contribution of electronic \revision{polarizabilities} to
the constant $C$ is described as $C=4\epsilon\sigma^6$, with $\epsilon$ and
$\sigma$ the usual Lennard-Jones parameters (an additional Keesom like term that
has been neglected here contributes to the $n=0$ static term only). Whence, the electronic Hamaker constant
for the growth of a water film in between ice and vapor simplifies
to:\cite{benet19}
\begin{equation}
   A^e_{iwa} = 4\pi^2 \epsilon \sigma^6 (\rho_v - \rho_w)(\rho_i - \rho_w)
   \label{ska}
\end{equation} 
where the superscript emphasizes that this expression accounts only for dispersion interactions due to electronic polarizabilities.
Since the density of ice is smaller than that of water, we find readily
that $A^e_{wia}>0$, in agreement with the far more involved DLP theory. 
For the related system of ice growing between liquid water and air, the
Hamaker function is the same as above,
albeit with the interchange of ice and water labels. 
Accordingly,  we find exactly the same result, but
with opposite sign, which is also consistent with results from DLP theory. 

\Eq{ska0}-\ref{ska} above show  that the positive sign of $A_{iwa}$ results from the fact that $\rho_i<\rho_w$,  so that the 
absence of complete surface premelting in ice is actually one more of water's
anomalies, a result in agreement with expectations by Nozieres,\cite{nozieres92} but at odds with claims by Fukuta.\cite{fukuta87} On the contrary, noble gases such as Neon and Argon, where the solid density is larger than that of the liquid phase, exhibit surface melting.
The result of \Eq{ska}, firmly rooted in the theory of
wetting,\cite{dietrich86,dietrich91} contrasts with attempts to determine intermolecular forces in premelting films 
with no account  of density differences between the involved
phases.\cite{fukuta87,makkonen97}

A compilation of the electronic Hamaker
constants for  interactions involving ice, water and air  obtained from \Eq{ska0} for different non-polarizable water models
are shown in table 
\ref{table:hamaker_comparison}.\cite{jorgensen83,berendsen87,abascal05,abascal05b,henriques16}.
The results are compared with predictions from DLP theory. The electronic
Hamaker constants from DLP are calculated using the parametrized dielectric
response stemming from electronic contributions only (i.e., oscillators with
absorption frequencies larger than  the \revision{near IR}.

For interactions of condensed water phases across air,
Table~\ref{table:hamaker_comparison} shows that the old
generation of water models (TIP4P and SPC/E) appear to perform
rather well, with values of electronic Hamaker constants rather close
to predictions from DLP theory. Surprisingly, the new generation of
force fields appear to overestimate considerably the dispersion interactions,
with TIP4P/2005 performing significantly better than TIP4P/Ice and TIP4P-D.
On the other hand, for interactions between a condensed phase and air
(ice/water/air and water/ice/air), all force
fields predict interactions that are too weak compared with DLP.

\begin{widetext}

\begin{table}
\begin{tabular}{l|ccccc|c}
 model & TIP4P & SPC/E & TIP4P/2005 & TIP4P/Ice & TIP4P-D &  
 DLP \\
 \hline
 $A_{iwa}^e/$~zJ & 3.83 & 3.92 & 4.62 & 5.33 & 5.65 & 9.57 \\
 $A_{wia}^e/$~zJ &-3.52 &-3.63 &-4.24 &-4.90 &-5.19 & -7.59 \\
 $A_{waw}^e/$~zJ & 46.7 & 47.9 & 56.3 & 65.04 & 68.8 & 49.1 \\
 $A_{iai}^e/$~zJ & 39.3 & 40.4 & 47.4 & 54.8 &  58.0 & 32.4 \\
 $A_{iaw}^e/$~zJ & 42.8 & 44.0 & 51.7 & 59.7 &  63.2 & 39.8 \\
\end{tabular}
\caption{Electronic contribution to Hamaker constants in point-charge models. Results for the models are obtained using \Eq{ska}, 
   with densities evaluated at the experimental triple point of water.
}
	\label{table:hamaker_comparison}
\end{table}

\end{widetext}

The direction for improvement of force field based on this comparison
appears to be to keep Lennard-Jones parameters similar to those
of TIP4P, but with an increased dipole moment. This is roughly the
direction taken in the development of TIP4P/2005.
Of course, a quantitative comparison between DLP cannot be taken too far,
because electronic and dipolar terms in empirical force fields are not fully
meaningful, and the DLP predictions are also somewhat subject to uncertainties
of the dielectric parametrization.

However, it is pleasing to find that currently accepted force fields appear to provide a correct qualitative description of surface dispersion
forces. Based on this observation, we expect that the long range behavior
of the interface potential predicted by empirical models used in our recent
work is a reliable proxy
for the physics of premelting films.\cite{benet19,llombart20,llombart20b}

\section{Discussion}

Having settled the role of van der Waals forces on the interface potential we
are now on good position to discuss the important problems of surface melting, regelation
and surface freezing.

\subsection{Surface premelting}

In the case of surface premelting, the surface free energy of the ice/vapor
interface as mediated by a premelting film of thickness $\hh$, is given by:
\begin{equation}
   \omega_{iv}(\hh) = \gamma_{iw} + \gamma_{wv} + g(\hh)
   \label{oiv}
\end{equation} 
where $g(\hh)$ is the interface potential describing the free energy cost of
a premelting film as a function of film thickness. In the limit that $\hh\to
\infty$, $g(\hh)=0$, and the surface free energy is just the sum of
ice/water and water/vapor surface tensions.

The equilibrium value of the film thickness, $\hh_e$ is obtained by 
minimization of the free energy, $\omega_{iv}(\hh)$, such that 
$dg(\hh)/d\hh = 0$,
whereupon, one obtains the equilibrium interface tension of the
ice/vapor interface as:
\begin{equation}
\gamma_{iv}=\omega_{iv}(\hh_e)
\label{giv}
\end{equation} 
This result acknowledges explicitly the fact that the ice/vapor surface tension
may be mediated by a finite premelting film of adsorbed water. Of course, its
properties need not be exactly as those of bulk water.

Three different situations are possible. The first corresponds
to the case where the absolute minimum of $g(\hh)$ occurs at $\hh=0$. In that
case, the structure of the ice/vapor interface would be that of a perfectly
terminated ice slab in contact with air. The second one occurs in the
opposite situation, where the absolute minimum is at $\hh\to\infty$. In
that case, $g(\hh)\to 0$ by construction, so that $\gamma_{iv}=\gamma_{iw} +
\gamma_{wv}$ exactly. This equality is the condition for wetting of
the ice/vapor interface by an intruding macroscopic film of liquid water,
which in this case is known also as {\em surface melting}. Finally,
a third situation can arise if an absolute minimum of $g(\hh)$ exists
for finite values of $\hh_e$. This is a situation of incomplete melting,
whereby the equilibrium ice/vapor interface is mediated by a stable premelting
film of finite thickness.

Based on the calculations of the previous section, we see that
$g_{vdw}(\hh)$  is a \revision{monotonic} and negative function, with an absolute minimum
at $\hh=0$. Accordingly, we confirm, in agreement with recent work 
and unpublished
results by ourselves,\cite{luengo19,luengo20,fiedler20,luengo21,li22} 
that van der
Waals forces conspire against the surface premelting of ice,  favoring
a perfectly terminated ice/vapor surface instead.  

\begin{figure}[htb!]
        \includegraphics[width=0.48\textwidth,height=0.35\textwidth,keepaspectratio]{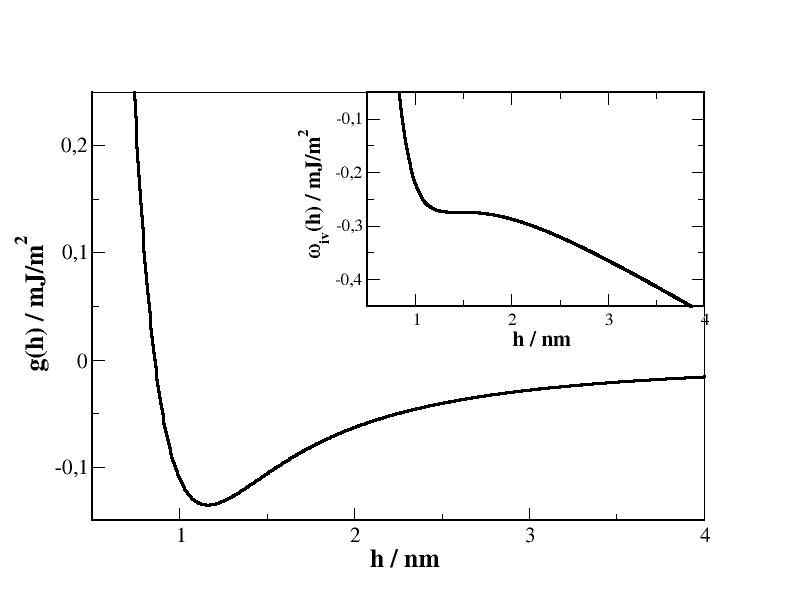}
	  \caption{Interface potential for ice premelting suggested in this work.
	     Notice the model is valid for crystal planes above their roughening
	     transition. Inset: The corresponding surface free energy at the
	     surface spinodal vapor pressure becomes unstable and the premelting
	     film thereof undergoes complete wetting of the underlying ice
	     substrate.
        \label{fig:g_iwa}
     }
\end{figure}

In practice, DLP theory is not accurate in the limit of $\hh\to 0$,
because it assumes structureless interfaces
(this can be readily
understood without any acquaintance of DLP, merely by noticing that its
only input are bulk dielectric properties). At short range, $g(\hh)$
is dominated by interactions arising from the distortion of
the bulk density profile, which, at low temperatures decays on the scale of
the molecular diameter. Therefore, the full interface potential
is expressed as:
\begin{equation}
      g(\hh) = g_{sr}(\hh) + g_{vdw}(\hh)
\end{equation} 
Here, $g_{sr}(\hh)$ is the short range contribution. For ice premelting,
we have recently showed that the qualitative behavior of this term  follows 
expectations from liquid state theory,\cite{chernov88,henderson94}
and may be described as:\cite{llombart20}
\begin{equation}
 g_{sr}(\hh) = 
 B_2 \exp(-\kappa_2 \hh) - B_1 \exp(-\kappa_1 \hh) \cos(q_o\hh + \phi)
 \label{gsr}
\end{equation} 
where $B_i$ are positive constants; $\kappa_i$ are  inverse decay lengths,
$q_o$ is dictated by the wave-length of packing correlations in the liquid
(i.e. the maximum of the liquid's structure factor) and $\phi$ is
the phase. This relation improves the result of field theory, which
provides just the first term of the right hand side, but shares, in
common with the above result the expectations that
$B_2>0$.\cite{lipowsky89,limmer14}

The actual value of the parameters depends somewhat on
the ice facet, but  a common feature they all share is that $B_2$
is significantly larger than $B_1$. This implies that for film thicknesses
on the order of $\kappa_2^{-1}$, $g_{sr}(\hh)$ is
positive,\cite{limmer14} while the
oscillatory term could make $g_{sr}(\hh)$ negative at larger distances
provided $\kappa_1 < \kappa_2$, a condition that is expected to hold
only for faces below their roughening transition.\cite{chernov88,llombart20} 
At the triple point, only the basal face meets this condition.

Overall, it follows  that at short range,
packing correlations repel the liquid-vapor interface away from the solid
substrate, promoting surface melting; at long range, van der Waals forces bind the
liquid-vapor interface, and inhibit the build up of a large liquid film.
Therefore, the interface potential must display an absolute minimum of
negative energy at intermediate thicknesses, such that $g(\hh_e) < 0$,
and the ice surface thus exhibits a liquid premelting film of finite
thickness at the triple point. Accordingly, from \Eq{oiv} and \Eq{giv},
we see that $\gamma_{iv} < \gamma_{iw} + \gamma_{wv}$. i.e.,
water does not wet ice at the triple point, and must form liquid
droplets of finite contact angle, in agreement with recurrent reports 
in the literature.\cite{knight67,knight71,elbaum91,elbaum93,gonda99,murata16}
The presence of the oscillatory term in \Eq{gsr} means that there could
be an additional shallow minima at larger thickness in the basal facet,
thus explaining the existence of two different incomplete wetting states
that have been reported in experiments.\cite{murata16}

Our considerations differ with the qualitative model of interfacial forces
suggested in Ref.\cite{murata16}. These authors observed growth and spread
of steps and terraces on the basal facet of ice, and argued that the presence
of steps must imply a smooth surface inconsistent with a highly disordered
premelting layer. Accordingly, they hypothesized an interface potential 
with an absolute minimum at $\hh=0$, implying a bare ice surface as the
stable state of the basal face. However, we have shown using both
theoretical models and computer simulations that a smooth facet with
steps can exist even in the presence of a premelting
film,\cite{benet16,benet19,llombart20,sibley21} such that our theoretical
model of the interface potential and the experimental observation of
steps and terraces are mutually consistent.

A simplified qualitative form of the interface potential, adequate
in the region $g(\hh)<0$ that 
characterizes the primary minimum  may be obtained from the 
leading term of \Eq{gsr} and the non-retarded 
van der Waals contribution:
\begin{equation}
  g(\hh) = B e^{-\kappa \hh} - \frac{A_{iwa}(0)}{12\pi \hh^2}
\end{equation} 
This result is of the same form as anticipated by Limmer and Chandler,\cite{limmer14} 
but notice that our results show the positive sign of the Hamaker constant
$A_{iwa}(0)$ stems 
mainly from high frequency contributions of the dielectric response (rather than
from the static term).

This simple model of interface potential allow us to clarify many of the speculations
on surface melting discussed in the
literature,\cite{jellinek67,kuroda82,nenow86,kuroda90,petrenko94,dash95,rosenberg05,dash06,slater19,nagata19}
as well as our own.\cite{benet19}

Using our estimates for the Hamaker constant in table \ref{table:hamaker_constants}, together with
$B=0.07$~J/m$^2$ and $\kappa=6.2~nm^{-1}$ obtained for the prism facet
in our previous work,\cite{llombart20}, we  find a free energy minimum
of $g(\hh_e)=-0.135$~mJ$/$m$^2$, and an equilibrium film thickness
of barely $\hh_e\approx 1.16$~nm.  The presence of this thin premelting film
is not in conflict with the formation of liquid droplets (c.f.
Ref.\cite{knight96,makkonen97}).
By \revision{plugging} \Eq{giv} into Young's equation,
$\gamma_{iv}=\gamma_{iw}+\gamma_{wv}\cos\theta$, we find that the shallow
minimum of the interface potential provides:
\begin{equation}
   \cos\theta = 1 + \frac{g(\hh_e)}{\gamma_{wv}}
\end{equation} 
Using our model interface potential into this formula, together with
$\gamma_{wv}=75.7$~mJ$/$m$^2$,\cite{fletcher70}
we predict a contact angle of $\theta = 3.4^{\circ}$.
For comparison, first experimental measures of about 12$^o$,\cite{knight67}
have thereafter been revised to smaller than 5$^o$,\cite{knight71}
with estimates of 2$^o$,\cite{elbaum93,murata16} or even less than 1$^o$ 
depending on the ice facet.\cite{ketcham69,elbaum93}

These considerations are relevant exactly at the triple point, 
where ice, water and vapor have all exactly the same chemical potential,
so that one phase can grow from the other at free cost. As soon as
one moves away from the triple point, the formation of the premelting
film picks an extra free energy cost, and the free energy in \Eq{oiv} becomes:
\begin{equation}
   \omega_{iv}(\hh) = \gamma_{iw} + \gamma_{wv} + g(\hh) - \Delta p_{wv}(\mu)\hh
   \label{oivp}
\end{equation} 
where $\Delta p_{wv}$ is  the difference between bulk liquid and vapor pressures
at the system's chemical potential. Assuming that the vapor sets
the total chemical potential, as is usual in experiments, we can estimate 
this as $\Delta p_{wv}=\rho_w k_BT\ln p/p_{wv}(T)$, where $p$ is the imposed 
vapor pressure, $\rho_w$ is the bulk liquid density  and $p_{wv}(T)$ is the 
water-vapor saturation pressure at the 
systems temperature. Whence, as the vapor pressure $p$ is raised above
$p_{wv}$, the free energy $\omega_{iv}(\hh)$ picks a linear term proportional to $\hh$,
and the equilibrium film thickness is displaced to larger values of
the film thickness. Assuming a metastable equilibrium of ice with
supersaturated vapor (this can be achieved for very small
growth rates of ice close to the triple point, c.f.~\onlinecite{sibley21}), the film thickness
may be obtained by extremalization of the above equation. This leads
readily to an expression for the vapor pressure in equilibrium with a
quasi-liquid layer
of thickness $\hh$:\cite{llombart20}
\begin{equation}
p = p_{wv}(T) e^{-\frac{\Pi(\hh)}{\rho_w k_BT}}
\label{kd}
\end{equation} 
where $\Pi(\hh)=-d g(\hh)/d \hh$ is the disjoining
pressure.\cite{derjaguin87,henderson05}
i.e., as $\hh$ increases, $\Pi(\hh)$ becomes more negative, and the vapor
pressure in equilibrium with the premelting film increases. 

Eventually, however, $\Delta p_{wv}$ is sufficiently large that the linear
term in \Eq{oivp} washes out completely the minimum of the interface 
potential. This corresponds to the limit where the disjoining pressure attains
its absolute minimum. i.e. a spinodal point is reached
and a premelting film can no longer be stabilized. 

Using our model potential, we find for the spinodal limit of the
premelting film $\hh_{sp}=1.44$~nm, which corresponds to a disjoining pressure
of $\Pi(\hh_{sp})=-1.12$~bar. Whence, the maximum vapor pressure than can be 
attained before the ice surface becomes wet
is estimated from \Eq{kd} as $p_{sp} \approx 1.0009 p_{wv}(T)$. This explains
why experiments often report very thick wetting films close to the
triple point. As soon as the vapor pressure is slightly above the liquid-vapor
coexistence curve, the premelting
film becomes unstable and can grow without bounds on top of the bulk
ice. For temperatures close to the triple point, this requires 
an exquisite control of the vapor pressure. Once the spinodal is traversed,
ice grown from vapor actually freezes from the
condensed wetting film that lies above.\cite{sibley21}

In practice, both vapor condensation and ice growth occur
simultaneously.\cite{kuroda82,kuroda90,neshyba16,mohandesi18} This results in an increase of the actual spinodal pressure 
in a complicated manner that depends on the precise growth mechanism.\cite{sibley21}
Accordingly, our calculation based on \Eq{kd} provide a lower estimate of the
pressure where premelting films become unstable.

\Eq{kd}  does become exact at thermodynamic
equilibrium, which is strictly realized for ice in contact with vapor along the
sublimation line, such that $p$ is equal to the saturation pressure of 
vapor over ice, $p_{iv}$. In this limit, using Clausius-Clapeyron, we find 
that $\Delta p_{iv} = \frac{\rho_w \Delta H_{iw}}{T_t}(T - T_t)$, leading to
the well known result for the premelting thickness as a function of 
temperature along the sublimation line:\cite{dash06}
\begin{equation}
   \frac{\rho_w \Delta H_{iw}}{T_t}(T - T_t) = \Pi(\hh)
\end{equation} 
Accordingly, at the triple point the condition of equilibrium is the
vanishing of $\Pi(\hh)$.  In the absence of a binding term in
the disjoining pressure, as often assumed,\cite{nagata19} 
this condition is met only for $\hh\to\infty$. However, because of
the van der Waals contribution, the condition of vanishing disjoining
pressure is met at a finite equilibrium film thickness $\hh_e$ corresponding to
the minimum of the interface potential $g(\hh)$.

Our results for the surface free energy of the equilibrium ice/vapor interface
(i.e., with a mediated premelting film), allow us to answer a general question
relevant in atmospheric physics. Consider one has at some low temperature a bulk
ice phase in equilibrium with its vapor, and gradually increases the temperature
along the ice-vapor coexistence line up to the triple point. In that moment, 
all three phases have exactly the same bulk free energy per molecule, and the
only factor inhibiting the condensation of a bulk water phase is the surface
free energy. The question then is posed, where will a bulk flat phase grow
preferentially? Will it be within the bulk vapor phase, within the bulk ice 
phase or intruding into the ice/vapor interface? To solve this, recall that the
equality of bulk free energies imposed at the triple point
 requires one to assume the interfaces that are formed have \revision{strictly} zero
curvature. Therefore, we refer here to the formation of bulk flat phases,
parallel to the current ice/vapor interface. The formation of a macroscopic
water phase within the vapor will then cost $2\gamma_{wv}$ per unit surface.
Similarly, the formation of the water phase within the bulk ice phase
will require $2\gamma_{iw}$. If, on the other hand, the bulk water phase
forms at the ice/vapor interface, the cost is
$\gamma_{iw}+\gamma_{wv}-\gamma_{iv}$. \revision{Bearing} in mind approximate estimates of the
surface free energies of ca. $\gamma_{iv} > \gamma_{wv} > \gamma_{iw}$, we
readily rule out the formation of water within the vapor phase. Furthermore,
recalling from \Eq{oiv}-\ref{giv} 
that $-\gamma_{iv}=\gamma_{iw}+\gamma_{wv}+g(\hh_e)$,
we see that the formation of condensed water at the ice/vapor interface
will  cost $-g(\hh_e)$, which barely amounts to $1.35\times 10^{-4}$ J/m~$^2$, much less than
any of the other cases. So if a large macroscopic condensed water phase is to
form, it will grow in between ice and vapor. However, the state of minimal
free energy is in fact a bulk ice phase coexisting with a finite film
of equilibrium thickness $\hh_e$ between a bulk vapor phase. Accordingly,
the melting of a perfect ice monocrystal is actually a weakly activated process, as postulated by
Knight many years ago,\cite{knight71} and is actually the expected situation for materials with incomplete surface premelting.\cite{dash95,metois89} However, our discussion
above shows that it will suffice to increase the pressure barely a few Pascal
above the
triple phase for the macroscopic bulk water phase to become the preferred state.

\subsection{Regelation}

Describing a famous experiment,\cite{faraday60} Faraday noted that 
"two pieces of thawing ice, if put together, adhere and become one".
This and some other observations
helped formulate the hypothesis of ice premelting for the first time,
despite great difficulties to understand how freezing could occur
in regions of the bulk phase diagram where water is the preferred
phase.\cite{jellinek67} 

In this regard, notice however that van der Waals forces always
conspire in favor of two equal bulk materials to adhere. In this
case, assuming ice is covered with a premelting layer, and two such ice samples
are brought together, a liquid bridge between the ice slabs will form
spontaneously.  In conditions favoring water over vapor, 
i.e. above the liquid-vapor coexistence curve, water will not evaporate. 
However, the liquid layer can vanish by freezing, since,
as described in the previous section, interactions in an
ice/water/ice system favor attraction of the bulk
bodies. i.e.: shrinking of the intervening liquid layer by freezing.

In practice, because the involved Hamaker constants are rather small, the
process is mainly driven by bulk and surface free energies, rather than by the
van der Waals forces. In the language of wetting physics it may be described
essentially as a  phenomenon of capillary freezing. 

To be more specific, consider two large spherical ice
balls, of radius $R$. Then, the force of attraction between the balls will 
be given, according to the Derjaguin approximation,\cite{israelachvili11}  
by $F(d) = \pi R ( \omega_i(d) - \omega_v(\infty) )$, 
where $\omega_i(d)$ is the free energy of the
ice bridge joining two {\em planar} bulk ice slabs, and $\omega_v(\infty)$ is 
the free energy of two {\em planar} ice slabs separated by a macroscopic bulk
vapor phase. The former is just the bulk free energy
of forming ice from the vapor, $\omega_i(d) = -\Delta p_{iv} d$.
The latter is  the cost of the two isolated ice/vapor interfaces,
$\omega_v(\infty) = 2\omega_{iv}(\hh_e)$, with $\hh_e$, the equilibrium layer
thickness at ambient pressure as dictated by minimization of 
\Eq{oivp}. Whence, the total free energy
cost of forming the ice bridge is:
\begin{equation}
   F(d) = 2 \pi R [ \Delta p_{wv} \hh_e - \frac{1}{2}\Delta p_{iv} d -  
                    \gamma_{iw} - \gamma_{wv} - g(\hh_e) ]
\end{equation} 
We see that at the triple point, where $\Delta p_{wv}=\Delta p_{iv}=0$, the
formation of an ice bridge is a favorable process that is driven by the
surface free energies. 

In order to show why regelation occurs exceptionally for ice, let us simplify
the problem and choose quite naturally a bridge length $d=2\hh_e$. Then,
assuming the vapor is an ideal gas, and taking into account Clausius-Clapeyron,
we find that the condition for the formation of an ice bridge as a function
of vapor pressure and temperature is:
\begin{equation}
   \ln \frac{p}{p_t} \le \frac{\gamma_{iw}+\gamma_{wv}}{k_BT 
   (\rho_w-\rho_i)\hh_e} + \frac{\rho_i \Delta H_{iv} - \rho_w\Delta
   H_{lv}}{(\rho_w-\rho_i)} \left ( \frac{1}{T} - \frac{1}{T_t} \right )
	\label{regelation}
\end{equation} 
Due to the anomalous properties of ice, $\rho_w > \rho_i$, the first term in the
right hand side is positive. This means that, at $T=T_t$, where
the second term vanishes, the inequality can be met even at $p/p_t >1$. 
i.e., an ice bridge can form in conditions where the liquid is the preferred
phase. Or in Faraday's own words
''at a place were liquefaction was proceeding, congelation suddenly
occurs''.\cite{faraday60}

On the contrary, for ordinary materials, where the density of the solid phase is
larger than that of the liquid phase, this term is negative. Whence, the ice
bridge could form at pressures $p/p_t < 1$. However, at $T=T_t$ and $p<p_t$,
the solid does not exhibit a
noticeable premelting layer, because it is found in a region of the phase
diagram where the bulk liquid phase is not favored, i.e. sublimation occurs instead 
(recall the slope of the melting curve is positive for the usual case of solid
density larger than that of liquid).
Therefore, the kinetics of regelation is much slower in this case, since the bridge
must form from the vapor phase, rather than from a premelting layer.

As regards the famous controversy between Faraday and Thomson on the origin of
regelation,\cite{faraday60,thomson62} we see that both surface premelting and 
the negative slope of the melting line play a role in the overall free energy 
balance embodied in \Eq{regelation}, but pressure melting is not required for
regelation to occur. The capillary freezing described here, however, refers to
the regelation between two parcels of ice at atmospheric pressure as considered
by Faraday,\cite{faraday60} which can be
very different from the regelation mechanism invoked to explain glacier
motion and wire regelation.\cite{dash06,wettlaufer06}

As a final remark we note that in the region of higher pressures where
the condition of capillary freezing is satisfied, \Eq{regelation}, it
actually becomes more favorable for plain capillary condensation to occur. 
\revision{Bearing} this
in mind one finds that the stricter condition for regelation 
(i.e., such that an ice bridge is more favorable than a liquid bridge) satisfies
\Eq{regelation} with the term $\gamma_{iw}+\gamma_{wv}$ replaced by
$\gamma_{iw}$ alone.

\subsection{Surface freezing}

It is a matter of everyday experience that large ice crystals
 form on the surface of water. Similar observations may be made experimentally
for small but fully nucleated crystals of mm size.\cite{perezdiaz16} Such
observations are easily explained from buoyancy. However,
some experiments and theoretical studies have suggested that  tiny crystals might
not actually reach the surface as a result of buoyancy, but are actually
nucleated in-situ at the liquid-vapor interface.\cite{tabazadeh02,shaw05}
i.e., that the liquid-vapor interface could promote the nucleation of ice
by orders of magnitude compared to bulk.\cite{li13,hussain21}

The problem again is one of wetting physics: we consider the growth of an ice
film at a liquid-vapor interface at coexistence as the system is cooled 
down to the triple point. 
A bulk planar phase could then be formed with the same bulk free energy cost
as the vapor and liquid. Creating bulk ice in the midst of the vapor phase
costs $2\gamma_{iv}$; the cost of creating the same phase within bulk water
costs $2\gamma_{iw}$; while growing ice in between air and water has
a cost of $\gamma_{iw}+\gamma_{iv} - \gamma_{wv}$. But again, because the
equilibrium ice-air interface is actually covered by a premelting film,
use of Eq. \ref{oiv}-\ref{giv} shows that the total energy cost of growing the bulk ice
phase between bulk water and vapor phases is just $g(\hh_e)<0$. Therefore,
our calculations confirm that it is favorable for ice to grow at the 
water-vapor interface among all other choices. \revision{Bear} in mind that this does
not mean actually strict 'surface freezing', understood as the formation of
ice atop the liquid-vapor surface. Instead, because the ice-vapor surface that
is formed is covered by a premelting layer, the actual picture is that of a
bulk ice phase formed at a distance $\hh_e$ below water. However, this occurs
only in a very small range of vapor pressures. Indeed, since the premelting
film is destabilized and \revision{grows unbounded} as soon as the pressure is larger
than the surface spinodal pressure, saturating the interface above
$p_s \approx 1.0009 p_{wv}(T)  $ will promote condensation of water and ice would then be \revision{buried}
by a thick water film. Eventually, if the ice volume is large enough, it
will experience buoyancy forces, and the final outcome is the result of  a balance
between surface interactions and the bulk buoyancy force.\cite{thiyam18}




\section{Conclusions}

In this work we have combined results from wetting physics, quantum Density
Functional Theory and Lifshitz theory of van der Waals forces in order to
assess the role of  molecular interactions  at a number of relevant interfaces 
involving ice and liquid water.

For the long standing problem of ice
premelting,\cite{jellinek67,nenow84,petrenko94,dash95,rosenberg05,dash06,slater19,nagata19}
our results show that van 
der Waals forces inhibit the growth of thick liquid films and prevent ice from 
surface melting. On the contrary, short range structural forces promote 
wetting. The balance between these
competing forces results in a finite equilibrium premelting thickness on the
order of the nanometer.  Our theoretical results are
consistent with computer
simulations,\cite{vega06,neshyba09,neshyba16,kling18,qiu18,louden18,llombart19,llombart20,llombart20b} and  a large body of widely different
experimental techniques.\cite{bluhm02,sadtchenko02,constantin18,mitsui19}

Combining our model of intermolecular forces with results of wetting physics, we
are able to assess the role of vapor pressure on the premelting behavior.
Our results show that the premelting layer
can become unstable by increasing the vapor pressure just a few Pascal above
water-vapor saturation.  This implies that unless an exquisite pressure control is exercised,
ice will readily surface melt in a water supersaturated atmosphere. This adds an additional mechanism for surface melting \revision{apart from} impurities,\cite{wettlaufer99} and
explains recurrent \revision{observations} of very thick wetting layers (c.f. \cite{petrenko94,slater19}).  We believe this finding is particularly significant for studies of atmospheric ice, including ice growth and gas adsorption.\cite{libbrecht22}

Our results also provide insight into the related problem of surface freezing.
We show that the most stable site for ice to nucleate 
at under-saturation is immersed at a distance of roughly one nanometer below the
water-vapor interface, in agreement with suggestions and recent simulation studies.\cite{tabazadeh02,shaw05,li13,hussain21} However,
increasing saturation above the water-vapor coexistence
line promotes  the growth of a thick wetting film above ice. As
a result  the surface enhancement effect on ice nucleation is lost in a
supersaturated atmosphere.

Finally, we show that the property of ice regelation, understood as the ability
of thawing ice parcels to adhere by freezing can be described as a process of
capillary freezing at conditions in the phase diagram where liquid water is the
preferred phase.

We have shown that all of these observations --incomplete premelting, enhanced subsurface nucleation and regelation of thawing ice-- have  their origin in the negative
slope of the freezing line and can therefore be added to the large list of water anomalies.

\section*{Supplementary Material}

See supplementary material for
details on modeling dielectric functions; details on DFT methodology; and tables with parameters.

\begin{acknowledgments}
We would like to acknowledge helpful discussions with Pablo Llombart and
Eva G.  Noya.  This research was funded by the Spanish 
Agencia Estatal de Investigaci\'on under grant PIP2020-115722GB-C21.
The authors thankfully acknowledges the computer resources at Canig\'o
supercomputer and technical support provided by Consorci de Serveis
Universitaris de Catalunya-CSUC under grant FI-2019-3-0014
from the Spanish Network of Supercomputing (RES). 
F.I.R. thanks the Government of Principado de Asturias for its 
FICYT grant number AYUD/2021/58773; as well as the Consejería de Educación, 
Juventud y Deporte de la Comunidad de Madrid and the European Social Funds
for funding under grant PEJD-2018-POST/IND-8623.
\end{acknowledgments}

\section*{Author Information}

\subsection*{Contributions}

J.L.M. modeled dielectric functions and calculated van der Waals forces. F.I.R.
performed DFT calculations. L.G.M. interpreted results and wrote manuscript.

\subsection*{Corresponding author}

	   Correspondence to: lgmac@quim.ucm.es

\bibliography{referenc}

\begin{thebibliography}{151}%
\makeatletter
\providecommand \@ifxundefined [1]{%
 \@ifx{#1\undefined}
}%
\providecommand \@ifnum [1]{%
 \ifnum #1\expandafter \@firstoftwo
 \else \expandafter \@secondoftwo
 \fi
}%
\providecommand \@ifx [1]{%
 \ifx #1\expandafter \@firstoftwo
 \else \expandafter \@secondoftwo
 \fi
}%
\providecommand \natexlab [1]{#1}%
\providecommand \enquote  [1]{``#1''}%
\providecommand \bibnamefont  [1]{#1}%
\providecommand \bibfnamefont [1]{#1}%
\providecommand \citenamefont [1]{#1}%
\providecommand \href@noop [0]{\@secondoftwo}%
\providecommand \href [0]{\begingroup \@sanitize@url \@href}%
\providecommand \@href[1]{\@@startlink{#1}\@@href}%
\providecommand \@@href[1]{\endgroup#1\@@endlink}%
\providecommand \@sanitize@url [0]{\catcode `\\12\catcode `\$12\catcode
  `\&12\catcode `\#12\catcode `\^12\catcode `\_12\catcode `\%12\relax}%
\providecommand \@@startlink[1]{}%
\providecommand \@@endlink[0]{}%
\providecommand \url  [0]{\begingroup\@sanitize@url \@url }%
\providecommand \@url [1]{\endgroup\@href {#1}{\urlprefix }}%
\providecommand \urlprefix  [0]{URL }%
\providecommand \Eprint [0]{\href }%
\providecommand \doibase [0]{https://doi.org/}%
\providecommand \selectlanguage [0]{\@gobble}%
\providecommand \bibinfo  [0]{\@secondoftwo}%
\providecommand \bibfield  [0]{\@secondoftwo}%
\providecommand \translation [1]{[#1]}%
\providecommand \BibitemOpen [0]{}%
\providecommand \bibitemStop [0]{}%
\providecommand \bibitemNoStop [0]{.\EOS\space}%
\providecommand \EOS [0]{\spacefactor3000\relax}%
\providecommand \BibitemShut  [1]{\csname bibitem#1\endcsname}%
\let\auto@bib@innerbib\@empty
\bibitem [{\citenamefont {Bj{\"o}rneholm}\ \emph {et~al.}(2016)\citenamefont
  {Bj{\"o}rneholm}, \citenamefont {Hansen}, \citenamefont {Hodgson},
  \citenamefont {Liu}, \citenamefont {Limmer}, \citenamefont {Michaelides},
  \citenamefont {Pedevilla}, \citenamefont {Rossmeisl}, \citenamefont {Shen},
  \citenamefont {Tocci}, \citenamefont {Tyrode}, \citenamefont {Walz},
  \citenamefont {Werner},\ and\ \citenamefont {Bluhm}}]{bjorneholm16}%
  \BibitemOpen
  \bibfield  {author} {\bibinfo {author} {\bibfnamefont {O.}~\bibnamefont
  {Bj{\"o}rneholm}}, \bibinfo {author} {\bibfnamefont {M.~H.}\ \bibnamefont
  {Hansen}}, \bibinfo {author} {\bibfnamefont {A.}~\bibnamefont {Hodgson}},
  \bibinfo {author} {\bibfnamefont {L.-M.}\ \bibnamefont {Liu}}, \bibinfo
  {author} {\bibfnamefont {D.~T.}\ \bibnamefont {Limmer}}, \bibinfo {author}
  {\bibfnamefont {A.}~\bibnamefont {Michaelides}}, \bibinfo {author}
  {\bibfnamefont {P.}~\bibnamefont {Pedevilla}}, \bibinfo {author}
  {\bibfnamefont {J.}~\bibnamefont {Rossmeisl}}, \bibinfo {author}
  {\bibfnamefont {H.}~\bibnamefont {Shen}}, \bibinfo {author} {\bibfnamefont
  {G.}~\bibnamefont {Tocci}}, \bibinfo {author} {\bibfnamefont
  {E.}~\bibnamefont {Tyrode}}, \bibinfo {author} {\bibfnamefont {M.-M.}\
  \bibnamefont {Walz}}, \bibinfo {author} {\bibfnamefont {J.}~\bibnamefont
  {Werner}},\ and\ \bibinfo {author} {\bibfnamefont {H.}~\bibnamefont
  {Bluhm}},\ }\bibfield  {title} {\bibinfo {title} {Water at interfaces},\
  }\href {https://doi.org/10.1021/acs.chemrev.6b00045} {\bibfield  {journal}
  {\bibinfo  {journal} {Chem. Rev.}\ }\textbf {\bibinfo {volume} {116}},\
  \bibinfo {pages} {7698} (\bibinfo {year} {2016})},\ \bibinfo {note} {pMID:
  27232062},\ \Eprint
  {https://arxiv.org/abs/https://doi.org/10.1021/acs.chemrev.6b00045}
  {https://doi.org/10.1021/acs.chemrev.6b00045} \BibitemShut {NoStop}%
\bibitem [{\citenamefont {Bartels-Rausch}\ \emph {et~al.}(2012)\citenamefont
  {Bartels-Rausch}, \citenamefont {Bergeron}, \citenamefont {Cartwright},
  \citenamefont {Escribano}, \citenamefont {Finney}, \citenamefont {Grothe},
  \citenamefont {Guti\'errez}, \citenamefont {Haapala}, \citenamefont {Kuhs},
  \citenamefont {Pettersson}, \citenamefont {Price}, \citenamefont
  {Sainz-D\'{\i}az}, \citenamefont {Stokes}, \citenamefont {Strazzulla},
  \citenamefont {Thomson}, \citenamefont {Trinks},\ and\ \citenamefont
  {Uras-Aytemiz}}]{bartels12}%
  \BibitemOpen
  \bibfield  {author} {\bibinfo {author} {\bibfnamefont {T.}~\bibnamefont
  {Bartels-Rausch}}, \bibinfo {author} {\bibfnamefont {V.}~\bibnamefont
  {Bergeron}}, \bibinfo {author} {\bibfnamefont {J.~H.~E.}\ \bibnamefont
  {Cartwright}}, \bibinfo {author} {\bibfnamefont {R.}~\bibnamefont
  {Escribano}}, \bibinfo {author} {\bibfnamefont {J.~L.}\ \bibnamefont
  {Finney}}, \bibinfo {author} {\bibfnamefont {H.}~\bibnamefont {Grothe}},
  \bibinfo {author} {\bibfnamefont {P.~J.}\ \bibnamefont {Guti\'errez}},
  \bibinfo {author} {\bibfnamefont {J.}~\bibnamefont {Haapala}}, \bibinfo
  {author} {\bibfnamefont {W.~F.}\ \bibnamefont {Kuhs}}, \bibinfo {author}
  {\bibfnamefont {J.~B.~C.}\ \bibnamefont {Pettersson}}, \bibinfo {author}
  {\bibfnamefont {S.~D.}\ \bibnamefont {Price}}, \bibinfo {author}
  {\bibfnamefont {C.~I.}\ \bibnamefont {Sainz-D\'{\i}az}}, \bibinfo {author}
  {\bibfnamefont {D.~J.}\ \bibnamefont {Stokes}}, \bibinfo {author}
  {\bibfnamefont {G.}~\bibnamefont {Strazzulla}}, \bibinfo {author}
  {\bibfnamefont {E.~S.}\ \bibnamefont {Thomson}}, \bibinfo {author}
  {\bibfnamefont {H.}~\bibnamefont {Trinks}},\ and\ \bibinfo {author}
  {\bibfnamefont {N.}~\bibnamefont {Uras-Aytemiz}},\ }\bibfield  {title}
  {\bibinfo {title} {Ice structures, patterns, and processes: A view across the
  icefields},\ }\href {https://doi.org/10.1103/RevModPhys.84.885} {\bibfield
  {journal} {\bibinfo  {journal} {Rev. Mod. Phys.}\ }\textbf {\bibinfo {volume}
  {84}},\ \bibinfo {pages} {885} (\bibinfo {year} {2012})}\BibitemShut
  {NoStop}%
\bibitem [{\citenamefont {Libbrecht}(2022)}]{libbrecht22}%
  \BibitemOpen
  \bibfield  {author} {\bibinfo {author} {\bibfnamefont {K.~G.}\ \bibnamefont
  {Libbrecht}},\ }\href@noop {} {\emph {\bibinfo {title} {Snow Crystals}}}\
  (\bibinfo  {publisher} {Princeton University Press},\ \bibinfo {year}
  {2022})\BibitemShut {NoStop}%
\bibitem [{\citenamefont {Pruppacher}\ and\ \citenamefont
  {Klett}(2010)}]{pruppacher10}%
  \BibitemOpen
  \bibfield  {author} {\bibinfo {author} {\bibfnamefont {H.~R.}\ \bibnamefont
  {Pruppacher}}\ and\ \bibinfo {author} {\bibfnamefont {J.~D.}\ \bibnamefont
  {Klett}},\ }\href@noop {} {\emph {\bibinfo {title} {Microphysics of Clouds
  and Precipitation}}}\ (\bibinfo  {publisher} {Springer},\ \bibinfo {address}
  {Heidelberg},\ \bibinfo {year} {2010})\BibitemShut {NoStop}%
\bibitem [{\citenamefont {Weyl}(1951)}]{weyl51}%
  \BibitemOpen
  \bibfield  {author} {\bibinfo {author} {\bibfnamefont {W.}~\bibnamefont
  {Weyl}},\ }\bibfield  {title} {\bibinfo {title} {Surface structure of water
  and some of its physical and chemical manifestations},\ }\href
  {https://doi.org/https://doi.org/10.1016/0095-8522(51)90011-6} {\bibfield
  {journal} {\bibinfo  {journal} {J. Colloid. Sci.}\ }\textbf {\bibinfo
  {volume} {6}},\ \bibinfo {pages} {389} (\bibinfo {year} {1951})}\BibitemShut
  {NoStop}%
\bibitem [{\citenamefont {Lipowsky}(1982)}]{lipowsky82}%
  \BibitemOpen
  \bibfield  {author} {\bibinfo {author} {\bibfnamefont {R.}~\bibnamefont
  {Lipowsky}},\ }\bibfield  {title} {\bibinfo {title} {Critical surface
  phenomena at first-order bulk transitions},\ }\href@noop {} {\bibfield
  {journal} {\bibinfo  {journal} {Phys. Rev. Lett.}\ }\textbf {\bibinfo
  {volume} {49}},\ \bibinfo {pages} {1575} (\bibinfo {year}
  {1982})}\BibitemShut {NoStop}%
\bibitem [{\citenamefont {Dietrich}(1988)}]{dietrich88}%
  \BibitemOpen
  \bibfield  {author} {\bibinfo {author} {\bibfnamefont {S.}~\bibnamefont
  {Dietrich}},\ }\bibfield  {title} {\bibinfo {title} {Wetting phenomena},\
  }in\ \href@noop {} {\emph {\bibinfo {booktitle} {Phase Transitions and
  Critical Phenomena}}},\ Vol.~\bibinfo {volume} {12},\ \bibinfo {editor}
  {edited by\ \bibinfo {editor} {\bibfnamefont {C.}~\bibnamefont {Domb}}\ and\
  \bibinfo {editor} {\bibfnamefont {J.~L.}\ \bibnamefont {Lebowitz}}}\
  (\bibinfo  {publisher} {Academic},\ \bibinfo {address} {New York},\ \bibinfo
  {year} {1988})\ pp.\ \bibinfo {pages} {1--89}\BibitemShut {NoStop}%
\bibitem [{\citenamefont {Schick}(1990)}]{schick90}%
  \BibitemOpen
  \bibfield  {author} {\bibinfo {author} {\bibfnamefont {M.}~\bibnamefont
  {Schick}},\ }\bibfield  {title} {\bibinfo {title} {Introduction to wetting
  phenomena},\ }in\ \href@noop {} {\emph {\bibinfo {booktitle} {Liquids at
  Interfaces}}},\ \bibinfo {series and number} {Les Houches Lecture Notes}\
  (\bibinfo  {publisher} {Elsevier},\ \bibinfo {address} {Amsterdam},\ \bibinfo
  {year} {1990})\ pp.\ \bibinfo {pages} {1--89}\BibitemShut {NoStop}%
\bibitem [{\citenamefont {Jellinek}(1967)}]{jellinek67}%
  \BibitemOpen
  \bibfield  {author} {\bibinfo {author} {\bibfnamefont {H.}~\bibnamefont
  {Jellinek}},\ }\bibfield  {title} {\bibinfo {title} {Liquid-like (transition)
  layer on ice},\ }\href
  {https://doi.org/https://doi.org/10.1016/0021-9797(67)90022-7} {\bibfield
  {journal} {\bibinfo  {journal} {J. Colloid. Interface Sci.}\ }\textbf
  {\bibinfo {volume} {25}},\ \bibinfo {pages} {192} (\bibinfo {year}
  {1967})}\BibitemShut {NoStop}%
\bibitem [{\citenamefont {Nenow}(1984)}]{nenow84}%
  \BibitemOpen
  \bibfield  {author} {\bibinfo {author} {\bibfnamefont {D.}~\bibnamefont
  {Nenow}},\ }\bibfield  {title} {\bibinfo {title} {Surface premelting},\
  }\href {https://doi.org/https://doi.org/10.1016/0146-3535(84)90081-9}
  {\bibfield  {journal} {\bibinfo  {journal} {Prog. Cryst. Growth Charact.
  Mater.}\ }\textbf {\bibinfo {volume} {9}},\ \bibinfo {pages} {185} (\bibinfo
  {year} {1984})}\BibitemShut {NoStop}%
\bibitem [{\citenamefont {Petrenko}(1994)}]{petrenko94}%
  \BibitemOpen
  \bibfield  {author} {\bibinfo {author} {\bibfnamefont {V.~F.}\ \bibnamefont
  {Petrenko}},\ }\href@noop {} {\emph {\bibinfo {title} {The Surface of
  Ice}}},\ \bibinfo {organization} {Cold Regions Research and Engineering
  Laboratory (Special Report 94-22)} (\bibinfo {year} {1994})\BibitemShut
  {NoStop}%
\bibitem [{\citenamefont {Dash}\ \emph {et~al.}(1995)\citenamefont {Dash},
  \citenamefont {Fu},\ and\ \citenamefont {Wettlaufer}}]{dash95}%
  \BibitemOpen
  \bibfield  {author} {\bibinfo {author} {\bibfnamefont {J.~G.}\ \bibnamefont
  {Dash}}, \bibinfo {author} {\bibfnamefont {H.}~\bibnamefont {Fu}},\ and\
  \bibinfo {author} {\bibfnamefont {J.~S.}\ \bibnamefont {Wettlaufer}},\
  }\bibfield  {title} {\bibinfo {title} {The premelting of ice and its
  environmental consequences},\ }\href
  {https://doi.org/10.1088/0034-4885/58/1/003} {\bibfield  {journal} {\bibinfo
  {journal} {Reports on Progress in Physics}\ }\textbf {\bibinfo {volume}
  {58}},\ \bibinfo {pages} {115} (\bibinfo {year} {1995})}\BibitemShut
  {NoStop}%
\bibitem [{\citenamefont {Rosenberg}(2005)}]{rosenberg05}%
  \BibitemOpen
  \bibfield  {author} {\bibinfo {author} {\bibfnamefont {R.}~\bibnamefont
  {Rosenberg}},\ }\bibfield  {title} {\bibinfo {title} {Why is ice slippery?},\
  }\href@noop {} {\bibfield  {journal} {\bibinfo  {journal} {Phys. Today}\
  }\textbf {\bibinfo {volume} {58}},\ \bibinfo {pages} {50} (\bibinfo {year}
  {2005})}\BibitemShut {NoStop}%
\bibitem [{\citenamefont {Dash}\ \emph {et~al.}(2006)\citenamefont {Dash},
  \citenamefont {Rempel},\ and\ \citenamefont {Wettlaufer}}]{dash06}%
  \BibitemOpen
  \bibfield  {author} {\bibinfo {author} {\bibfnamefont {J.~G.}\ \bibnamefont
  {Dash}}, \bibinfo {author} {\bibfnamefont {A.~W.}\ \bibnamefont {Rempel}},\
  and\ \bibinfo {author} {\bibfnamefont {J.~S.}\ \bibnamefont {Wettlaufer}},\
  }\bibfield  {title} {\bibinfo {title} {The physics of premelted ice and its
  geophysical consequences},\ }\href@noop {} {\bibfield  {journal} {\bibinfo
  {journal} {Rev. Mod. Phys.}\ }\textbf {\bibinfo {volume} {78}},\ \bibinfo
  {pages} {695} (\bibinfo {year} {2006})}\BibitemShut {NoStop}%
\bibitem [{\citenamefont {Slater}\ and\ \citenamefont
  {Michaelides}(2019)}]{slater19}%
  \BibitemOpen
  \bibfield  {author} {\bibinfo {author} {\bibfnamefont {B.}~\bibnamefont
  {Slater}}\ and\ \bibinfo {author} {\bibfnamefont {A.}~\bibnamefont
  {Michaelides}},\ }\bibfield  {title} {\bibinfo {title} {Surface premelting of
  water ice},\ }\href@noop {} {\bibfield  {journal} {\bibinfo  {journal} {Nat.
  Rev. Chem}\ }\textbf {\bibinfo {volume} {3}},\ \bibinfo {pages} {172}
  (\bibinfo {year} {2019})}\BibitemShut {NoStop}%
\bibitem [{\citenamefont {Nagata}\ \emph {et~al.}(2019)\citenamefont {Nagata},
  \citenamefont {Hama}, \citenamefont {Backus}, \citenamefont {Mezger},
  \citenamefont {Bonn}, \citenamefont {Bonn},\ and\ \citenamefont
  {Sazaki}}]{nagata19}%
  \BibitemOpen
  \bibfield  {author} {\bibinfo {author} {\bibfnamefont {Y.}~\bibnamefont
  {Nagata}}, \bibinfo {author} {\bibfnamefont {T.}~\bibnamefont {Hama}},
  \bibinfo {author} {\bibfnamefont {E.~H.~G.}\ \bibnamefont {Backus}}, \bibinfo
  {author} {\bibfnamefont {M.}~\bibnamefont {Mezger}}, \bibinfo {author}
  {\bibfnamefont {D.}~\bibnamefont {Bonn}}, \bibinfo {author} {\bibfnamefont
  {M.}~\bibnamefont {Bonn}},\ and\ \bibinfo {author} {\bibfnamefont
  {G.}~\bibnamefont {Sazaki}},\ }\bibfield  {title} {\bibinfo {title} {The
  surface of ice under equilibrium and nonequilibrium conditions},\ }\href
  {https://doi.org/10.1021/acs.accounts.8b00615} {\bibfield  {journal}
  {\bibinfo  {journal} {Acc. Chem. Res.}\ }\textbf {\bibinfo {volume} {52}},\
  \bibinfo {pages} {1006} (\bibinfo {year} {2019})}\BibitemShut {NoStop}%
\bibitem [{\citenamefont {Elbaum}(1991)}]{elbaum91}%
  \BibitemOpen
  \bibfield  {author} {\bibinfo {author} {\bibfnamefont {M.}~\bibnamefont
  {Elbaum}},\ }\bibfield  {title} {\bibinfo {title} {Roughening transition
  observed on the prism facet of ice},\ }\href
  {https://doi.org/10.1103/PhysRevLett.67.2982} {\bibfield  {journal} {\bibinfo
   {journal} {Phys. Rev. Lett.}\ }\textbf {\bibinfo {volume} {67}},\ \bibinfo
  {pages} {2982} (\bibinfo {year} {1991})}\BibitemShut {NoStop}%
\bibitem [{\citenamefont {Wettlaufer}(1999)}]{wettlaufer99}%
  \BibitemOpen
  \bibfield  {author} {\bibinfo {author} {\bibfnamefont {J.}~\bibnamefont
  {Wettlaufer}},\ }\bibfield  {title} {\bibinfo {title} {Impurity effects in
  the premelting of ice},\ }\href {https://doi.org/10.1103/PhysRevLett.82.2516}
  {\bibfield  {journal} {\bibinfo  {journal} {Phys. Rev. Lett.}\ }\textbf
  {\bibinfo {volume} {82}},\ \bibinfo {pages} {2516} (\bibinfo {year}
  {1999})}\BibitemShut {NoStop}%
\bibitem [{\citenamefont {Li}\ and\ \citenamefont {Somorjai}(2007)}]{li07b}%
  \BibitemOpen
  \bibfield  {author} {\bibinfo {author} {\bibfnamefont {Y.}~\bibnamefont
  {Li}}\ and\ \bibinfo {author} {\bibfnamefont {G.~A.}\ \bibnamefont
  {Somorjai}},\ }\bibfield  {title} {\bibinfo {title} {Surface premelting of
  ice},\ }\href {https://doi.org/10.1021/jp071102f} {\bibfield  {journal}
  {\bibinfo  {journal} {J. Phys. Chem. C}\ }\textbf {\bibinfo {volume} {111}},\
  \bibinfo {pages} {9631} (\bibinfo {year} {2007})}\BibitemShut {NoStop}%
\bibitem [{\citenamefont {Llombart}\ \emph
  {et~al.}(2020{\natexlab{a}})\citenamefont {Llombart}, \citenamefont {Noya},
  \citenamefont {Sibley}, \citenamefont {Archer},\ and\ \citenamefont
  {MacDowell}}]{llombart20}%
  \BibitemOpen
  \bibfield  {author} {\bibinfo {author} {\bibfnamefont {P.}~\bibnamefont
  {Llombart}}, \bibinfo {author} {\bibfnamefont {E.~G.}\ \bibnamefont {Noya}},
  \bibinfo {author} {\bibfnamefont {D.~N.}\ \bibnamefont {Sibley}}, \bibinfo
  {author} {\bibfnamefont {A.~J.}\ \bibnamefont {Archer}},\ and\ \bibinfo
  {author} {\bibfnamefont {L.~G.}\ \bibnamefont {MacDowell}},\ }\bibfield
  {title} {\bibinfo {title} {Rounded layering transitions on the surface of
  ice},\ }\href {https://doi.org/10.1103/PhysRevLett.124.065702} {\bibfield
  {journal} {\bibinfo  {journal} {Phys. Rev. Lett.}\ }\textbf {\bibinfo
  {volume} {124}},\ \bibinfo {pages} {065702} (\bibinfo {year}
  {2020}{\natexlab{a}})}\BibitemShut {NoStop}%
\bibitem [{\citenamefont {Sibley}\ \emph {et~al.}(2021)\citenamefont {Sibley},
  \citenamefont {Llombart}, \citenamefont {Noya}, \citenamefont {Archer},\ and\
  \citenamefont {MacDowell}}]{sibley21}%
  \BibitemOpen
  \bibfield  {author} {\bibinfo {author} {\bibfnamefont {D.}~\bibnamefont
  {Sibley}}, \bibinfo {author} {\bibfnamefont {P.}~\bibnamefont {Llombart}},
  \bibinfo {author} {\bibfnamefont {E.~G.}\ \bibnamefont {Noya}}, \bibinfo
  {author} {\bibfnamefont {A.}~\bibnamefont {Archer}},\ and\ \bibinfo {author}
  {\bibfnamefont {L.~G.}\ \bibnamefont {MacDowell}},\ }\bibfield  {title}
  {\bibinfo {title} {How ice grows from premelting films and liquid droplets},\
  }\href@noop {} {\bibfield  {journal} {\bibinfo  {journal} {Nat. Commun.}\
  }\textbf {\bibinfo {volume} {12}},\ \bibinfo {pages} {239} (\bibinfo {year}
  {2021})}\BibitemShut {NoStop}%
\bibitem [{\citenamefont {Lipowsky}\ \emph {et~al.}(1989)\citenamefont
  {Lipowsky}, \citenamefont {Breuer}, \citenamefont {Prince},\ and\
  \citenamefont {Bonzel}}]{lipowsky89}%
  \BibitemOpen
  \bibfield  {author} {\bibinfo {author} {\bibfnamefont {R.}~\bibnamefont
  {Lipowsky}}, \bibinfo {author} {\bibfnamefont {U.}~\bibnamefont {Breuer}},
  \bibinfo {author} {\bibfnamefont {K.~C.}\ \bibnamefont {Prince}},\ and\
  \bibinfo {author} {\bibfnamefont {H.~P.}\ \bibnamefont {Bonzel}},\ }\bibfield
   {title} {\bibinfo {title} {Multicomponent order parameter for surface
  melting},\ }\href {https://doi.org/10.1103/PhysRevLett.62.913} {\bibfield
  {journal} {\bibinfo  {journal} {Phys. Rev. Lett.}\ }\textbf {\bibinfo
  {volume} {62}},\ \bibinfo {pages} {913} (\bibinfo {year} {1989})}\BibitemShut
  {NoStop}%
\bibitem [{\citenamefont {Limmer}\ and\ \citenamefont
  {Chandler}(2014)}]{limmer14}%
  \BibitemOpen
  \bibfield  {author} {\bibinfo {author} {\bibfnamefont {D.~T.}\ \bibnamefont
  {Limmer}}\ and\ \bibinfo {author} {\bibfnamefont {D.}~\bibnamefont
  {Chandler}},\ }\bibfield  {title} {\bibinfo {title} {Premelting,
  fluctuations, and coarse-graining of water-ice interfaces},\ }\href
  {https://doi.org/http://dx.doi.org/10.1063/1.4895399} {\bibfield  {journal}
  {\bibinfo  {journal} {J. Chem. Phys.}\ }\textbf {\bibinfo {volume} {141}},\
  \bibinfo {pages} {18C505} (\bibinfo {year} {2014})}\BibitemShut {NoStop}%
\bibitem [{\citenamefont {Li}\ \emph {et~al.}(2019)\citenamefont {Li},
  \citenamefont {Bier}, \citenamefont {Mars}, \citenamefont {Weiss},
  \citenamefont {Dippel}, \citenamefont {Gutowski}, \citenamefont
  {Honkimäki},\ and\ \citenamefont {Mezger}}]{li19}%
  \BibitemOpen
  \bibfield  {author} {\bibinfo {author} {\bibfnamefont {H.}~\bibnamefont
  {Li}}, \bibinfo {author} {\bibfnamefont {M.}~\bibnamefont {Bier}}, \bibinfo
  {author} {\bibfnamefont {J.}~\bibnamefont {Mars}}, \bibinfo {author}
  {\bibfnamefont {H.}~\bibnamefont {Weiss}}, \bibinfo {author} {\bibfnamefont
  {A.-C.}\ \bibnamefont {Dippel}}, \bibinfo {author} {\bibfnamefont
  {O.}~\bibnamefont {Gutowski}}, \bibinfo {author} {\bibfnamefont
  {V.}~\bibnamefont {Honkimäki}},\ and\ \bibinfo {author} {\bibfnamefont
  {M.}~\bibnamefont {Mezger}},\ }\bibfield  {title} {\bibinfo {title}
  {Interfacial premelting of ice in nano composite materials},\ }\href
  {https://doi.org/10.1039/C8CP05604H} {\bibfield  {journal} {\bibinfo
  {journal} {Phys. Chem. Chem. Phys.}\ }\textbf {\bibinfo {volume} {21}},\
  \bibinfo {pages} {3734} (\bibinfo {year} {2019})}\BibitemShut {NoStop}%
\bibitem [{\citenamefont {Chernov}\ and\ \citenamefont
  {Mikheev}(1988)}]{chernov88}%
  \BibitemOpen
  \bibfield  {author} {\bibinfo {author} {\bibfnamefont {A.~A.}\ \bibnamefont
  {Chernov}}\ and\ \bibinfo {author} {\bibfnamefont {L.~V.}\ \bibnamefont
  {Mikheev}},\ }\bibfield  {title} {\bibinfo {title} {Wetting of solid surfaces
  by a structured simple liquid: Effect of fluctuations},\ }\href
  {https://doi.org/10.1103/PhysRevLett.60.2488} {\bibfield  {journal} {\bibinfo
   {journal} {Phys. Rev. Lett.}\ }\textbf {\bibinfo {volume} {60}},\ \bibinfo
  {pages} {2488} (\bibinfo {year} {1988})}\BibitemShut {NoStop}%
\bibitem [{\citenamefont {Henderson}(1994)}]{henderson94}%
  \BibitemOpen
  \bibfield  {author} {\bibinfo {author} {\bibfnamefont {J.~R.}\ \bibnamefont
  {Henderson}},\ }\bibfield  {title} {\bibinfo {title} {Wetting phenomena and
  the decay of correlations at fluid interfaces},\ }\href
  {https://doi.org/10.1103/PhysRevE.50.4836} {\bibfield  {journal} {\bibinfo
  {journal} {Phys. Rev. E}\ }\textbf {\bibinfo {volume} {50}},\ \bibinfo
  {pages} {4836} (\bibinfo {year} {1994})}\BibitemShut {NoStop}%
\bibitem [{\citenamefont {Henderson}(2005)}]{henderson05}%
  \BibitemOpen
  \bibfield  {author} {\bibinfo {author} {\bibfnamefont {J.~R.}\ \bibnamefont
  {Henderson}},\ }\bibfield  {title} {\bibinfo {title} {Statistical mechanics
  of the disjoining pressure of a planar film},\ }\href@noop {} {\bibfield
  {journal} {\bibinfo  {journal} {Phys. Rev. E}\ }\textbf {\bibinfo {volume}
  {72}},\ \bibinfo {pages} {051602} (\bibinfo {year} {2005})}\BibitemShut
  {NoStop}%
\bibitem [{\citenamefont {Elbaum}\ and\ \citenamefont
  {Schick}(1991{\natexlab{a}})}]{elbaum91b}%
  \BibitemOpen
  \bibfield  {author} {\bibinfo {author} {\bibfnamefont {M.}~\bibnamefont
  {Elbaum}}\ and\ \bibinfo {author} {\bibfnamefont {M.}~\bibnamefont
  {Schick}},\ }\bibfield  {title} {\bibinfo {title} {Application of the theory
  of dispersion forces to the surface melting of ice},\ }\href@noop {}
  {\bibfield  {journal} {\bibinfo  {journal} {Phys. Rev. Lett.}\ }\textbf
  {\bibinfo {volume} {66}},\ \bibinfo {pages} {1713} (\bibinfo {year}
  {1991}{\natexlab{a}})}\BibitemShut {NoStop}%
\bibitem [{\citenamefont {Parsegian}(2005)}]{parsegian05}%
  \BibitemOpen
  \bibfield  {author} {\bibinfo {author} {\bibfnamefont {V.~A.}\ \bibnamefont
  {Parsegian}},\ }\href@noop {} {\emph {\bibinfo {title} {Van der Waals
  Forces}}}\ (\bibinfo  {publisher} {Cambridge University Press},\ \bibinfo
  {address} {Cambridge},\ \bibinfo {year} {2005})\ pp.\ \bibinfo {pages}
  {1--311}\BibitemShut {NoStop}%
\bibitem [{\citenamefont {Parsegian}\ and\ \citenamefont
  {Weiss}(1981)}]{parsegian81}%
  \BibitemOpen
  \bibfield  {author} {\bibinfo {author} {\bibfnamefont {V.~A.}\ \bibnamefont
  {Parsegian}}\ and\ \bibinfo {author} {\bibfnamefont {G.~H.}\ \bibnamefont
  {Weiss}},\ }\bibfield  {title} {\bibinfo {title} {Spectroscopic parameters
  for computation of van der waals forces},\ }\href@noop {} {\bibfield
  {journal} {\bibinfo  {journal} {J. Colloid. Interface Sci.}\ }\textbf
  {\bibinfo {volume} {81}},\ \bibinfo {pages} {285} (\bibinfo {year}
  {1981})}\BibitemShut {NoStop}%
\bibitem [{\citenamefont {Roth}\ and\ \citenamefont {Lenhoff}(1996)}]{roth96}%
  \BibitemOpen
  \bibfield  {author} {\bibinfo {author} {\bibfnamefont {C.~M.}\ \bibnamefont
  {Roth}}\ and\ \bibinfo {author} {\bibfnamefont {A.~M.}\ \bibnamefont
  {Lenhoff}},\ }\bibfield  {title} {\bibinfo {title} {Improved parametric
  representation of water dielectric data for lifshit z theory calculations},\
  }\href {https://doi.org/https://doi.org/10.1006/jcis.1996.0261} {\bibfield
  {journal} {\bibinfo  {journal} {J. Colloid. Interface Sci.}\ }\textbf
  {\bibinfo {volume} {179}},\ \bibinfo {pages} {637 } (\bibinfo {year}
  {1996})}\BibitemShut {NoStop}%
\bibitem [{\citenamefont {Dagastine}\ \emph {et~al.}(2000)\citenamefont
  {Dagastine}, \citenamefont {Prieve},\ and\ \citenamefont
  {White}}]{dagastine00}%
  \BibitemOpen
  \bibfield  {author} {\bibinfo {author} {\bibfnamefont {R.~R.}\ \bibnamefont
  {Dagastine}}, \bibinfo {author} {\bibfnamefont {D.~C.}\ \bibnamefont
  {Prieve}},\ and\ \bibinfo {author} {\bibfnamefont {L.~R.}\ \bibnamefont
  {White}},\ }\bibfield  {title} {\bibinfo {title} {The dielectric function for
  water and its application to van der waals forces},\ }\href
  {https://doi.org/https://doi.org/10.1006/jcis.2000.7164} {\bibfield
  {journal} {\bibinfo  {journal} {J. Colloid. Interface Sci.}\ }\textbf
  {\bibinfo {volume} {231}},\ \bibinfo {pages} {351 } (\bibinfo {year}
  {2000})}\BibitemShut {NoStop}%
\bibitem [{\citenamefont {Fernández-Varea}\ and\ \citenamefont
  {Garcia-Molina}(2000)}]{fernandez00}%
  \BibitemOpen
  \bibfield  {author} {\bibinfo {author} {\bibfnamefont {J.~M.}\ \bibnamefont
  {Fernández-Varea}}\ and\ \bibinfo {author} {\bibfnamefont {R.}~\bibnamefont
  {Garcia-Molina}},\ }\bibfield  {title} {\bibinfo {title} {Hamaker constants
  of systems involving water obtained from a dielectric function that fulfills
  the f sum rule},\ }\href
  {https://doi.org/https://doi.org/10.1006/jcis.2000.7140} {\bibfield
  {journal} {\bibinfo  {journal} {J. Colloid. Interface Sci.}\ }\textbf
  {\bibinfo {volume} {231}},\ \bibinfo {pages} {394} (\bibinfo {year}
  {2000})}\BibitemShut {NoStop}%
\bibitem [{\citenamefont {Wang}\ and\ \citenamefont {Nguyen}(2017)}]{wang17}%
  \BibitemOpen
  \bibfield  {author} {\bibinfo {author} {\bibfnamefont {J.}~\bibnamefont
  {Wang}}\ and\ \bibinfo {author} {\bibfnamefont {A.~V.}\ \bibnamefont
  {Nguyen}},\ }\bibfield  {title} {\bibinfo {title} {A review on data and
  predictions of water dielectric spectra for calcul ations of van der waals
  surface forces},\ }\href
  {https://doi.org/https://doi.org/10.1016/j.cis.2017.10.004} {\bibfield
  {journal} {\bibinfo  {journal} {Adv. Colloid Interface Sci.}\ }\textbf
  {\bibinfo {volume} {250}},\ \bibinfo {pages} {54 } (\bibinfo {year}
  {2017})}\BibitemShut {NoStop}%
\bibitem [{\citenamefont {Luengo}(2019)}]{luengo19}%
  \BibitemOpen
  \bibfield  {author} {\bibinfo {author} {\bibfnamefont {J.}~\bibnamefont
  {Luengo}},\ }\href@noop {} {\bibinfo {title} {Fuerzas de van der waals en la
  superficie del hielo}},\ \bibinfo {howpublished} {Degree Thesis} (\bibinfo
  {year} {2019}),\ \bibinfo {note} {universidad Complutense de
  Madrid}\BibitemShut {NoStop}%
\bibitem [{\citenamefont {Luengo}\ and\ \citenamefont
  {MacDowell}(2020)}]{luengo20}%
  \BibitemOpen
  \bibfield  {author} {\bibinfo {author} {\bibfnamefont {J.}~\bibnamefont
  {Luengo}}\ and\ \bibinfo {author} {\bibfnamefont {L.}~\bibnamefont
  {MacDowell}},\ }\emph {\bibinfo {title} {Van der Waals Forces at Ice Surfaces
  with Atmospheric Interest}},\ \href@noop {} {Master's thesis},\ \bibinfo
  {school} {Facultad de Ciencias} (\bibinfo {year} {2020})\BibitemShut
  {NoStop}%
\bibitem [{\citenamefont {Fiedler}\ \emph {et~al.}(2020)\citenamefont
  {Fiedler}, \citenamefont {Bostr{\"o}m}, \citenamefont {Persson},
  \citenamefont {Brevik}, \citenamefont {Corkery}, \citenamefont {Buhmann},\
  and\ \citenamefont {Parsons}}]{fiedler20}%
  \BibitemOpen
  \bibfield  {author} {\bibinfo {author} {\bibfnamefont {J.}~\bibnamefont
  {Fiedler}}, \bibinfo {author} {\bibfnamefont {M.}~\bibnamefont
  {Bostr{\"o}m}}, \bibinfo {author} {\bibfnamefont {C.}~\bibnamefont
  {Persson}}, \bibinfo {author} {\bibfnamefont {I.}~\bibnamefont {Brevik}},
  \bibinfo {author} {\bibfnamefont {R.}~\bibnamefont {Corkery}}, \bibinfo
  {author} {\bibfnamefont {S.~Y.}\ \bibnamefont {Buhmann}},\ and\ \bibinfo
  {author} {\bibfnamefont {D.~F.}\ \bibnamefont {Parsons}},\ }\bibfield
  {title} {\bibinfo {title} {Full-spectrum high-resolution modeling of the
  dielectric function of water},\ }\href
  {https://doi.org/10.1021/acs.jpcb.0c00410} {\bibfield  {journal} {\bibinfo
  {journal} {J. Phys. Chem. B}\ }\textbf {\bibinfo {volume} {124}},\ \bibinfo
  {pages} {3103} (\bibinfo {year} {2020})},\ \bibinfo {note} {pMID: 32208624},\
  \Eprint {https://arxiv.org/abs/https://doi.org/10.1021/acs.jpcb.0c00410}
  {https://doi.org/10.1021/acs.jpcb.0c00410} \BibitemShut {NoStop}%
\bibitem [{\citenamefont {Luengo-M{\'a}rquez}\ and\ \citenamefont
  {MacDowell}(2021)}]{luengo21}%
  \BibitemOpen
  \bibfield  {author} {\bibinfo {author} {\bibfnamefont {J.}~\bibnamefont
  {Luengo-M{\'a}rquez}}\ and\ \bibinfo {author} {\bibfnamefont {L.~G.}\
  \bibnamefont {MacDowell}},\ }\bibfield  {title} {\bibinfo {title} {Lifshitz
  theory of wetting films at three phase coexistence: The case of ice
  nucleation on silver iodide (agi)},\ }\href@noop {} {\bibfield  {journal}
  {\bibinfo  {journal} {Journal of Colloid and Interface Science}\ }\textbf
  {\bibinfo {volume} {590}},\ \bibinfo {pages} {527} (\bibinfo {year}
  {2021})}\BibitemShut {NoStop}%
\bibitem [{\citenamefont {Gudarzi}\ and\ \citenamefont
  {Aboutalebi}(2021)}]{gudarzi21}%
  \BibitemOpen
  \bibfield  {author} {\bibinfo {author} {\bibfnamefont {M.~M.}\ \bibnamefont
  {Gudarzi}}\ and\ \bibinfo {author} {\bibfnamefont {S.~H.}\ \bibnamefont
  {Aboutalebi}},\ }\bibfield  {title} {\bibinfo {title} {Self-consistent
  dielectric functions of materials: Toward accurate computation of casimir van
  der waals forces},\ }\href {https://doi.org/10.1126/sciadv.abg2272}
  {\bibfield  {journal} {\bibinfo  {journal} {Sci. Adv.}\ }\textbf {\bibinfo
  {volume} {7}},\ \bibinfo {pages} {eabg2272} (\bibinfo {year} {2021})},\
  \Eprint
  {https://arxiv.org/abs/https://www.science.org/doi/pdf/10.1126/sciadv.abg2272}
  {https://www.science.org/doi/pdf/10.1126/sciadv.abg2272} \BibitemShut
  {NoStop}%
\bibitem [{\citenamefont {Zelsmann}(1995)}]{zelsmann95}%
  \BibitemOpen
  \bibfield  {author} {\bibinfo {author} {\bibfnamefont {H.~R.}\ \bibnamefont
  {Zelsmann}},\ }\bibfield  {title} {\bibinfo {title} {Temperature dependence
  of the optical constants for liquid {H2O} and {D2O} in the far {IR} region},\
  }\href@noop {} {\bibfield  {journal} {\bibinfo  {journal} {J. Mol.
  Structure.}\ }\textbf {\bibinfo {volume} {350}},\ \bibinfo {pages} {95}
  (\bibinfo {year} {1995})}\BibitemShut {NoStop}%
\bibitem [{\citenamefont {Segelstein}(1981)}]{segelstein81}%
  \BibitemOpen
  \bibfield  {author} {\bibinfo {author} {\bibfnamefont {D.~J.}\ \bibnamefont
  {Segelstein}},\ }\emph {\bibinfo {title} {The complex refractive index of
  water}},\ \href@noop {} {Ph.D. thesis},\ \bibinfo  {school} {University of
  Missouri--Kansas City} (\bibinfo {year} {1981})\BibitemShut {NoStop}%
\bibitem [{\citenamefont {Wieliczka}\ \emph {et~al.}(1989)\citenamefont
  {Wieliczka}, \citenamefont {Weng},\ and\ \citenamefont
  {Querry}}]{wieliczka89}%
  \BibitemOpen
  \bibfield  {author} {\bibinfo {author} {\bibfnamefont {D.~M.}\ \bibnamefont
  {Wieliczka}}, \bibinfo {author} {\bibfnamefont {S.}~\bibnamefont {Weng}},\
  and\ \bibinfo {author} {\bibfnamefont {M.~R.}\ \bibnamefont {Querry}},\
  }\bibfield  {title} {\bibinfo {title} {Wedge shaped cell for highly absorbent
  liquids: infrared optical constants of water},\ }\href@noop {} {\bibfield
  {journal} {\bibinfo  {journal} {Applied optics}\ }\textbf {\bibinfo {volume}
  {28}},\ \bibinfo {pages} {1714} (\bibinfo {year} {1989})}\BibitemShut
  {NoStop}%
\bibitem [{\citenamefont {Bertie}\ and\ \citenamefont {Lan}(1996)}]{bertie96}%
  \BibitemOpen
  \bibfield  {author} {\bibinfo {author} {\bibfnamefont {J.~E.}\ \bibnamefont
  {Bertie}}\ and\ \bibinfo {author} {\bibfnamefont {Z.}~\bibnamefont {Lan}},\
  }\bibfield  {title} {\bibinfo {title} {Infrared intensities of liquids xx:
  The intensity of the oh stretching band of liquid water revisited, and the
  best current values of the optical constants of h2o(l) at 25~c between 15,000
  and 1 cm-1},\ }\href
  {http://www.osapublishing.org/as/abstract.cfm?URI=as-50-8-1047} {\bibfield
  {journal} {\bibinfo  {journal} {Appl. Spectrosc.}\ }\textbf {\bibinfo
  {volume} {50}},\ \bibinfo {pages} {1047} (\bibinfo {year}
  {1996})}\BibitemShut {NoStop}%
\bibitem [{\citenamefont {Heller}\ \emph {et~al.}(1974)\citenamefont {Heller},
  \citenamefont {Hamm}, \citenamefont {Birkhoff},\ and\ \citenamefont
  {Painter}}]{heller74}%
  \BibitemOpen
  \bibfield  {author} {\bibinfo {author} {\bibfnamefont {J.~M.}\ \bibnamefont
  {Heller}}, \bibinfo {author} {\bibfnamefont {R.~N.}\ \bibnamefont {Hamm}},
  \bibinfo {author} {\bibfnamefont {R.~D.}\ \bibnamefont {Birkhoff}},\ and\
  \bibinfo {author} {\bibfnamefont {L.~R.}\ \bibnamefont {Painter}},\
  }\bibfield  {title} {\bibinfo {title} {Collective oscillation in liquid
  water},\ }\href {https://doi.org/10.1063/1.1681563} {\bibfield  {journal}
  {\bibinfo  {journal} {J. Chem. Phys.}\ }\textbf {\bibinfo {volume} {60}},\
  \bibinfo {pages} {3483} (\bibinfo {year} {1974})},\ \Eprint
  {https://arxiv.org/abs/https://doi.org/10.1063/1.1681563}
  {https://doi.org/10.1063/1.1681563} \BibitemShut {NoStop}%
\bibitem [{\citenamefont {Hayashi}\ and\ \citenamefont
  {Hiraoka}(2015)}]{hayashi15}%
  \BibitemOpen
  \bibfield  {author} {\bibinfo {author} {\bibfnamefont {H.}~\bibnamefont
  {Hayashi}}\ and\ \bibinfo {author} {\bibfnamefont {N.}~\bibnamefont
  {Hiraoka}},\ }\bibfield  {title} {\bibinfo {title} {Accurate measurements of
  dielectric and optical functions of liquid water and liquid benzene in the
  vuv region (1--100 ev) using small-angle inelastic x-ray scattering},\
  }\href@noop {} {\bibfield  {journal} {\bibinfo  {journal} {The Journal of
  Physical Chemistry B}\ }\textbf {\bibinfo {volume} {119}},\ \bibinfo {pages}
  {5609} (\bibinfo {year} {2015})}\BibitemShut {NoStop}%
\bibitem [{\citenamefont {Buckley}(1958)}]{buckley58}%
  \BibitemOpen
  \bibfield  {author} {\bibinfo {author} {\bibfnamefont {F.}~\bibnamefont
  {Buckley}},\ }\bibfield  {title} {\bibinfo {title} {Tables of dielectric
  dispersion data for pure liquids and dilute solutions},\ }\href@noop {}
  {\bibfield  {journal} {\bibinfo  {journal} {Natl. Bur. Stand. Circ.}\
  }\textbf {\bibinfo {volume} {589}},\ \bibinfo {pages} {7} (\bibinfo {year}
  {1958})}\BibitemShut {NoStop}%
\bibitem [{\citenamefont {Warren}\ and\ \citenamefont
  {Brandt}(2008)}]{warren08}%
  \BibitemOpen
  \bibfield  {author} {\bibinfo {author} {\bibfnamefont {S.~G.}\ \bibnamefont
  {Warren}}\ and\ \bibinfo {author} {\bibfnamefont {R.~E.}\ \bibnamefont
  {Brandt}},\ }\bibfield  {title} {\bibinfo {title} {Optical constants of ice
  from the ultraviolet to the microwave: A revised compilation},\ }\href@noop
  {} {\bibfield  {journal} {\bibinfo  {journal} {J. Geophys. Research}\
  }\textbf {\bibinfo {volume} {113}},\ \bibinfo {pages} {D14220} (\bibinfo
  {year} {2008})}\BibitemShut {NoStop}%
\bibitem [{\citenamefont {Seki}\ \emph {et~al.}(1981)\citenamefont {Seki},
  \citenamefont {Kobayashi},\ and\ \citenamefont {Nakahara}}]{seki81}%
  \BibitemOpen
  \bibfield  {author} {\bibinfo {author} {\bibfnamefont {M.}~\bibnamefont
  {Seki}}, \bibinfo {author} {\bibfnamefont {K.}~\bibnamefont {Kobayashi}},\
  and\ \bibinfo {author} {\bibfnamefont {J.}~\bibnamefont {Nakahara}},\
  }\bibfield  {title} {\bibinfo {title} {Optical spectra of hexagonal ice},\
  }\href {https://doi.org/10.1143/JPSJ.50.2643} {\bibfield  {journal} {\bibinfo
   {journal} {J. Phys. Soc. Jpn.}\ }\textbf {\bibinfo {volume} {50}},\ \bibinfo
  {pages} {2643} (\bibinfo {year} {1981})},\ \Eprint
  {https://arxiv.org/abs/https://doi.org/10.1143/JPSJ.50.2643}
  {https://doi.org/10.1143/JPSJ.50.2643} \BibitemShut {NoStop}%
\bibitem [{\citenamefont {Auty}\ and\ \citenamefont {Cole}(1952)}]{auty52}%
  \BibitemOpen
  \bibfield  {author} {\bibinfo {author} {\bibfnamefont {R.~P.}\ \bibnamefont
  {Auty}}\ and\ \bibinfo {author} {\bibfnamefont {R.~H.}\ \bibnamefont
  {Cole}},\ }\bibfield  {title} {\bibinfo {title} {Dielectric properties of ice
  and solid d2o},\ }\href@noop {} {\bibfield  {journal} {\bibinfo  {journal}
  {J. Chem. Phys.}\ }\textbf {\bibinfo {volume} {20}},\ \bibinfo {pages} {1309}
  (\bibinfo {year} {1952})}\BibitemShut {NoStop}%
\bibitem [{\citenamefont {Dzyaloshinskii}\ \emph {et~al.}(1961)\citenamefont
  {Dzyaloshinskii}, \citenamefont {Lifshitz},\ and\ \citenamefont
  {Pitaevskii}}]{dzyaloshinskii61}%
  \BibitemOpen
  \bibfield  {author} {\bibinfo {author} {\bibfnamefont {I.~E.}\ \bibnamefont
  {Dzyaloshinskii}}, \bibinfo {author} {\bibfnamefont {E.~M.}\ \bibnamefont
  {Lifshitz}},\ and\ \bibinfo {author} {\bibfnamefont {L.~P.}\ \bibnamefont
  {Pitaevskii}},\ }\bibfield  {title} {\bibinfo {title} {General theory of van
  der waals forces},\ }\href {http://stacks.iop.org/0038-5670/4/i=2/a=R01}
  {\bibfield  {journal} {\bibinfo  {journal} {Soviet Physics Uspekhi}\ }\textbf
  {\bibinfo {volume} {4}},\ \bibinfo {pages} {153} (\bibinfo {year}
  {1961})}\BibitemShut {NoStop}%
\bibitem [{\citenamefont {Ninham}\ \emph {et~al.}(1970)\citenamefont {Ninham},
  \citenamefont {Parsegian},\ and\ \citenamefont {Weiss}}]{ninham70b}%
  \BibitemOpen
  \bibfield  {author} {\bibinfo {author} {\bibfnamefont {B.~W.}\ \bibnamefont
  {Ninham}}, \bibinfo {author} {\bibfnamefont {V.~A.}\ \bibnamefont
  {Parsegian}},\ and\ \bibinfo {author} {\bibfnamefont {G.~H.}\ \bibnamefont
  {Weiss}},\ }\bibfield  {title} {\bibinfo {title} {On the macroscopic theory
  of temperature-dependent van der waals forces.},\ }\href@noop {} {\bibfield
  {journal} {\bibinfo  {journal} {J. Stat. Phys.}\ }\textbf {\bibinfo {volume}
  {2}},\ \bibinfo {pages} {323} (\bibinfo {year} {1970})}\BibitemShut {NoStop}%
\bibitem [{\citenamefont {Kresse}\ and\ \citenamefont {Hafner}(1993)}]{vasp1}%
  \BibitemOpen
  \bibfield  {author} {\bibinfo {author} {\bibfnamefont {G.}~\bibnamefont
  {Kresse}}\ and\ \bibinfo {author} {\bibfnamefont {J.}~\bibnamefont
  {Hafner}},\ }\bibfield  {title} {\bibinfo {title} {Vienna ab-initio
  simulation package},\ }\href@noop {} {\bibfield  {journal} {\bibinfo
  {journal} {Phys. Rev. B}\ }\textbf {\bibinfo {volume} {47}},\ \bibinfo
  {pages} {558} (\bibinfo {year} {1993})}\BibitemShut {NoStop}%
\bibitem [{\citenamefont {Kresse}\ and\ \citenamefont
  {Furthm{\"u}ller}(1996{\natexlab{a}})}]{vasp2}%
  \BibitemOpen
  \bibfield  {author} {\bibinfo {author} {\bibfnamefont {G.}~\bibnamefont
  {Kresse}}\ and\ \bibinfo {author} {\bibfnamefont {J.}~\bibnamefont
  {Furthm{\"u}ller}},\ }\bibfield  {title} {\bibinfo {title} {Efficiency of
  ab-initio total energy calculations for metals and semiconductors using a
  plane-wave basis set},\ }\href@noop {} {\bibfield  {journal} {\bibinfo
  {journal} {Computational Materials Science}\ }\textbf {\bibinfo {volume}
  {6}},\ \bibinfo {pages} {15} (\bibinfo {year}
  {1996}{\natexlab{a}})}\BibitemShut {NoStop}%
\bibitem [{\citenamefont {Kresse}\ and\ \citenamefont
  {Furthm{\"u}ller}(1996{\natexlab{b}})}]{vasp3}%
  \BibitemOpen
  \bibfield  {author} {\bibinfo {author} {\bibfnamefont {G.}~\bibnamefont
  {Kresse}}\ and\ \bibinfo {author} {\bibfnamefont {J.}~\bibnamefont
  {Furthm{\"u}ller}},\ }\bibfield  {title} {\bibinfo {title} {Efficient
  iterative schemes for ab initio total-energy calculations using a plane-wave
  basis set},\ }\href@noop {} {\bibfield  {journal} {\bibinfo  {journal} {Phys.
  Rev. B}\ }\textbf {\bibinfo {volume} {54}},\ \bibinfo {pages} {11169}
  (\bibinfo {year} {1996}{\natexlab{b}})}\BibitemShut {NoStop}%
\bibitem [{\citenamefont {Perdew}\ \emph {et~al.}(1996)\citenamefont {Perdew},
  \citenamefont {Burke},\ and\ \citenamefont {Ernzerhof}}]{perdew96}%
  \BibitemOpen
  \bibfield  {author} {\bibinfo {author} {\bibfnamefont {J.~P.}\ \bibnamefont
  {Perdew}}, \bibinfo {author} {\bibfnamefont {K.}~\bibnamefont {Burke}},\ and\
  \bibinfo {author} {\bibfnamefont {M.}~\bibnamefont {Ernzerhof}},\ }\bibfield
  {title} {\bibinfo {title} {Generalized gradient approximation made simple},\
  }\href@noop {} {\bibfield  {journal} {\bibinfo  {journal} {PRL}\ }\textbf
  {\bibinfo {volume} {77}},\ \bibinfo {pages} {3865} (\bibinfo {year}
  {1996})}\BibitemShut {NoStop}%
\bibitem [{\citenamefont {Shishkin}\ and\ \citenamefont
  {Kresse}(2006)}]{shishkin06}%
  \BibitemOpen
  \bibfield  {author} {\bibinfo {author} {\bibfnamefont {M.}~\bibnamefont
  {Shishkin}}\ and\ \bibinfo {author} {\bibfnamefont {G.}~\bibnamefont
  {Kresse}},\ }\bibfield  {title} {\bibinfo {title} {Implementation and
  performance of the frequency-dependent g w method within the paw framework},\
  }\href@noop {} {\bibfield  {journal} {\bibinfo  {journal} {Phys. Rev. B}\
  }\textbf {\bibinfo {volume} {74}},\ \bibinfo {pages} {035101} (\bibinfo
  {year} {2006})}\BibitemShut {NoStop}%
\bibitem [{\citenamefont {Fuchs}\ \emph {et~al.}(2007)\citenamefont {Fuchs},
  \citenamefont {Furthm{\"u}ller}, \citenamefont {Bechstedt}, \citenamefont
  {Shishkin},\ and\ \citenamefont {Kresse}}]{fuchs07}%
  \BibitemOpen
  \bibfield  {author} {\bibinfo {author} {\bibfnamefont {F.}~\bibnamefont
  {Fuchs}}, \bibinfo {author} {\bibfnamefont {J.}~\bibnamefont
  {Furthm{\"u}ller}}, \bibinfo {author} {\bibfnamefont {F.}~\bibnamefont
  {Bechstedt}}, \bibinfo {author} {\bibfnamefont {M.}~\bibnamefont
  {Shishkin}},\ and\ \bibinfo {author} {\bibfnamefont {G.}~\bibnamefont
  {Kresse}},\ }\bibfield  {title} {\bibinfo {title} {Quasiparticle band
  structure based on a generalized kohn-sham scheme},\ }\href@noop {}
  {\bibfield  {journal} {\bibinfo  {journal} {Phys. Rev. B}\ }\textbf {\bibinfo
  {volume} {76}},\ \bibinfo {pages} {115109} (\bibinfo {year}
  {2007})}\BibitemShut {NoStop}%
\bibitem [{\citenamefont {Gajdo{\v{s}}}\ \emph {et~al.}(2006)\citenamefont
  {Gajdo{\v{s}}}, \citenamefont {Hummer}, \citenamefont {Kresse}, \citenamefont
  {Furthm{\"u}ller},\ and\ \citenamefont {Bechstedt}}]{gajdos06}%
  \BibitemOpen
  \bibfield  {author} {\bibinfo {author} {\bibfnamefont {M.}~\bibnamefont
  {Gajdo{\v{s}}}}, \bibinfo {author} {\bibfnamefont {K.}~\bibnamefont
  {Hummer}}, \bibinfo {author} {\bibfnamefont {G.}~\bibnamefont {Kresse}},
  \bibinfo {author} {\bibfnamefont {J.}~\bibnamefont {Furthm{\"u}ller}},\ and\
  \bibinfo {author} {\bibfnamefont {F.}~\bibnamefont {Bechstedt}},\ }\bibfield
  {title} {\bibinfo {title} {Linear optical properties in the
  projector-augmented wave methodology},\ }\href@noop {} {\bibfield  {journal}
  {\bibinfo  {journal} {Phys. Rev. B}\ }\textbf {\bibinfo {volume} {73}},\
  \bibinfo {pages} {045112} (\bibinfo {year} {2006})}\BibitemShut {NoStop}%
\bibitem [{\citenamefont {Nunes}\ and\ \citenamefont {Gonze}(2001)}]{nunes01}%
  \BibitemOpen
  \bibfield  {author} {\bibinfo {author} {\bibfnamefont {R.}~\bibnamefont
  {Nunes}}\ and\ \bibinfo {author} {\bibfnamefont {X.}~\bibnamefont {Gonze}},\
  }\bibfield  {title} {\bibinfo {title} {Berry-phase treatment of the
  homogeneous electric field perturbation in insulators},\ }\href@noop {}
  {\bibfield  {journal} {\bibinfo  {journal} {Phys. Rev. B}\ }\textbf {\bibinfo
  {volume} {63}},\ \bibinfo {pages} {155107} (\bibinfo {year}
  {2001})}\BibitemShut {NoStop}%
\bibitem [{\citenamefont {Benet}\ \emph {et~al.}(2016)\citenamefont {Benet},
  \citenamefont {Llombart}, \citenamefont {Sanz},\ and\ \citenamefont
  {MacDowell}}]{benet16}%
  \BibitemOpen
  \bibfield  {author} {\bibinfo {author} {\bibfnamefont {J.}~\bibnamefont
  {Benet}}, \bibinfo {author} {\bibfnamefont {P.}~\bibnamefont {Llombart}},
  \bibinfo {author} {\bibfnamefont {E.}~\bibnamefont {Sanz}},\ and\ \bibinfo
  {author} {\bibfnamefont {L.~G.}\ \bibnamefont {MacDowell}},\ }\bibfield
  {title} {\bibinfo {title} {Premelting-induced smoothening of the ice-vapor
  interface},\ }\href {https://doi.org/10.1103/PhysRevLett.117.096101}
  {\bibfield  {journal} {\bibinfo  {journal} {Phys. Rev. Lett.}\ }\textbf
  {\bibinfo {volume} {117}},\ \bibinfo {pages} {096101} (\bibinfo {year}
  {2016})}\BibitemShut {NoStop}%
\bibitem [{\citenamefont {Benet}\ \emph {et~al.}(2019)\citenamefont {Benet},
  \citenamefont {Llombart}, \citenamefont {Sanz},\ and\ \citenamefont
  {MacDowell}}]{benet19}%
  \BibitemOpen
  \bibfield  {author} {\bibinfo {author} {\bibfnamefont {J.}~\bibnamefont
  {Benet}}, \bibinfo {author} {\bibfnamefont {P.}~\bibnamefont {Llombart}},
  \bibinfo {author} {\bibfnamefont {E.}~\bibnamefont {Sanz}},\ and\ \bibinfo
  {author} {\bibfnamefont {L.~G.}\ \bibnamefont {MacDowell}},\ }\bibfield
  {title} {\bibinfo {title} {Structure and fluctuations of the premelted liquid
  film of ice at the triple point},\ }\href
  {https://doi.org/10.1080/00268976.2019.1583388} {\bibfield  {journal}
  {\bibinfo  {journal} {Mol. Phys.}\ }\textbf {\bibinfo {volume} {117}},\
  \bibinfo {pages} {2846} (\bibinfo {year} {2019})}\BibitemShut {NoStop}%
\bibitem [{\citenamefont {Llombart}\ \emph {et~al.}(2019)\citenamefont
  {Llombart}, \citenamefont {Bergua}, \citenamefont {Noya},\ and\ \citenamefont
  {MacDowell}}]{llombart19}%
  \BibitemOpen
  \bibfield  {author} {\bibinfo {author} {\bibfnamefont {P.}~\bibnamefont
  {Llombart}}, \bibinfo {author} {\bibfnamefont {R.~M.}\ \bibnamefont
  {Bergua}}, \bibinfo {author} {\bibfnamefont {E.~G.}\ \bibnamefont {Noya}},\
  and\ \bibinfo {author} {\bibfnamefont {L.~G.}\ \bibnamefont {MacDowell}},\
  }\bibfield  {title} {\bibinfo {title} {Structure and water attachment rates
  of ice in the atmosphere: Role of nitrogen},\ }\href@noop {} {\bibfield
  {journal} {\bibinfo  {journal} {Phys. Chem. Chem. Phys}\ }\textbf {\bibinfo
  {volume} {21}},\ \bibinfo {pages} {19594} (\bibinfo {year}
  {2019})}\BibitemShut {NoStop}%
\bibitem [{\citenamefont {Llombart}\ \emph
  {et~al.}(2020{\natexlab{b}})\citenamefont {Llombart}, \citenamefont {Noya},\
  and\ \citenamefont {MacDowell}}]{llombart20b}%
  \BibitemOpen
  \bibfield  {author} {\bibinfo {author} {\bibfnamefont {P.}~\bibnamefont
  {Llombart}}, \bibinfo {author} {\bibfnamefont {E.~G.}\ \bibnamefont {Noya}},\
  and\ \bibinfo {author} {\bibfnamefont {L.~G.}\ \bibnamefont {MacDowell}},\
  }\bibfield  {title} {\bibinfo {title} {Surface phase transitions and crystal
  habits of ice in the atmosphere},\ }\bibfield  {journal} {\bibinfo  {journal}
  {Sci. Adv.}\ }\textbf {\bibinfo {volume} {6}},\ \href
  {https://doi.org/10.1126/sciadv.aay9322} {10.1126/sciadv.aay9322} (\bibinfo
  {year} {2020}{\natexlab{b}}),\ \Eprint
  {https://arxiv.org/abs/https://advances.sciencemag.org/content/6/21/eaay9322.full.pdf}
  {https://advances.sciencemag.org/content/6/21/eaay9322.full.pdf} \BibitemShut
  {NoStop}%
\bibitem [{\citenamefont {Tabazadeh}\ \emph {et~al.}(2002)\citenamefont
  {Tabazadeh}, \citenamefont {Djikaev},\ and\ \citenamefont
  {Reiss}}]{tabazadeh02}%
  \BibitemOpen
  \bibfield  {author} {\bibinfo {author} {\bibfnamefont {A.}~\bibnamefont
  {Tabazadeh}}, \bibinfo {author} {\bibfnamefont {Y.~S.}\ \bibnamefont
  {Djikaev}},\ and\ \bibinfo {author} {\bibfnamefont {H.}~\bibnamefont
  {Reiss}},\ }\bibfield  {title} {\bibinfo {title} {Surface crystallization of
  supercooled water in clouds},\ }\href
  {https://doi.org/10.1073/pnas.252640699} {\bibfield  {journal} {\bibinfo
  {journal} {Proceedings of the National Academy of Sciences}\ }\textbf
  {\bibinfo {volume} {99}},\ \bibinfo {pages} {15873} (\bibinfo {year}
  {2002})},\ \Eprint
  {https://arxiv.org/abs/https://www.pnas.org/content/99/25/15873.full.pdf}
  {https://www.pnas.org/content/99/25/15873.full.pdf} \BibitemShut {NoStop}%
\bibitem [{\citenamefont {Shaw}\ \emph {et~al.}(2005)\citenamefont {Shaw},
  \citenamefont {Durant},\ and\ \citenamefont {Mi}}]{shaw05}%
  \BibitemOpen
  \bibfield  {author} {\bibinfo {author} {\bibfnamefont {R.~A.}\ \bibnamefont
  {Shaw}}, \bibinfo {author} {\bibfnamefont {A.~J.}\ \bibnamefont {Durant}},\
  and\ \bibinfo {author} {\bibfnamefont {Y.}~\bibnamefont {Mi}},\ }\bibfield
  {title} {\bibinfo {title} {Heterogeneous surface crystallization observed in
  undercooled water},\ }\href {https://doi.org/10.1021/jp0506336} {\bibfield
  {journal} {\bibinfo  {journal} {J. Phys. Chem. B}\ }\textbf {\bibinfo
  {volume} {109}},\ \bibinfo {pages} {9865} (\bibinfo {year} {2005})},\
  \bibinfo {note} {pMID: 16852192},\ \Eprint
  {https://arxiv.org/abs/https://doi.org/10.1021/jp0506336}
  {https://doi.org/10.1021/jp0506336} \BibitemShut {NoStop}%
\bibitem [{\citenamefont {Li}\ \emph {et~al.}(2013)\citenamefont {Li},
  \citenamefont {Donadio},\ and\ \citenamefont {Galli}}]{li13}%
  \BibitemOpen
  \bibfield  {author} {\bibinfo {author} {\bibfnamefont {T.}~\bibnamefont
  {Li}}, \bibinfo {author} {\bibfnamefont {D.}~\bibnamefont {Donadio}},\ and\
  \bibinfo {author} {\bibfnamefont {G.}~\bibnamefont {Galli}},\ }\bibfield
  {title} {\bibinfo {title} {Ice nucleation at the nanoscale probes no man’s
  land of water},\ }\href@noop {} {\bibfield  {journal} {\bibinfo  {journal}
  {Nuovo Cimento}\ }\textbf {\bibinfo {volume} {4}},\ \bibinfo {pages} {1887}
  (\bibinfo {year} {2013})}\BibitemShut {NoStop}%
\bibitem [{\citenamefont {Hussain}\ and\ \citenamefont
  {Haji-Akbari}(2021)}]{hussain21}%
  \BibitemOpen
  \bibfield  {author} {\bibinfo {author} {\bibfnamefont {S.}~\bibnamefont
  {Hussain}}\ and\ \bibinfo {author} {\bibfnamefont {A.}~\bibnamefont
  {Haji-Akbari}},\ }\bibfield  {title} {\bibinfo {title} {Role of nanoscale
  interfacial proximity in contact freezing in water},\ }\href
  {https://doi.org/10.1021/jacs.0c10663} {\bibfield  {journal} {\bibinfo
  {journal} {J. Am. Chem. Soc.}\ }\textbf {\bibinfo {volume} {143}},\ \bibinfo
  {pages} {2272} (\bibinfo {year} {2021})},\ \bibinfo {note} {pMID: 33507741},\
  \Eprint {https://arxiv.org/abs/https://doi.org/10.1021/jacs.0c10663}
  {https://doi.org/10.1021/jacs.0c10663} \BibitemShut {NoStop}%
\bibitem [{\citenamefont {Jorgensen}\ \emph {et~al.}(1983)\citenamefont
  {Jorgensen}, \citenamefont {Chandrasekhar}, \citenamefont {Madura},
  \citenamefont {Impey},\ and\ \citenamefont {Klein}}]{jorgensen83}%
  \BibitemOpen
  \bibfield  {author} {\bibinfo {author} {\bibfnamefont {W.~L.}\ \bibnamefont
  {Jorgensen}}, \bibinfo {author} {\bibfnamefont {J.}~\bibnamefont
  {Chandrasekhar}}, \bibinfo {author} {\bibfnamefont {J.~D.}\ \bibnamefont
  {Madura}}, \bibinfo {author} {\bibfnamefont {R.~W.}\ \bibnamefont {Impey}},\
  and\ \bibinfo {author} {\bibfnamefont {M.~L.}\ \bibnamefont {Klein}},\
  }\bibfield  {title} {\bibinfo {title} {Comparison of simple potential
  functions for simulating liquid water},\ }\href@noop {} {\bibfield  {journal}
  {\bibinfo  {journal} {J. Phys. Chem.}\ }\textbf {\bibinfo {volume} {79}},\
  \bibinfo {pages} {926} (\bibinfo {year} {1983})}\BibitemShut {NoStop}%
\bibitem [{\citenamefont {Berendsen}\ \emph {et~al.}(1987)\citenamefont
  {Berendsen}, \citenamefont {Grigera},\ and\ \citenamefont
  {Straatsma}}]{berendsen87}%
  \BibitemOpen
  \bibfield  {author} {\bibinfo {author} {\bibfnamefont {H.~J.~C.}\
  \bibnamefont {Berendsen}}, \bibinfo {author} {\bibfnamefont {J.~R.}\
  \bibnamefont {Grigera}},\ and\ \bibinfo {author} {\bibfnamefont {T.~P.}\
  \bibnamefont {Straatsma}},\ }\bibfield  {title} {\bibinfo {title} {The
  missing term in effective pair potentials},\ }\href@noop {} {\bibfield
  {journal} {\bibinfo  {journal} {J. Phys. Chem.}\ }\textbf {\bibinfo {volume}
  {91}},\ \bibinfo {pages} {6269} (\bibinfo {year} {1987})}\BibitemShut
  {NoStop}%
\bibitem [{\citenamefont {Abascal}\ \emph {et~al.}(2005)\citenamefont
  {Abascal}, \citenamefont {Sanz}, \citenamefont {Fernandez},\ and\
  \citenamefont {Vega}}]{abascal05}%
  \BibitemOpen
  \bibfield  {author} {\bibinfo {author} {\bibfnamefont {J.~L.~F.}\
  \bibnamefont {Abascal}}, \bibinfo {author} {\bibfnamefont {E.}~\bibnamefont
  {Sanz}}, \bibinfo {author} {\bibfnamefont {R.~G.}\ \bibnamefont
  {Fernandez}},\ and\ \bibinfo {author} {\bibfnamefont {C.}~\bibnamefont
  {Vega}},\ }\bibfield  {title} {\bibinfo {title} {A potential model for the
  study of ices and amorphous water: {TIP4P/Ice}},\ }\href@noop {} {\bibfield
  {journal} {\bibinfo  {journal} {J. Chem. Phys.}\ }\textbf {\bibinfo {volume}
  {122}},\ \bibinfo {pages} {234511} (\bibinfo {year} {2005})}\BibitemShut
  {NoStop}%
\bibitem [{\citenamefont {Abascal}\ and\ \citenamefont
  {Vega}(2005)}]{abascal05b}%
  \BibitemOpen
  \bibfield  {author} {\bibinfo {author} {\bibfnamefont {J.~L.~F.}\
  \bibnamefont {Abascal}}\ and\ \bibinfo {author} {\bibfnamefont
  {C.}~\bibnamefont {Vega}},\ }\bibfield  {title} {\bibinfo {title} {A general
  purpose model for the condensed phases of water: Tip4p/2005},\ }\href@noop {}
  {\bibfield  {journal} {\bibinfo  {journal} {J. Chem. Phys.}\ }\textbf
  {\bibinfo {volume} {123}},\ \bibinfo {pages} {234505} (\bibinfo {year}
  {2005})}\BibitemShut {NoStop}%
\bibitem [{\citenamefont {Henriques}\ and\ \citenamefont
  {Skepö}(2016)}]{henriques16}%
  \BibitemOpen
  \bibfield  {author} {\bibinfo {author} {\bibfnamefont {J.}~\bibnamefont
  {Henriques}}\ and\ \bibinfo {author} {\bibfnamefont {M.}~\bibnamefont
  {Skepö}},\ }\bibfield  {title} {\bibinfo {title} {Molecular dynamics
  simulations of intrinsically disordered proteins: On the accuracy of the
  tip4p-d water model and the representativeness of protein disorder models},\
  }\href {https://doi.org/10.1021/acs.jctc.6b00429} {\bibfield  {journal}
  {\bibinfo  {journal} {Journal of Chemical Theory and Computation}\ }\textbf
  {\bibinfo {volume} {12}},\ \bibinfo {pages} {3407} (\bibinfo {year}
  {2016})},\ \bibinfo {note} {pMID: 27243806},\ \Eprint
  {https://arxiv.org/abs/https://doi.org/10.1021/acs.jctc.6b00429}
  {https://doi.org/10.1021/acs.jctc.6b00429} \BibitemShut {NoStop}%
\bibitem [{\citenamefont {Cohen-Tannoudji}\ \emph {et~al.}(2005)\citenamefont
  {Cohen-Tannoudji}, \citenamefont {Diu},\ and\ \citenamefont
  {Laloe}}]{cohen05b}%
  \BibitemOpen
  \bibfield  {author} {\bibinfo {author} {\bibfnamefont {C.}~\bibnamefont
  {Cohen-Tannoudji}}, \bibinfo {author} {\bibfnamefont {B.}~\bibnamefont
  {Diu}},\ and\ \bibinfo {author} {\bibfnamefont {F.}~\bibnamefont {Laloe}},\
  }\href@noop {} {\emph {\bibinfo {title} {Quantum Mechanics, Volume 2}}}\
  (\bibinfo  {publisher} {Wiley-VCH},\ \bibinfo {address} {Paris},\ \bibinfo
  {year} {2005})\BibitemShut {NoStop}%
\bibitem [{\citenamefont {Hamaker}(1937)}]{hamaker37}%
  \BibitemOpen
  \bibfield  {author} {\bibinfo {author} {\bibfnamefont {H.}~\bibnamefont
  {Hamaker}},\ }\bibfield  {title} {\bibinfo {title} {The london—van der
  waals attraction between spherical particles},\ }\href
  {https://doi.org/https://doi.org/10.1016/S0031-8914(37)80203-7} {\bibfield
  {journal} {\bibinfo  {journal} {Physica}\ }\textbf {\bibinfo {volume} {4}},\
  \bibinfo {pages} {1058} (\bibinfo {year} {1937})}\BibitemShut {NoStop}%
\bibitem [{\citenamefont {Gregory}(1981)}]{gregory81}%
  \BibitemOpen
  \bibfield  {author} {\bibinfo {author} {\bibfnamefont {J.}~\bibnamefont
  {Gregory}},\ }\bibfield  {title} {\bibinfo {title} {Approximate expressions
  for retarded van der waals interaction},\ }\href
  {https://doi.org/https://doi.org/10.1016/0021-9797(81)90018-7} {\bibfield
  {journal} {\bibinfo  {journal} {J. Colloid. Interface Sci.}\ }\textbf
  {\bibinfo {volume} {83}},\ \bibinfo {pages} {138 } (\bibinfo {year}
  {1981})}\BibitemShut {NoStop}%
\bibitem [{\citenamefont {Israelachvili}(1991)}]{israelachvili11}%
  \BibitemOpen
  \bibfield  {author} {\bibinfo {author} {\bibfnamefont {J.~N.}\ \bibnamefont
  {Israelachvili}},\ }\href@noop {} {\emph {\bibinfo {title} {Intermolecular
  and Surfaces Forces}}},\ \bibinfo {edition} {3rd}\ ed.\ (\bibinfo
  {publisher} {Academic Press},\ \bibinfo {address} {London},\ \bibinfo {year}
  {1991})\ pp.\ \bibinfo {pages} {1--674}\BibitemShut {NoStop}%
\bibitem [{\citenamefont {MacDowell}(2019)}]{macdowell19}%
  \BibitemOpen
  \bibfield  {author} {\bibinfo {author} {\bibfnamefont {L.~G.}\ \bibnamefont
  {MacDowell}},\ }\bibfield  {title} {\bibinfo {title} {Surface van der waals
  forces in a nutshell},\ }\href {https://doi.org/10.1063/1.5089019} {\bibfield
   {journal} {\bibinfo  {journal} {J. Chem. Phys.}\ }\textbf {\bibinfo {volume}
  {150}},\ \bibinfo {pages} {081101} (\bibinfo {year} {2019})}\BibitemShut
  {NoStop}%
\bibitem [{\citenamefont {Ninham}\ and\ \citenamefont
  {Parsegian}(1970)}]{ninham70}%
  \BibitemOpen
  \bibfield  {author} {\bibinfo {author} {\bibfnamefont {B.~W.}\ \bibnamefont
  {Ninham}}\ and\ \bibinfo {author} {\bibfnamefont {V.~A.}\ \bibnamefont
  {Parsegian}},\ }\bibfield  {title} {\bibinfo {title} {Van der waals forces.
  special characteristics in lipid-water systems and a general method of
  calculation based on the lifshitz theory.},\ }\href@noop {} {\bibfield
  {journal} {\bibinfo  {journal} {Biophys. J.}\ }\textbf {\bibinfo {volume}
  {10}},\ \bibinfo {pages} {646} (\bibinfo {year} {1970})}\BibitemShut
  {NoStop}%
\bibitem [{\citenamefont {Dingfelder}\ \emph {et~al.}(1998)\citenamefont
  {Dingfelder}, \citenamefont {Hantke}, \citenamefont {Inokuti},\ and\
  \citenamefont {Paretzke}}]{dingfelder98}%
  \BibitemOpen
  \bibfield  {author} {\bibinfo {author} {\bibfnamefont {M.}~\bibnamefont
  {Dingfelder}}, \bibinfo {author} {\bibfnamefont {D.}~\bibnamefont {Hantke}},
  \bibinfo {author} {\bibfnamefont {M.}~\bibnamefont {Inokuti}},\ and\ \bibinfo
  {author} {\bibfnamefont {H.~G.}\ \bibnamefont {Paretzke}},\ }\bibfield
  {title} {\bibinfo {title} {Electron inelastic-scattering cross sections in
  liquid water},\ }\href
  {https://doi.org/https://doi.org/10.1016/S0969-806X(97)00317-4} {\bibfield
  {journal} {\bibinfo  {journal} {Radiation Phys. Chem.}\ }\textbf {\bibinfo
  {volume} {53}},\ \bibinfo {pages} {1} (\bibinfo {year} {1998})}\BibitemShut
  {NoStop}%
\bibitem [{\citenamefont {Emfietzoglou}\ \emph {et~al.}(2007)\citenamefont
  {Emfietzoglou}, \citenamefont {Nikjoo}, \citenamefont {Petsalakis},\ and\
  \citenamefont {Pathak}}]{emfietzoglou07}%
  \BibitemOpen
  \bibfield  {author} {\bibinfo {author} {\bibfnamefont {D.}~\bibnamefont
  {Emfietzoglou}}, \bibinfo {author} {\bibfnamefont {H.}~\bibnamefont
  {Nikjoo}}, \bibinfo {author} {\bibfnamefont {I.}~\bibnamefont {Petsalakis}},\
  and\ \bibinfo {author} {\bibfnamefont {A.}~\bibnamefont {Pathak}},\
  }\bibfield  {title} {\bibinfo {title} {A consistent dielectric response model
  for water ice over the whole energy–momentum plane},\ }\href
  {https://doi.org/https://doi.org/10.1016/j.nimb.2006.11.105} {\bibfield
  {journal} {\bibinfo  {journal} {Nuclear Inst. Meth. Phys.}\ }\textbf
  {\bibinfo {volume} {256}},\ \bibinfo {pages} {141 } (\bibinfo {year}
  {2007})},\ \bibinfo {note} {atomic Collisions in Solids}\BibitemShut
  {NoStop}%
\bibitem [{\citenamefont {Brendel}\ and\ \citenamefont
  {Bormann}(1992)}]{brendel92}%
  \BibitemOpen
  \bibfield  {author} {\bibinfo {author} {\bibfnamefont {R.}~\bibnamefont
  {Brendel}}\ and\ \bibinfo {author} {\bibfnamefont {D.}~\bibnamefont
  {Bormann}},\ }\bibfield  {title} {\bibinfo {title} {An infrared dielectric
  function model for amorphous solids},\ }\href
  {https://doi.org/10.1063/1.350737} {\bibfield  {journal} {\bibinfo  {journal}
  {Journal of Applied Physics}\ }\textbf {\bibinfo {volume} {71}},\ \bibinfo
  {pages} {1} (\bibinfo {year} {1992})},\ \Eprint
  {https://arxiv.org/abs/https://doi.org/10.1063/1.350737}
  {https://doi.org/10.1063/1.350737} \BibitemShut {NoStop}%
\bibitem [{\citenamefont {Orosco}\ and\ \citenamefont
  {Coimbra}(2018)}]{orosco18b}%
  \BibitemOpen
  \bibfield  {author} {\bibinfo {author} {\bibfnamefont {J.}~\bibnamefont
  {Orosco}}\ and\ \bibinfo {author} {\bibfnamefont {C.~F.~M.}\ \bibnamefont
  {Coimbra}},\ }\bibfield  {title} {\bibinfo {title} {On a causal dispersion
  model for the optical properties of metals},\ }\href
  {https://doi.org/10.1364/AO.57.005333} {\bibfield  {journal} {\bibinfo
  {journal} {Appl. Opt.}\ }\textbf {\bibinfo {volume} {57}},\ \bibinfo {pages}
  {5333} (\bibinfo {year} {2018})}\BibitemShut {NoStop}%
\bibitem [{\citenamefont {Lide}(1994)}]{lide94}%
  \BibitemOpen
  \bibfield  {author} {\bibinfo {author} {\bibfnamefont {D.}~\bibnamefont
  {Lide}},\ }\href@noop {} {\emph {\bibinfo {title} {Handbook of Chemistry and
  Physics}}}\ (\bibinfo  {publisher} {CRC Press},\ \bibinfo {year}
  {1994})\BibitemShut {NoStop}%
\bibitem [{\citenamefont {Li}\ \emph {et~al.}(2022)\citenamefont {Li},
  \citenamefont {Milton}, \citenamefont {Brevik}, \citenamefont {Malyi},
  \citenamefont {Thiyam}, \citenamefont {Persson}, \citenamefont {Parsons},\
  and\ \citenamefont {Bostrom}}]{li22}%
  \BibitemOpen
  \bibfield  {author} {\bibinfo {author} {\bibfnamefont {Y.}~\bibnamefont
  {Li}}, \bibinfo {author} {\bibfnamefont {K.~A.}\ \bibnamefont {Milton}},
  \bibinfo {author} {\bibfnamefont {I.}~\bibnamefont {Brevik}}, \bibinfo
  {author} {\bibfnamefont {O.~I.}\ \bibnamefont {Malyi}}, \bibinfo {author}
  {\bibfnamefont {P.}~\bibnamefont {Thiyam}}, \bibinfo {author} {\bibfnamefont
  {C.}~\bibnamefont {Persson}}, \bibinfo {author} {\bibfnamefont {D.~F.}\
  \bibnamefont {Parsons}},\ and\ \bibinfo {author} {\bibfnamefont
  {M.}~\bibnamefont {Bostrom}},\ }\href@noop {} {\bibfield  {journal} {\bibinfo
   {journal} {Phys. Rev. B}\ } (\bibinfo {year} {2022})},\ \bibinfo {note} {in
  press}\BibitemShut {NoStop}%
\bibitem [{\citenamefont {Luengo-M{\'a}rquez}\ and\ \citenamefont
  {MacDowell}(2022)}]{luengo22}%
  \BibitemOpen
  \bibfield  {author} {\bibinfo {author} {\bibfnamefont {J.}~\bibnamefont
  {Luengo-M{\'a}rquez}}\ and\ \bibinfo {author} {\bibfnamefont {L.~G.}\
  \bibnamefont {MacDowell}},\ }\bibfield  {title} {\bibinfo {title} {Analytical
  theory for the crossover from retarded to non-retarded interactions between
  metal plates},\ }\href@noop {} {\bibfield  {journal} {\bibinfo  {journal}
  {Journal of Physics: Condensed Matter}\ }\textbf {\bibinfo {volume} {34}},\
  \bibinfo {pages} {275701} (\bibinfo {year} {2022})}\BibitemShut {NoStop}%
\bibitem [{\citenamefont {Elbaum}\ \emph {et~al.}(1993)\citenamefont {Elbaum},
  \citenamefont {Lipson},\ and\ \citenamefont {Dash}}]{elbaum93}%
  \BibitemOpen
  \bibfield  {author} {\bibinfo {author} {\bibfnamefont {M.}~\bibnamefont
  {Elbaum}}, \bibinfo {author} {\bibfnamefont {S.~G.}\ \bibnamefont {Lipson}},\
  and\ \bibinfo {author} {\bibfnamefont {J.~G.}\ \bibnamefont {Dash}},\
  }\bibfield  {title} {\bibinfo {title} {Optical study of surface melting on
  ice},\ }\href@noop {} {\bibfield  {journal} {\bibinfo  {journal} {J. Cryst.
  Growth}\ }\textbf {\bibinfo {volume} {129}},\ \bibinfo {pages} {491}
  (\bibinfo {year} {1993})}\BibitemShut {NoStop}%
\bibitem [{\citenamefont {Tanner}(2013)}]{tanner13}%
  \BibitemOpen
  \bibfield  {author} {\bibinfo {author} {\bibfnamefont {D.~B.}\ \bibnamefont
  {Tanner}},\ }\href@noop {} {\emph {\bibinfo {title} {Optical Effects in
  Solids}}}\ (\bibinfo  {publisher} {Cambridge University Press},\ \bibinfo
  {address} {Cambridge},\ \bibinfo {year} {2013})\BibitemShut {NoStop}%
\bibitem [{\citenamefont {Wilen}\ \emph {et~al.}(1995)\citenamefont {Wilen},
  \citenamefont {Wettlaufer}, \citenamefont {Elbaum},\ and\ \citenamefont
  {Schick}}]{wilen95}%
  \BibitemOpen
  \bibfield  {author} {\bibinfo {author} {\bibfnamefont {L.~A.}\ \bibnamefont
  {Wilen}}, \bibinfo {author} {\bibfnamefont {J.~S.}\ \bibnamefont
  {Wettlaufer}}, \bibinfo {author} {\bibfnamefont {M.}~\bibnamefont {Elbaum}},\
  and\ \bibinfo {author} {\bibfnamefont {M.}~\bibnamefont {Schick}},\
  }\bibfield  {title} {\bibinfo {title} {Dispersion-force effects in
  interfacial premelting of ice},\ }\href
  {https://doi.org/10.1103/PhysRevB.52.12426} {\bibfield  {journal} {\bibinfo
  {journal} {Phys. Rev. B}\ }\textbf {\bibinfo {volume} {52}},\ \bibinfo
  {pages} {12426} (\bibinfo {year} {1995})}\BibitemShut {NoStop}%
\bibitem [{\citenamefont {French}\ \emph {et~al.}(2010)\citenamefont {French},
  \citenamefont {Parsegian}, \citenamefont {Podgornik}, \citenamefont {Rajter},
  \citenamefont {Jagota}, \citenamefont {Luo}, \citenamefont {Asthagiri},
  \citenamefont {Chaudhury}, \citenamefont {Chiang}, \citenamefont {Granick},
  \citenamefont {Kalinin}, \citenamefont {Kardar}, \citenamefont {Kjellander},
  \citenamefont {Langreth}, \citenamefont {Lewis}, \citenamefont {Lustig},
  \citenamefont {Wesolowski}, \citenamefont {Wettlaufer}, \citenamefont
  {Ching}, \citenamefont {Finnis}, \citenamefont {Houlihan}, \citenamefont {von
  Lilienfeld}, \citenamefont {van Oss},\ and\ \citenamefont {Zemb}}]{french10}%
  \BibitemOpen
  \bibfield  {author} {\bibinfo {author} {\bibfnamefont {R.~H.}\ \bibnamefont
  {French}}, \bibinfo {author} {\bibfnamefont {V.~A.}\ \bibnamefont
  {Parsegian}}, \bibinfo {author} {\bibfnamefont {R.}~\bibnamefont
  {Podgornik}}, \bibinfo {author} {\bibfnamefont {R.~F.}\ \bibnamefont
  {Rajter}}, \bibinfo {author} {\bibfnamefont {A.}~\bibnamefont {Jagota}},
  \bibinfo {author} {\bibfnamefont {J.}~\bibnamefont {Luo}}, \bibinfo {author}
  {\bibfnamefont {D.}~\bibnamefont {Asthagiri}}, \bibinfo {author}
  {\bibfnamefont {M.~K.}\ \bibnamefont {Chaudhury}}, \bibinfo {author}
  {\bibfnamefont {Y.-m.}\ \bibnamefont {Chiang}}, \bibinfo {author}
  {\bibfnamefont {S.}~\bibnamefont {Granick}}, \bibinfo {author} {\bibfnamefont
  {S.}~\bibnamefont {Kalinin}}, \bibinfo {author} {\bibfnamefont
  {M.}~\bibnamefont {Kardar}}, \bibinfo {author} {\bibfnamefont
  {R.}~\bibnamefont {Kjellander}}, \bibinfo {author} {\bibfnamefont {D.~C.}\
  \bibnamefont {Langreth}}, \bibinfo {author} {\bibfnamefont {J.}~\bibnamefont
  {Lewis}}, \bibinfo {author} {\bibfnamefont {S.}~\bibnamefont {Lustig}},
  \bibinfo {author} {\bibfnamefont {D.}~\bibnamefont {Wesolowski}}, \bibinfo
  {author} {\bibfnamefont {J.~S.}\ \bibnamefont {Wettlaufer}}, \bibinfo
  {author} {\bibfnamefont {W.-Y.}\ \bibnamefont {Ching}}, \bibinfo {author}
  {\bibfnamefont {M.}~\bibnamefont {Finnis}}, \bibinfo {author} {\bibfnamefont
  {F.}~\bibnamefont {Houlihan}}, \bibinfo {author} {\bibfnamefont {O.~A.}\
  \bibnamefont {von Lilienfeld}}, \bibinfo {author} {\bibfnamefont {C.~J.}\
  \bibnamefont {van Oss}},\ and\ \bibinfo {author} {\bibfnamefont
  {T.}~\bibnamefont {Zemb}},\ }\bibfield  {title} {\bibinfo {title} {Long range
  interactions in nanoscale science},\ }\href
  {https://doi.org/10.1103/RevModPhys.82.1887} {\bibfield  {journal} {\bibinfo
  {journal} {Rev. Mod. Phys.}\ }\textbf {\bibinfo {volume} {82}},\ \bibinfo
  {pages} {1887} (\bibinfo {year} {2010})}\BibitemShut {NoStop}%
\bibitem [{\citenamefont {Bostr{\"o}m}\ \emph {et~al.}(2017)\citenamefont
  {Bostr{\"o}m}, \citenamefont {Malyi}, \citenamefont {Parashar}, \citenamefont
  {Shajesh}, \citenamefont {Thiyam}, \citenamefont {Milton}, \citenamefont
  {Persson}, \citenamefont {Parsons},\ and\ \citenamefont
  {Brevik}}]{bostrom17}%
  \BibitemOpen
  \bibfield  {author} {\bibinfo {author} {\bibfnamefont {M.}~\bibnamefont
  {Bostr{\"o}m}}, \bibinfo {author} {\bibfnamefont {O.~I.}\ \bibnamefont
  {Malyi}}, \bibinfo {author} {\bibfnamefont {P.}~\bibnamefont {Parashar}},
  \bibinfo {author} {\bibfnamefont {K.~V.}\ \bibnamefont {Shajesh}}, \bibinfo
  {author} {\bibfnamefont {P.}~\bibnamefont {Thiyam}}, \bibinfo {author}
  {\bibfnamefont {K.~A.}\ \bibnamefont {Milton}}, \bibinfo {author}
  {\bibfnamefont {C.}~\bibnamefont {Persson}}, \bibinfo {author} {\bibfnamefont
  {D.~F.}\ \bibnamefont {Parsons}},\ and\ \bibinfo {author} {\bibfnamefont
  {I.}~\bibnamefont {Brevik}},\ }\bibfield  {title} {\bibinfo {title} {Lifshitz
  interaction can promote ice growth at water-silica interfaces},\ }\href
  {https://doi.org/10.1103/PhysRevB.95.155422} {\bibfield  {journal} {\bibinfo
  {journal} {Phys. Rev. B}\ }\textbf {\bibinfo {volume} {95}},\ \bibinfo
  {pages} {155422} (\bibinfo {year} {2017})}\BibitemShut {NoStop}%
\bibitem [{\citenamefont {Thiyam}\ \emph {et~al.}(2018)\citenamefont {Thiyam},
  \citenamefont {Fiedler}, \citenamefont {Buhmann}, \citenamefont {Persson},
  \citenamefont {Brevik}, \citenamefont {Boström},\ and\ \citenamefont
  {Parsons}}]{thiyam18}%
  \BibitemOpen
  \bibfield  {author} {\bibinfo {author} {\bibfnamefont {P.}~\bibnamefont
  {Thiyam}}, \bibinfo {author} {\bibfnamefont {J.}~\bibnamefont {Fiedler}},
  \bibinfo {author} {\bibfnamefont {S.~Y.}\ \bibnamefont {Buhmann}}, \bibinfo
  {author} {\bibfnamefont {C.}~\bibnamefont {Persson}}, \bibinfo {author}
  {\bibfnamefont {I.}~\bibnamefont {Brevik}}, \bibinfo {author} {\bibfnamefont
  {M.}~\bibnamefont {Boström}},\ and\ \bibinfo {author} {\bibfnamefont
  {D.~F.}\ \bibnamefont {Parsons}},\ }\bibfield  {title} {\bibinfo {title} {Ice
  particles sink below the water surface due to a balance of salt, van der
  waals, and buoyancy forces},\ }\href
  {https://doi.org/10.1021/acs.jpcc.8b02351} {\bibfield  {journal} {\bibinfo
  {journal} {J. Phys. Chem. C}\ }\textbf {\bibinfo {volume} {122}},\ \bibinfo
  {pages} {15311} (\bibinfo {year} {2018})},\ \Eprint
  {https://arxiv.org/abs/https://doi.org/10.1021/acs.jpcc.8b02351}
  {https://doi.org/10.1021/acs.jpcc.8b02351} \BibitemShut {NoStop}%
\bibitem [{\citenamefont {Esteso}\ \emph {et~al.}(2020)\citenamefont {Esteso},
  \citenamefont {Carretero-Palacios}, \citenamefont {MacDowell}, \citenamefont
  {Fiedler}, \citenamefont {Parsons}, \citenamefont {Spallek}, \citenamefont
  {M{\'\i}guez}, \citenamefont {Persson}, \citenamefont {Buhmann},
  \citenamefont {Brevik},\ and\ \citenamefont {Bostr{\"o}m}}]{esteso20}%
  \BibitemOpen
  \bibfield  {author} {\bibinfo {author} {\bibfnamefont {V.}~\bibnamefont
  {Esteso}}, \bibinfo {author} {\bibfnamefont {S.}~\bibnamefont
  {Carretero-Palacios}}, \bibinfo {author} {\bibfnamefont {L.~G.}\ \bibnamefont
  {MacDowell}}, \bibinfo {author} {\bibfnamefont {J.}~\bibnamefont {Fiedler}},
  \bibinfo {author} {\bibfnamefont {D.~F.}\ \bibnamefont {Parsons}}, \bibinfo
  {author} {\bibfnamefont {F.}~\bibnamefont {Spallek}}, \bibinfo {author}
  {\bibfnamefont {H.}~\bibnamefont {M{\'\i}guez}}, \bibinfo {author}
  {\bibfnamefont {C.}~\bibnamefont {Persson}}, \bibinfo {author} {\bibfnamefont
  {S.~Y.}\ \bibnamefont {Buhmann}}, \bibinfo {author} {\bibfnamefont
  {I.}~\bibnamefont {Brevik}},\ and\ \bibinfo {author} {\bibfnamefont
  {M.}~\bibnamefont {Bostr{\"o}m}},\ }\bibfield  {title} {\bibinfo {title}
  {Premelting of ice adsorbed on a rock surface},\ }\href@noop {} {\bibfield
  {journal} {\bibinfo  {journal} {Phys. Chem. Chem. Phys}\ }\textbf {\bibinfo
  {volume} {22}},\ \bibinfo {pages} {11362} (\bibinfo {year}
  {2020})}\BibitemShut {NoStop}%
\bibitem [{\citenamefont {Derjaguin}(1987)}]{derjaguin87}%
  \BibitemOpen
  \bibfield  {author} {\bibinfo {author} {\bibfnamefont {B.}~\bibnamefont
  {Derjaguin}},\ }\bibfield  {title} {\bibinfo {title} {Modern state of the
  investigation of long-range surface forces},\ }\href@noop {} {\bibfield
  {journal} {\bibinfo  {journal} {Langmuir}\ }\textbf {\bibinfo {volume} {3}},\
  \bibinfo {pages} {601} (\bibinfo {year} {1987})}\BibitemShut {NoStop}%
\bibitem [{\citenamefont {Elbaum}\ and\ \citenamefont
  {Schick}(1991{\natexlab{b}})}]{elbaum91c}%
  \BibitemOpen
  \bibfield  {author} {\bibinfo {author} {\bibfnamefont {M.}~\bibnamefont
  {Elbaum}}\ and\ \bibinfo {author} {\bibfnamefont {M.}~\bibnamefont
  {Schick}},\ }\bibfield  {title} {\bibinfo {title} {On the failure of water to
  freeze from its surface},\ }\href@noop {} {\bibfield  {journal} {\bibinfo
  {journal} {J. Phys. I}\ }\textbf {\bibinfo {volume} {1}},\ \bibinfo {pages}
  {1665} (\bibinfo {year} {1991}{\natexlab{b}})}\BibitemShut {NoStop}%
\bibitem [{\citenamefont {Dietrich}\ and\ \citenamefont
  {Schick}(1986)}]{dietrich86}%
  \BibitemOpen
  \bibfield  {author} {\bibinfo {author} {\bibfnamefont {S.}~\bibnamefont
  {Dietrich}}\ and\ \bibinfo {author} {\bibfnamefont {M.}~\bibnamefont
  {Schick}},\ }\bibfield  {title} {\bibinfo {title} {Order of wetting
  transitions},\ }\href {https://doi.org/10.1103/PhysRevB.33.4952} {\bibfield
  {journal} {\bibinfo  {journal} {Phys. Rev. B}\ }\textbf {\bibinfo {volume}
  {33}},\ \bibinfo {pages} {4952} (\bibinfo {year} {1986})}\BibitemShut
  {NoStop}%
\bibitem [{\citenamefont {Dietrich}\ and\ \citenamefont
  {Napi\'orkowski}(1991)}]{dietrich91}%
  \BibitemOpen
  \bibfield  {author} {\bibinfo {author} {\bibfnamefont {S.}~\bibnamefont
  {Dietrich}}\ and\ \bibinfo {author} {\bibfnamefont {M.}~\bibnamefont
  {Napi\'orkowski}},\ }\bibfield  {title} {\bibinfo {title} {Analytic results
  for wetting transitions in the presence of van der waals tails},\ }\href
  {https://doi.org/10.1103/PhysRevA.43.1861} {\bibfield  {journal} {\bibinfo
  {journal} {Phys. Rev. A}\ }\textbf {\bibinfo {volume} {43}},\ \bibinfo
  {pages} {1861} (\bibinfo {year} {1991})}\BibitemShut {NoStop}%
\bibitem [{\citenamefont {MacDowell}(2017)}]{macdowell17}%
  \BibitemOpen
  \bibfield  {author} {\bibinfo {author} {\bibfnamefont {L.~G.}\ \bibnamefont
  {MacDowell}},\ }\bibfield  {title} {\bibinfo {title} {Capillary wave theory
  of adsorbed liquid films and the structure of the liquid-vapor interface},\
  }\href@noop {} {\bibfield  {journal} {\bibinfo  {journal} {Phys. Rev. E}\
  }\textbf {\bibinfo {volume} {96}},\ \bibinfo {pages} {022801} (\bibinfo
  {year} {2017})}\BibitemShut {NoStop}%
\bibitem [{\citenamefont {Nozieres}(1992)}]{nozieres92}%
  \BibitemOpen
  \bibfield  {author} {\bibinfo {author} {\bibfnamefont {P.}~\bibnamefont
  {Nozieres}},\ }\bibinfo {title} {Solids far from equilibrium}\ (\bibinfo
  {publisher} {Cambridge University Press},\ \bibinfo {year} {1992})\
  Chap.~\bibinfo {chapter} {1}, pp.\ \bibinfo {pages} {1--152}\BibitemShut
  {NoStop}%
\bibitem [{\citenamefont {Fukuta}(1987)}]{fukuta87}%
  \BibitemOpen
  \bibfield  {author} {\bibinfo {author} {\bibfnamefont {N.}~\bibnamefont
  {Fukuta}},\ }\bibfield  {title} {\bibinfo {title} {An origin of the
  equilibrium liquid-like layer on ice},\ }\href@noop {} {\bibfield  {journal}
  {\bibinfo  {journal} {J. Phys.(Paris)}\ }\textbf {\bibinfo {volume} {48}},\
  \bibinfo {pages} {C} (\bibinfo {year} {1987})}\BibitemShut {NoStop}%
\bibitem [{\citenamefont {Makkonen}(1997)}]{makkonen97}%
  \BibitemOpen
  \bibfield  {author} {\bibinfo {author} {\bibfnamefont {L.}~\bibnamefont
  {Makkonen}},\ }\bibfield  {title} {\bibinfo {title} {Surface melting of
  ice},\ }\href {https://doi.org/10.1021/jp963248c} {\bibfield  {journal}
  {\bibinfo  {journal} {J. Phys. Chem. B}\ }\textbf {\bibinfo {volume} {101}},\
  \bibinfo {pages} {6196} (\bibinfo {year} {1997})},\ \Eprint
  {https://arxiv.org/abs/https://doi.org/10.1021/jp963248c}
  {https://doi.org/10.1021/jp963248c} \BibitemShut {NoStop}%
\bibitem [{\citenamefont {Knight}(1967)}]{knight67}%
  \BibitemOpen
  \bibfield  {author} {\bibinfo {author} {\bibfnamefont {C.~A.}\ \bibnamefont
  {Knight}},\ }\bibfield  {title} {\bibinfo {title} {The contact angle of water
  on ice},\ }\href
  {https://doi.org/https://doi.org/10.1016/0021-9797(67)90031-8} {\bibfield
  {journal} {\bibinfo  {journal} {J. Colloid. Interface Sci.}\ }\textbf
  {\bibinfo {volume} {25}},\ \bibinfo {pages} {280} (\bibinfo {year}
  {1967})}\BibitemShut {NoStop}%
\bibitem [{\citenamefont {Knight}(1971)}]{knight71}%
  \BibitemOpen
  \bibfield  {author} {\bibinfo {author} {\bibfnamefont {C.~A.}\ \bibnamefont
  {Knight}},\ }\bibfield  {title} {\bibinfo {title} {Experiments on the contact
  angle of water on ice},\ }\href {https://doi.org/10.1080/14786437108216369}
  {\bibfield  {journal} {\bibinfo  {journal} {Phil. Magazine}\ }\textbf
  {\bibinfo {volume} {23}},\ \bibinfo {pages} {153} (\bibinfo {year} {1971})},\
  \Eprint {https://arxiv.org/abs/https://doi.org/10.1080/14786437108216369}
  {https://doi.org/10.1080/14786437108216369} \BibitemShut {NoStop}%
\bibitem [{\citenamefont {Gonda}\ \emph {et~al.}(1999)\citenamefont {Gonda},
  \citenamefont {Arai},\ and\ \citenamefont {Sei}}]{gonda99}%
  \BibitemOpen
  \bibfield  {author} {\bibinfo {author} {\bibfnamefont {T.}~\bibnamefont
  {Gonda}}, \bibinfo {author} {\bibfnamefont {T.}~\bibnamefont {Arai}},\ and\
  \bibinfo {author} {\bibfnamefont {T.}~\bibnamefont {Sei}},\ }\bibfield
  {title} {\bibinfo {title} {Experimental study on the melting process of ice
  crystals just below the melting point.},\ }\href@noop {} {\bibfield
  {journal} {\bibinfo  {journal} {Polar Meteorol. Glaciol.}\ }\textbf {\bibinfo
  {volume} {13}},\ \bibinfo {pages} {38} (\bibinfo {year} {1999})}\BibitemShut
  {NoStop}%
\bibitem [{\citenamefont {Murata}\ \emph {et~al.}(2016)\citenamefont {Murata},
  \citenamefont {Asakawa}, \citenamefont {Nagashima}, \citenamefont
  {Furukawa},\ and\ \citenamefont {Sazaki}}]{murata16}%
  \BibitemOpen
  \bibfield  {author} {\bibinfo {author} {\bibfnamefont {K.-i.}\ \bibnamefont
  {Murata}}, \bibinfo {author} {\bibfnamefont {H.}~\bibnamefont {Asakawa}},
  \bibinfo {author} {\bibfnamefont {K.}~\bibnamefont {Nagashima}}, \bibinfo
  {author} {\bibfnamefont {Y.}~\bibnamefont {Furukawa}},\ and\ \bibinfo
  {author} {\bibfnamefont {G.}~\bibnamefont {Sazaki}},\ }\bibfield  {title}
  {\bibinfo {title} {Thermodynamic origin of surface melting on ice crystals},\
  }\href@noop {} {\bibfield  {journal} {\bibinfo  {journal} {Proc. Natl. Acad.
  Sci. U.S.A.}\ }\textbf {\bibinfo {volume} {113}},\ \bibinfo {pages} {E6741}
  (\bibinfo {year} {2016})}\BibitemShut {NoStop}%
\bibitem [{\citenamefont {Kuroda}\ and\ \citenamefont
  {Lacmann}(1982)}]{kuroda82}%
  \BibitemOpen
  \bibfield  {author} {\bibinfo {author} {\bibfnamefont {T.}~\bibnamefont
  {Kuroda}}\ and\ \bibinfo {author} {\bibfnamefont {R.}~\bibnamefont
  {Lacmann}},\ }\bibfield  {title} {\bibinfo {title} {Growth kinetics of ice
  from the vapour phase and its growth forms},\ }\href
  {https://doi.org/http://dx.doi.org/10.1016/0022-0248(82)90028-8} {\bibfield
  {journal} {\bibinfo  {journal} {J. Cryst. Growth}\ }\textbf {\bibinfo
  {volume} {56}},\ \bibinfo {pages} {189} (\bibinfo {year} {1982})}\BibitemShut
  {NoStop}%
\bibitem [{\citenamefont {Nenow}\ and\ \citenamefont
  {Trayanov}(1986)}]{nenow86}%
  \BibitemOpen
  \bibfield  {author} {\bibinfo {author} {\bibfnamefont {D.}~\bibnamefont
  {Nenow}}\ and\ \bibinfo {author} {\bibfnamefont {A.}~\bibnamefont
  {Trayanov}},\ }\bibfield  {title} {\bibinfo {title} {Thermodynamics of
  crystal surfaces with quasi-liquid layer},\ }\href
  {https://doi.org/http://dx.doi.org/10.1016/0022-0248(86)90557-9} {\bibfield
  {journal} {\bibinfo  {journal} {J. Cryst. Growth}\ }\textbf {\bibinfo
  {volume} {79}},\ \bibinfo {pages} {801} (\bibinfo {year} {1986})}\BibitemShut
  {NoStop}%
\bibitem [{\citenamefont {Kuroda}\ and\ \citenamefont
  {Gonda}(1990)}]{kuroda90}%
  \BibitemOpen
  \bibfield  {author} {\bibinfo {author} {\bibfnamefont {T.}~\bibnamefont
  {Kuroda}}\ and\ \bibinfo {author} {\bibfnamefont {T.}~\bibnamefont {Gonda}},\
  }\bibfield  {title} {\bibinfo {title} {Vapor growth mechanism of a crystal
  surface covered with a quasi-liquid-layer-effect of self-diffusion
  coefficient of the quasi-liquid-layer on the growth rate},\ }\href@noop {}
  {\bibfield  {journal} {\bibinfo  {journal} {J. Cryst. Growth}\ }\textbf
  {\bibinfo {volume} {99}},\ \bibinfo {pages} {83} (\bibinfo {year}
  {1990})}\BibitemShut {NoStop}%
\bibitem [{\citenamefont {Knight}(1996)}]{knight96}%
  \BibitemOpen
  \bibfield  {author} {\bibinfo {author} {\bibfnamefont {C.~A.}\ \bibnamefont
  {Knight}},\ }\bibfield  {title} {\bibinfo {title} {Surface layers on ice},\
  }\href {https://doi.org/https://doi.org/10.1029/96JD00554} {\bibfield
  {journal} {\bibinfo  {journal} {J. Geophys. Res.: Atmos.}\ }\textbf {\bibinfo
  {volume} {101}},\ \bibinfo {pages} {12921} (\bibinfo {year} {1996})},\
  \Eprint
  {https://arxiv.org/abs/https://agupubs.onlinelibrary.wiley.com/doi/pdf/10.1029/96JD00554}
  {https://agupubs.onlinelibrary.wiley.com/doi/pdf/10.1029/96JD00554}
  \BibitemShut {NoStop}%
\bibitem [{\citenamefont {Fletcher}(1970)}]{fletcher70}%
  \BibitemOpen
  \bibfield  {author} {\bibinfo {author} {\bibfnamefont {N.~H.}\ \bibnamefont
  {Fletcher}},\ }\href {http://dx.doi.org/10.1017/CBO9780511735639} {\emph
  {\bibinfo {title} {The Chemical Physics of Ice}}}\ (\bibinfo  {publisher}
  {Cambridge University Press},\ \bibinfo {year} {1970})\ \bibinfo {note}
  {cambridge Books Online}\BibitemShut {NoStop}%
\bibitem [{\citenamefont {Ketcham}\ and\ \citenamefont
  {Hobbs}(1969)}]{ketcham69}%
  \BibitemOpen
  \bibfield  {author} {\bibinfo {author} {\bibfnamefont {W.~M.}\ \bibnamefont
  {Ketcham}}\ and\ \bibinfo {author} {\bibfnamefont {P.~V.}\ \bibnamefont
  {Hobbs}},\ }\bibfield  {title} {\bibinfo {title} {An experimental
  determination of the surface energies of ice},\ }\href
  {https://doi.org/10.1080/14786436908228641} {\bibfield  {journal} {\bibinfo
  {journal} {Phil. Magazine}\ }\textbf {\bibinfo {volume} {19}},\ \bibinfo
  {pages} {1161} (\bibinfo {year} {1969})},\ \Eprint
  {https://arxiv.org/abs/https://doi.org/10.1080/14786436908228641}
  {https://doi.org/10.1080/14786436908228641} \BibitemShut {NoStop}%
\bibitem [{\citenamefont {Neshyba}\ \emph {et~al.}(2016)\citenamefont
  {Neshyba}, \citenamefont {Adams}, \citenamefont {Reed}, \citenamefont
  {Rowe},\ and\ \citenamefont {Gladich}}]{neshyba16}%
  \BibitemOpen
  \bibfield  {author} {\bibinfo {author} {\bibfnamefont {S.}~\bibnamefont
  {Neshyba}}, \bibinfo {author} {\bibfnamefont {J.}~\bibnamefont {Adams}},
  \bibinfo {author} {\bibfnamefont {K.}~\bibnamefont {Reed}}, \bibinfo {author}
  {\bibfnamefont {P.~M.}\ \bibnamefont {Rowe}},\ and\ \bibinfo {author}
  {\bibfnamefont {I.}~\bibnamefont {Gladich}},\ }\bibfield  {title} {\bibinfo
  {title} {A quasi-liquid mediated continuum model of faceted ice dynamics},\
  }\href {https://doi.org/10.1002/2016JD025458} {\bibfield  {journal} {\bibinfo
   {journal} {J. Geophys. Res.: Atmos.}\ }\textbf {\bibinfo {volume} {121}},\
  \bibinfo {pages} {14,035} (\bibinfo {year} {2016})}\BibitemShut {NoStop}%
\bibitem [{\citenamefont {Mohandesi}\ and\ \citenamefont
  {Kusalik}(2018)}]{mohandesi18}%
  \BibitemOpen
  \bibfield  {author} {\bibinfo {author} {\bibfnamefont {A.}~\bibnamefont
  {Mohandesi}}\ and\ \bibinfo {author} {\bibfnamefont {P.~G.}\ \bibnamefont
  {Kusalik}},\ }\bibfield  {title} {\bibinfo {title} {Probing ice growth from
  vapor phase: A molecular dynamics simulation approach},\ }\href
  {https://doi.org/https://doi.org/10.1016/j.jcrysgro.2017.11.022} {\bibfield
  {journal} {\bibinfo  {journal} {J. Cryst. Growth}\ }\textbf {\bibinfo
  {volume} {483}},\ \bibinfo {pages} {156 } (\bibinfo {year}
  {2018})}\BibitemShut {NoStop}%
\bibitem [{\citenamefont {M{\'e}tois}\ and\ \citenamefont
  {Heyraud}(1989)}]{metois89}%
  \BibitemOpen
  \bibfield  {author} {\bibinfo {author} {\bibfnamefont {J.}~\bibnamefont
  {M{\'e}tois}}\ and\ \bibinfo {author} {\bibfnamefont {J.}~\bibnamefont
  {Heyraud}},\ }\bibfield  {title} {\bibinfo {title} {{The overheating of lead
  crystals}},\ }\href {https://doi.org/10.1051/jphys:0198900500210317500}
  {\bibfield  {journal} {\bibinfo  {journal} {J. Phys.(Paris)}\ }\textbf
  {\bibinfo {volume} {50}},\ \bibinfo {pages} {3175} (\bibinfo {year}
  {1989})}\BibitemShut {NoStop}%
\bibitem [{\citenamefont {Faraday}(1860)}]{faraday60}%
  \BibitemOpen
  \bibfield  {author} {\bibinfo {author} {\bibfnamefont {M.}~\bibnamefont
  {Faraday}},\ }\bibfield  {title} {\bibinfo {title} {I. note on regelation},\
  }\href {https://doi.org/10.1098/rspl.1859.0082} {\bibfield  {journal}
  {\bibinfo  {journal} {PRSL}\ }\textbf {\bibinfo {volume} {10}},\ \bibinfo
  {pages} {440} (\bibinfo {year} {1860})},\ \Eprint
  {https://arxiv.org/abs/https://royalsocietypublishing.org/doi/pdf/10.1098/rspl.1859.0082}
  {https://royalsocietypublishing.org/doi/pdf/10.1098/rspl.1859.0082}
  \BibitemShut {NoStop}%
\bibitem [{\citenamefont {Thomson}\ and\ \citenamefont
  {Thomson}(1862)}]{thomson62}%
  \BibitemOpen
  \bibfield  {author} {\bibinfo {author} {\bibfnamefont {J.}~\bibnamefont
  {Thomson}}\ and\ \bibinfo {author} {\bibfnamefont {W.}~\bibnamefont
  {Thomson}},\ }\bibfield  {title} {\bibinfo {title} {Iii. note on professor
  faraday's recent experiments on 'regelation.'},\ }\href
  {https://doi.org/10.1098/rspl.1860.0041} {\bibfield  {journal} {\bibinfo
  {journal} {Proc. R. Soc. Lond.}\ }\textbf {\bibinfo {volume} {11}},\ \bibinfo
  {pages} {198} (\bibinfo {year} {1862})},\ \Eprint
  {https://arxiv.org/abs/https://royalsocietypublishing.org/doi/pdf/10.1098/rspl.1860.0041}
  {https://royalsocietypublishing.org/doi/pdf/10.1098/rspl.1860.0041}
  \BibitemShut {NoStop}%
\bibitem [{\citenamefont {Wettlaufer}\ and\ \citenamefont
  {Worster}(2006)}]{wettlaufer06}%
  \BibitemOpen
  \bibfield  {author} {\bibinfo {author} {\bibfnamefont {J.}~\bibnamefont
  {Wettlaufer}}\ and\ \bibinfo {author} {\bibfnamefont {M.~G.}\ \bibnamefont
  {Worster}},\ }\bibfield  {title} {\bibinfo {title} {Premelting dynamics},\
  }\href {https://doi.org/10.1146/annurev.fluid.37.061903.175758} {\bibfield
  {journal} {\bibinfo  {journal} {Ann. Rev. Fluid. Mech.}\ }\textbf {\bibinfo
  {volume} {38}},\ \bibinfo {pages} {427} (\bibinfo {year} {2006})},\ \Eprint
  {https://arxiv.org/abs/https://doi.org/10.1146/annurev.fluid.37.061903.175758}
  {https://doi.org/10.1146/annurev.fluid.37.061903.175758} \BibitemShut
  {NoStop}%
\bibitem [{\citenamefont {P{\'e}rez-D{\'i}az}\ \emph
  {et~al.}(2016)\citenamefont {P{\'e}rez-D{\'i}az}, \citenamefont
  {{\'A}lvarez-Valenzuela},\ and\ \citenamefont
  {Rodr{\'i}guez-Celis}}]{perezdiaz16}%
  \BibitemOpen
  \bibfield  {author} {\bibinfo {author} {\bibfnamefont {J.}~\bibnamefont
  {P{\'e}rez-D{\'i}az}}, \bibinfo {author} {\bibfnamefont {M.}~\bibnamefont
  {{\'A}lvarez-Valenzuela}},\ and\ \bibinfo {author} {\bibfnamefont
  {F.}~\bibnamefont {Rodr{\'i}guez-Celis}},\ }\bibfield  {title} {\bibinfo
  {title} {Surface freezing of water},\ }\href
  {https://doi.org/10.1186/s40064-016-2196-3} {\bibfield  {journal} {\bibinfo
  {journal} {Springerplus}\ }\textbf {\bibinfo {volume} {5}},\ \bibinfo {pages}
  {629} (\bibinfo {year} {2016})}\BibitemShut {NoStop}%
\bibitem [{\citenamefont {Vega}\ \emph {et~al.}(2006)\citenamefont {Vega},
  \citenamefont {Martin-Conde},\ and\ \citenamefont {Patrykiejew}}]{vega06}%
  \BibitemOpen
  \bibfield  {author} {\bibinfo {author} {\bibfnamefont {C.}~\bibnamefont
  {Vega}}, \bibinfo {author} {\bibfnamefont {M.}~\bibnamefont {Martin-Conde}},\
  and\ \bibinfo {author} {\bibfnamefont {A.}~\bibnamefont {Patrykiejew}},\
  }\bibfield  {title} {\bibinfo {title} {Absence of superheating for ice ih
  with a free surface: a new method of determining the melting point of
  different water models},\ }\href@noop {} {\bibfield  {journal} {\bibinfo
  {journal} {Mol. Phys.}\ }\textbf {\bibinfo {volume} {104}},\ \bibinfo {pages}
  {3583} (\bibinfo {year} {2006})}\BibitemShut {NoStop}%
\bibitem [{\citenamefont {Neshyba}\ \emph {et~al.}(2009)\citenamefont
  {Neshyba}, \citenamefont {Nugent}, \citenamefont {Roeselova},\ and\
  \citenamefont {Jungwirth}}]{neshyba09}%
  \BibitemOpen
  \bibfield  {author} {\bibinfo {author} {\bibfnamefont {S.}~\bibnamefont
  {Neshyba}}, \bibinfo {author} {\bibfnamefont {E.}~\bibnamefont {Nugent}},
  \bibinfo {author} {\bibfnamefont {M.}~\bibnamefont {Roeselova}},\ and\
  \bibinfo {author} {\bibfnamefont {P.}~\bibnamefont {Jungwirth}},\ }\bibfield
  {title} {\bibinfo {title} {Molecular dynamics study of ice-vapor interactions
  via the quasi-liquid layer},\ }\href {https://doi.org/10.1021/jp810589a}
  {\bibfield  {journal} {\bibinfo  {journal} {J. Phys. Chem. C}\ }\textbf
  {\bibinfo {volume} {113}},\ \bibinfo {pages} {4597} (\bibinfo {year}
  {2009})}\BibitemShut {NoStop}%
\bibitem [{\citenamefont {Kling}\ \emph {et~al.}(2018)\citenamefont {Kling},
  \citenamefont {Kling},\ and\ \citenamefont {Donadio}}]{kling18}%
  \BibitemOpen
  \bibfield  {author} {\bibinfo {author} {\bibfnamefont {T.}~\bibnamefont
  {Kling}}, \bibinfo {author} {\bibfnamefont {F.}~\bibnamefont {Kling}},\ and\
  \bibinfo {author} {\bibfnamefont {D.}~\bibnamefont {Donadio}},\ }\bibfield
  {title} {\bibinfo {title} {Structure and dynamics of the quasi-liquid layer
  at the surface of ice from molecular simulations},\ }\href
  {https://doi.org/10.1021/acs.jpcc.8b07724} {\bibfield  {journal} {\bibinfo
  {journal} {J. Phys. Chem. C}\ }\textbf {\bibinfo {volume} {122}},\ \bibinfo
  {pages} {24780} (\bibinfo {year} {2018})},\ \Eprint
  {https://arxiv.org/abs/https://doi.org/10.1021/acs.jpcc.8b07724}
  {https://doi.org/10.1021/acs.jpcc.8b07724} \BibitemShut {NoStop}%
\bibitem [{\citenamefont {Qiu}\ and\ \citenamefont {Molinero}(2018)}]{qiu18}%
  \BibitemOpen
  \bibfield  {author} {\bibinfo {author} {\bibfnamefont {Y.}~\bibnamefont
  {Qiu}}\ and\ \bibinfo {author} {\bibfnamefont {V.}~\bibnamefont {Molinero}},\
  }\bibfield  {title} {\bibinfo {title} {Why is it so difficult to identify the
  onset of ice premelting?},\ }\href
  {https://doi.org/10.1021/acs.jpclett.8b02244} {\bibfield  {journal} {\bibinfo
   {journal} {J. Phys. Chem. Lett.}\ }\textbf {\bibinfo {volume} {9}},\
  \bibinfo {pages} {5179} (\bibinfo {year} {2018})}\BibitemShut {NoStop}%
\bibitem [{\citenamefont {Louden}\ and\ \citenamefont
  {Gezelter}(2018)}]{louden18}%
  \BibitemOpen
  \bibfield  {author} {\bibinfo {author} {\bibfnamefont {P.~B.}\ \bibnamefont
  {Louden}}\ and\ \bibinfo {author} {\bibfnamefont {J.~D.}\ \bibnamefont
  {Gezelter}},\ }\bibfield  {title} {\bibinfo {title} {Why is ice slippery?
  simulations of shear viscosity of the quasi-liquid layer on ice},\ }\href
  {https://doi.org/10.1021/acs.jpclett.8b01339} {\bibfield  {journal} {\bibinfo
   {journal} {J. Phys. Chem. Lett.}\ }\textbf {\bibinfo {volume} {9}},\
  \bibinfo {pages} {3686} (\bibinfo {year} {2018})},\ \bibinfo {note} {pMID:
  29916247},\ \Eprint
  {https://arxiv.org/abs/https://doi.org/10.1021/acs.jpclett.8b01339}
  {https://doi.org/10.1021/acs.jpclett.8b01339} \BibitemShut {NoStop}%
\bibitem [{\citenamefont {Bluhm}\ \emph {et~al.}(2002)\citenamefont {Bluhm},
  \citenamefont {Ogletree}, \citenamefont {Fadley}, \citenamefont {Hussain},\
  and\ \citenamefont {Salmeron}}]{bluhm02}%
  \BibitemOpen
  \bibfield  {author} {\bibinfo {author} {\bibfnamefont {H.}~\bibnamefont
  {Bluhm}}, \bibinfo {author} {\bibfnamefont {D.~F.}\ \bibnamefont {Ogletree}},
  \bibinfo {author} {\bibfnamefont {C.~S.}\ \bibnamefont {Fadley}}, \bibinfo
  {author} {\bibfnamefont {Z.}~\bibnamefont {Hussain}},\ and\ \bibinfo {author}
  {\bibfnamefont {M.}~\bibnamefont {Salmeron}},\ }\bibfield  {title} {\bibinfo
  {title} {The premelting of ice studied with photoelectron spectroscopy},\
  }\href@noop {} {\bibfield  {journal} {\bibinfo  {journal} {J. Phys.: Condens.
  Matter}\ }\textbf {\bibinfo {volume} {14}},\ \bibinfo {pages} {L227}
  (\bibinfo {year} {2002})}\BibitemShut {NoStop}%
\bibitem [{\citenamefont {Sadtchenko}\ and\ \citenamefont
  {Ewing}(2002)}]{sadtchenko02}%
  \BibitemOpen
  \bibfield  {author} {\bibinfo {author} {\bibfnamefont {V.}~\bibnamefont
  {Sadtchenko}}\ and\ \bibinfo {author} {\bibfnamefont {G.~E.}\ \bibnamefont
  {Ewing}},\ }\bibfield  {title} {\bibinfo {title} {Interfacial melting of thin
  ice films: An infrared study},\ }\href {https://doi.org/10.1063/1.1449947}
  {\bibfield  {journal} {\bibinfo  {journal} {J. Chem. Phys.}\ }\textbf
  {\bibinfo {volume} {116}},\ \bibinfo {pages} {4686} (\bibinfo {year}
  {2002})},\ \Eprint {https://arxiv.org/abs/https://doi.org/10.1063/1.1449947}
  {https://doi.org/10.1063/1.1449947} \BibitemShut {NoStop}%
\bibitem [{\citenamefont {Gelman~Constantin}\ \emph {et~al.}(2018)\citenamefont
  {Gelman~Constantin}, \citenamefont {Gianetti}, \citenamefont {Longinotti},\
  and\ \citenamefont {Corti}}]{constantin18}%
  \BibitemOpen
  \bibfield  {author} {\bibinfo {author} {\bibfnamefont {J.}~\bibnamefont
  {Gelman~Constantin}}, \bibinfo {author} {\bibfnamefont {M.~M.}\ \bibnamefont
  {Gianetti}}, \bibinfo {author} {\bibfnamefont {M.~P.}\ \bibnamefont
  {Longinotti}},\ and\ \bibinfo {author} {\bibfnamefont {H.~R.}\ \bibnamefont
  {Corti}},\ }\bibfield  {title} {\bibinfo {title} {The quasi-liquid layer of
  ice revisited: the role of temperature gradients and tip chemistry in afm
  studies},\ }\href {https://doi.org/10.5194/acp-18-14965-2018} {\bibfield
  {journal} {\bibinfo  {journal} {Adv. Chem. Phys.}\ }\textbf {\bibinfo
  {volume} {18}},\ \bibinfo {pages} {14965} (\bibinfo {year}
  {2018})}\BibitemShut {NoStop}%
\bibitem [{\citenamefont {Mitsui}\ and\ \citenamefont {Aoki}(2019)}]{mitsui19}%
  \BibitemOpen
  \bibfield  {author} {\bibinfo {author} {\bibfnamefont {T.}~\bibnamefont
  {Mitsui}}\ and\ \bibinfo {author} {\bibfnamefont {K.}~\bibnamefont {Aoki}},\
  }\bibfield  {title} {\bibinfo {title} {Fluctuation spectroscopy of surface
  melting of ice with and without impurities},\ }\href
  {https://doi.org/10.1103/PhysRevE.99.010801} {\bibfield  {journal} {\bibinfo
  {journal} {Phys. Rev. E}\ }\textbf {\bibinfo {volume} {99}},\ \bibinfo
  {pages} {010801} (\bibinfo {year} {2019})}\BibitemShut {NoStop}%
\bibitem [{\citenamefont {Pinkley}\ \emph {et~al.}(1977)\citenamefont
  {Pinkley}, \citenamefont {Sethna},\ and\ \citenamefont
  {Williams}}]{pinkley77}%
  \BibitemOpen
  \bibfield  {author} {\bibinfo {author} {\bibfnamefont {L.~W.}\ \bibnamefont
  {Pinkley}}, \bibinfo {author} {\bibfnamefont {P.~P.}\ \bibnamefont
  {Sethna}},\ and\ \bibinfo {author} {\bibfnamefont {D.}~\bibnamefont
  {Williams}},\ }\bibfield  {title} {\bibinfo {title} {Optical constants of
  water in the infrared: Influence of temperature},\ }\href
  {https://doi.org/10.1364/JOSA.67.000494} {\bibfield  {journal} {\bibinfo
  {journal} {J. Opt. Soc. Am.}\ }\textbf {\bibinfo {volume} {67}},\ \bibinfo
  {pages} {494} (\bibinfo {year} {1977})}\BibitemShut {NoStop}%
\bibitem [{\citenamefont {Hasted}\ \emph {et~al.}(1987)\citenamefont {Hasted},
  \citenamefont {Husain}, \citenamefont {Frescura},\ and\ \citenamefont
  {Birch}}]{hasted87}%
  \BibitemOpen
  \bibfield  {author} {\bibinfo {author} {\bibfnamefont {J.}~\bibnamefont
  {Hasted}}, \bibinfo {author} {\bibfnamefont {S.}~\bibnamefont {Husain}},
  \bibinfo {author} {\bibfnamefont {F.}~\bibnamefont {Frescura}},\ and\
  \bibinfo {author} {\bibfnamefont {J.}~\bibnamefont {Birch}},\ }\bibfield
  {title} {\bibinfo {title} {The temperature variation of the near millimetre
  wavelength optical constants of water},\ }\href@noop {} {\bibfield  {journal}
  {\bibinfo  {journal} {Infrared physics}\ }\textbf {\bibinfo {volume} {27}},\
  \bibinfo {pages} {11} (\bibinfo {year} {1987})}\BibitemShut {NoStop}%
\bibitem [{\citenamefont {Zasetsky}\ \emph {et~al.}(2004)\citenamefont
  {Zasetsky}, \citenamefont {Khalizov},\ and\ \citenamefont
  {Sloan}}]{zasetsky04}%
  \BibitemOpen
  \bibfield  {author} {\bibinfo {author} {\bibfnamefont {A.~Y.}\ \bibnamefont
  {Zasetsky}}, \bibinfo {author} {\bibfnamefont {A.~F.}\ \bibnamefont
  {Khalizov}},\ and\ \bibinfo {author} {\bibfnamefont {J.~J.}\ \bibnamefont
  {Sloan}},\ }\bibfield  {title} {\bibinfo {title} {Local order and dynamics in
  supercooled water: A study by ir spectroscopy and molecular dynamic
  simulations},\ }\href {https://doi.org/10.1063/1.1787494} {\bibfield
  {journal} {\bibinfo  {journal} {J. Chem. Phys.}\ }\textbf {\bibinfo {volume}
  {121}},\ \bibinfo {pages} {6941} (\bibinfo {year} {2004})},\ \Eprint
  {https://arxiv.org/abs/https://doi.org/10.1063/1.1787494}
  {https://doi.org/10.1063/1.1787494} \BibitemShut {NoStop}%
\bibitem [{\citenamefont {Wagner}\ \emph {et~al.}(2005)\citenamefont {Wagner},
  \citenamefont {Benz}, \citenamefont {Möhler}, \citenamefont {Saathoff},
  \citenamefont {Schnaiter},\ and\ \citenamefont {Schurath}}]{wagner05}%
  \BibitemOpen
  \bibfield  {author} {\bibinfo {author} {\bibfnamefont {R.}~\bibnamefont
  {Wagner}}, \bibinfo {author} {\bibfnamefont {S.}~\bibnamefont {Benz}},
  \bibinfo {author} {\bibfnamefont {O.}~\bibnamefont {Möhler}}, \bibinfo
  {author} {\bibfnamefont {H.}~\bibnamefont {Saathoff}}, \bibinfo {author}
  {\bibfnamefont {M.}~\bibnamefont {Schnaiter}},\ and\ \bibinfo {author}
  {\bibfnamefont {U.}~\bibnamefont {Schurath}},\ }\bibfield  {title} {\bibinfo
  {title} {Mid-infrared extinction spectra and optical constants of supercooled
  water droplets},\ }\href {https://doi.org/10.1021/jp051942z} {\bibfield
  {journal} {\bibinfo  {journal} {J. Phys. Chem. A}\ }\textbf {\bibinfo
  {volume} {109}},\ \bibinfo {pages} {7099} (\bibinfo {year} {2005})},\
  \bibinfo {note} {pMID: 16834073},\ \Eprint
  {https://arxiv.org/abs/https://doi.org/10.1021/jp051942z}
  {https://doi.org/10.1021/jp051942z} \BibitemShut {NoStop}%
\bibitem [{\citenamefont {Rowe}\ \emph {et~al.}(2020)\citenamefont {Rowe},
  \citenamefont {Fergoda},\ and\ \citenamefont {Neshyba}}]{rowe20}%
  \BibitemOpen
  \bibfield  {author} {\bibinfo {author} {\bibfnamefont {P.~M.}\ \bibnamefont
  {Rowe}}, \bibinfo {author} {\bibfnamefont {M.}~\bibnamefont {Fergoda}},\ and\
  \bibinfo {author} {\bibfnamefont {S.}~\bibnamefont {Neshyba}},\ }\bibfield
  {title} {\bibinfo {title} {Temperature-dependent optical properties of liquid
  water from 240 to 298 k},\ }\href
  {https://doi.org/https://doi.org/10.1029/2020JD032624} {\bibfield  {journal}
  {\bibinfo  {journal} {J. Geophys. Res.: Atmos.}\ }\textbf {\bibinfo {volume}
  {125}},\ \bibinfo {pages} {e2020JD032624} (\bibinfo {year}
  {2020})}\BibitemShut {NoStop}%
\bibitem [{\citenamefont {Dobbins}\ and\ \citenamefont
  {Peck}(1973)}]{dobbins73}%
  \BibitemOpen
  \bibfield  {author} {\bibinfo {author} {\bibfnamefont {H.~M.}\ \bibnamefont
  {Dobbins}}\ and\ \bibinfo {author} {\bibfnamefont {E.~R.}\ \bibnamefont
  {Peck}},\ }\bibfield  {title} {\bibinfo {title} {Change of refractive index
  of water as a function of temperature$\ast$},\ }\href
  {https://doi.org/10.1364/JOSA.63.000318} {\bibfield  {journal} {\bibinfo
  {journal} {J. Opt. Soc. Am.}\ }\textbf {\bibinfo {volume} {63}},\ \bibinfo
  {pages} {318} (\bibinfo {year} {1973})}\BibitemShut {NoStop}%
\bibitem [{\citenamefont {Irvine}\ and\ \citenamefont
  {Pollack}(1968)}]{irvine68}%
  \BibitemOpen
  \bibfield  {author} {\bibinfo {author} {\bibfnamefont {W.~M.}\ \bibnamefont
  {Irvine}}\ and\ \bibinfo {author} {\bibfnamefont {J.~B.}\ \bibnamefont
  {Pollack}},\ }\bibfield  {title} {\bibinfo {title} {Infrared optical
  properties of water and ice spheres},\ }\href@noop {} {\bibfield  {journal}
  {\bibinfo  {journal} {Icarus}\ }\textbf {\bibinfo {volume} {8}},\ \bibinfo
  {pages} {324} (\bibinfo {year} {1968})}\BibitemShut {NoStop}%
\bibitem [{\citenamefont {Warren}(1984)}]{warren84}%
  \BibitemOpen
  \bibfield  {author} {\bibinfo {author} {\bibfnamefont {S.~G.}\ \bibnamefont
  {Warren}},\ }\bibfield  {title} {\bibinfo {title} {Optical constants of ice
  from the ultraviolet to the microwave},\ }\href
  {https://doi.org/10.1364/AO.23.001206} {\bibfield  {journal} {\bibinfo
  {journal} {Appl. Opt.}\ }\textbf {\bibinfo {volume} {23}},\ \bibinfo {pages}
  {1206} (\bibinfo {year} {1984})}\BibitemShut {NoStop}%
\bibitem [{\citenamefont {Daniels}(1971)}]{daniels71}%
  \BibitemOpen
  \bibfield  {author} {\bibinfo {author} {\bibfnamefont {J.}~\bibnamefont
  {Daniels}},\ }\bibfield  {title} {\bibinfo {title} {Bestimmung der optischen
  konstanten von eis aus energie - verlustmessungen von schnellen elektronen},\
  }\href {https://doi.org/https://doi.org/10.1016/0030-4018(71)90012-5}
  {\bibfield  {journal} {\bibinfo  {journal} {Optics Comm.}\ }\textbf {\bibinfo
  {volume} {3}},\ \bibinfo {pages} {240 } (\bibinfo {year} {1971})}\BibitemShut
  {NoStop}%
\bibitem [{\citenamefont {Kobayashi}(1983)}]{kobayashi83}%
  \BibitemOpen
  \bibfield  {author} {\bibinfo {author} {\bibfnamefont {K.}~\bibnamefont
  {Kobayashi}},\ }\bibfield  {title} {\bibinfo {title} {Optical spectra and
  electronic structure of ice},\ }\href {https://doi.org/10.1021/j100244a065}
  {\bibfield  {journal} {\bibinfo  {journal} {J. Phys. Chem.}\ }\textbf
  {\bibinfo {volume} {87}},\ \bibinfo {pages} {4317} (\bibinfo {year}
  {1983})},\ \Eprint
  {https://arxiv.org/abs/https://doi.org/10.1021/j100244a065}
  {https://doi.org/10.1021/j100244a065} \BibitemShut {NoStop}%
\bibitem [{\citenamefont {Kofman}\ \emph {et~al.}(2019)\citenamefont {Kofman},
  \citenamefont {He}, \citenamefont {ten Kate},\ and\ \citenamefont
  {Linnartz}}]{kofman19}%
  \BibitemOpen
  \bibfield  {author} {\bibinfo {author} {\bibfnamefont {V.}~\bibnamefont
  {Kofman}}, \bibinfo {author} {\bibfnamefont {J.}~\bibnamefont {He}}, \bibinfo
  {author} {\bibfnamefont {I.~L.}\ \bibnamefont {ten Kate}},\ and\ \bibinfo
  {author} {\bibfnamefont {H.}~\bibnamefont {Linnartz}},\ }\bibfield  {title}
  {\bibinfo {title} {The refractive index of amorphous and crystalline water
  ice in the {UV}{\textendash}vis},\ }\href
  {https://doi.org/10.3847/1538-4357/ab0d89} {\bibfield  {journal} {\bibinfo
  {journal} {Astrophys. J.}\ }\textbf {\bibinfo {volume} {875}},\ \bibinfo
  {pages} {131} (\bibinfo {year} {2019})}\BibitemShut {NoStop}%
\bibitem [{\citenamefont {Nordlund}\ \emph {et~al.}(2008)\citenamefont
  {Nordlund}, \citenamefont {Odelius}, \citenamefont {Bluhm}, \citenamefont
  {Ogasawara}, \citenamefont {Pettersson},\ and\ \citenamefont
  {Nilsson}}]{nordlund08}%
  \BibitemOpen
  \bibfield  {author} {\bibinfo {author} {\bibfnamefont {D.}~\bibnamefont
  {Nordlund}}, \bibinfo {author} {\bibfnamefont {M.}~\bibnamefont {Odelius}},
  \bibinfo {author} {\bibfnamefont {H.}~\bibnamefont {Bluhm}}, \bibinfo
  {author} {\bibfnamefont {H.}~\bibnamefont {Ogasawara}}, \bibinfo {author}
  {\bibfnamefont {L.}~\bibnamefont {Pettersson}},\ and\ \bibinfo {author}
  {\bibfnamefont {A.}~\bibnamefont {Nilsson}},\ }\bibfield  {title} {\bibinfo
  {title} {Electronic structure effects in liquid water studied by
  photoelectron spectroscopy and density functional theory},\ }\href
  {https://doi.org/https://doi.org/10.1016/j.cplett.2008.04.096} {\bibfield
  {journal} {\bibinfo  {journal} {Chem. Phys. Lett.}\ }\textbf {\bibinfo
  {volume} {460}},\ \bibinfo {pages} {86} (\bibinfo {year} {2008})}\BibitemShut
  {NoStop}%
\bibitem [{\citenamefont {Buch}\ \emph {et~al.}(1998)\citenamefont {Buch},
  \citenamefont {Sandler},\ and\ \citenamefont {Sadlej}}]{buch98}%
  \BibitemOpen
  \bibfield  {author} {\bibinfo {author} {\bibfnamefont {V.}~\bibnamefont
  {Buch}}, \bibinfo {author} {\bibfnamefont {P.}~\bibnamefont {Sandler}},\ and\
  \bibinfo {author} {\bibfnamefont {J.}~\bibnamefont {Sadlej}},\ }\bibfield
  {title} {\bibinfo {title} {Simulations of {H$_2$O} solid, liquid and
  clusters, with an emphasis on ferroelectric ordering transition in hexagonal
  ice},\ }\href@noop {} {\bibfield  {journal} {\bibinfo  {journal} {J. Phys.
  Chem. B}\ }\textbf {\bibinfo {volume} {102}},\ \bibinfo {pages} {8641}
  (\bibinfo {year} {1998})}\BibitemShut {NoStop}%
\bibitem [{\citenamefont {MacDowell}\ \emph {et~al.}(2004)\citenamefont
  {MacDowell}, \citenamefont {Sanz}, \citenamefont {Vega},\ and\ \citenamefont
  {Abascal}}]{macdowell04b}%
  \BibitemOpen
  \bibfield  {author} {\bibinfo {author} {\bibfnamefont {L.~G.}\ \bibnamefont
  {MacDowell}}, \bibinfo {author} {\bibfnamefont {E.}~\bibnamefont {Sanz}},
  \bibinfo {author} {\bibfnamefont {C.}~\bibnamefont {Vega}},\ and\ \bibinfo
  {author} {\bibfnamefont {J.~L.~F.}\ \bibnamefont {Abascal}},\ }\bibfield
  {title} {\bibinfo {title} {Combinatorial entropy and phase diagram of
  partially ordered ice phases},\ }\href@noop {} {\bibfield  {journal}
  {\bibinfo  {journal} {J. Chem. Phys.}\ }\textbf {\bibinfo {volume} {121}},\
  \bibinfo {pages} {10145} (\bibinfo {year} {2004})}\BibitemShut {NoStop}%
\bibitem [{\citenamefont {Bischoff}\ \emph {et~al.}(2021)\citenamefont
  {Bischoff}, \citenamefont {Reshetnyak},\ and\ \citenamefont
  {Pasquarello}}]{bischoff21}%
  \BibitemOpen
  \bibfield  {author} {\bibinfo {author} {\bibfnamefont {T.}~\bibnamefont
  {Bischoff}}, \bibinfo {author} {\bibfnamefont {I.}~\bibnamefont
  {Reshetnyak}},\ and\ \bibinfo {author} {\bibfnamefont {A.}~\bibnamefont
  {Pasquarello}},\ }\bibfield  {title} {\bibinfo {title} {Band gaps of liquid
  water and hexagonal ice through advanced electronic-structure calculations},\
  }\href {https://doi.org/10.1103/PhysRevResearch.3.023182} {\bibfield
  {journal} {\bibinfo  {journal} {Phys. Rev. Research}\ }\textbf {\bibinfo
  {volume} {3}},\ \bibinfo {pages} {023182} (\bibinfo {year}
  {2021})}\BibitemShut {NoStop}%
\bibitem [{\citenamefont {de~Koning}\ \emph {et~al.}(2006)\citenamefont
  {de~Koning}, \citenamefont {Antonelli}, \citenamefont {da~Silva},\ and\
  \citenamefont {Fazzio}}]{koning06}%
  \BibitemOpen
  \bibfield  {author} {\bibinfo {author} {\bibfnamefont {M.}~\bibnamefont
  {de~Koning}}, \bibinfo {author} {\bibfnamefont {A.}~\bibnamefont
  {Antonelli}}, \bibinfo {author} {\bibfnamefont {A.~J.~R.}\ \bibnamefont
  {da~Silva}},\ and\ \bibinfo {author} {\bibfnamefont {A.}~\bibnamefont
  {Fazzio}},\ }\bibfield  {title} {\bibinfo {title} {Orientational defects in
  ice ih: An interpretation of electrical conductivity measurements},\ }\href
  {https://doi.org/10.1103/PhysRevLett.96.075501} {\bibfield  {journal}
  {\bibinfo  {journal} {Phys. Rev. Lett.}\ }\textbf {\bibinfo {volume} {96}},\
  \bibinfo {pages} {075501} (\bibinfo {year} {2006})}\BibitemShut {NoStop}%
\bibitem [{\citenamefont {Feibelman}(2008)}]{feibelman08}%
  \BibitemOpen
  \bibfield  {author} {\bibinfo {author} {\bibfnamefont {P.~J.}\ \bibnamefont
  {Feibelman}},\ }\bibfield  {title} {\bibinfo {title} {Lattice match in
  density functional calculations: ice ihvs.$\beta$-agi},\ }\href
  {https://doi.org/10.1039/B808482N} {\bibfield  {journal} {\bibinfo  {journal}
  {Phys. Chem. Chem. Phys.}\ }\textbf {\bibinfo {volume} {10}},\ \bibinfo
  {pages} {4688} (\bibinfo {year} {2008})}\BibitemShut {NoStop}%
\bibitem [{\citenamefont {Santos~Rego}\ and\ \citenamefont
  {de~Koning}(2020)}]{santos20}%
  \BibitemOpen
  \bibfield  {author} {\bibinfo {author} {\bibfnamefont {J.}~\bibnamefont
  {Santos~Rego}}\ and\ \bibinfo {author} {\bibfnamefont {M.}~\bibnamefont
  {de~Koning}},\ }\bibfield  {title} {\bibinfo {title} {Density-functional
  theory prediction of the elastic constants of ice ih},\ }\href
  {https://doi.org/10.1063/1.5142710} {\bibfield  {journal} {\bibinfo
  {journal} {J. Chem. Phys.}\ }\textbf {\bibinfo {volume} {152}},\ \bibinfo
  {pages} {084502} (\bibinfo {year} {2020})},\ \Eprint
  {https://arxiv.org/abs/https://doi.org/10.1063/1.5142710}
  {https://doi.org/10.1063/1.5142710} \BibitemShut {NoStop}%
\bibitem [{\citenamefont {Zhong}\ \emph {et~al.}(2020)\citenamefont {Zhong},
  \citenamefont {Yu}, \citenamefont {Dodia}, \citenamefont {Bonn},
  \citenamefont {Nagata},\ and\ \citenamefont {Ohto}}]{zhong20}%
  \BibitemOpen
  \bibfield  {author} {\bibinfo {author} {\bibfnamefont {K.}~\bibnamefont
  {Zhong}}, \bibinfo {author} {\bibfnamefont {C.-C.}\ \bibnamefont {Yu}},
  \bibinfo {author} {\bibfnamefont {M.}~\bibnamefont {Dodia}}, \bibinfo
  {author} {\bibfnamefont {M.}~\bibnamefont {Bonn}}, \bibinfo {author}
  {\bibfnamefont {Y.}~\bibnamefont {Nagata}},\ and\ \bibinfo {author}
  {\bibfnamefont {T.}~\bibnamefont {Ohto}},\ }\bibfield  {title} {\bibinfo
  {title} {Vibrational mode frequency correction of liquid water in density
  functional theory molecular dynamics simulations with van der waals
  correction},\ }\href {https://doi.org/10.1039/C9CP06335H} {\bibfield
  {journal} {\bibinfo  {journal} {Phys. Chem. Chem. Phys}\ }\textbf {\bibinfo
  {volume} {22}},\ \bibinfo {pages} {12785} (\bibinfo {year}
  {2020})}\BibitemShut {NoStop}%
\bibitem [{\citenamefont {Santra}\ \emph {et~al.}(2011)\citenamefont {Santra},
  \citenamefont {Klime\ifmmode~\check{s}\else \v{s}\fi{}}, \citenamefont
  {Alf\`e}, \citenamefont {Tkatchenko}, \citenamefont {Slater}, \citenamefont
  {Michaelides}, \citenamefont {Car},\ and\ \citenamefont
  {Scheffler}}]{santra11}%
  \BibitemOpen
  \bibfield  {author} {\bibinfo {author} {\bibfnamefont {B.}~\bibnamefont
  {Santra}}, \bibinfo {author} {\bibfnamefont {J.~c.~v.}\ \bibnamefont
  {Klime\ifmmode~\check{s}\else \v{s}\fi{}}}, \bibinfo {author} {\bibfnamefont
  {D.}~\bibnamefont {Alf\`e}}, \bibinfo {author} {\bibfnamefont
  {A.}~\bibnamefont {Tkatchenko}}, \bibinfo {author} {\bibfnamefont
  {B.}~\bibnamefont {Slater}}, \bibinfo {author} {\bibfnamefont
  {A.}~\bibnamefont {Michaelides}}, \bibinfo {author} {\bibfnamefont
  {R.}~\bibnamefont {Car}},\ and\ \bibinfo {author} {\bibfnamefont
  {M.}~\bibnamefont {Scheffler}},\ }\bibfield  {title} {\bibinfo {title}
  {Hydrogen bonds and van der waals forces in ice at ambient and high
  pressures},\ }\href {https://doi.org/10.1103/PhysRevLett.107.185701}
  {\bibfield  {journal} {\bibinfo  {journal} {Phys. Rev. Lett.}\ }\textbf
  {\bibinfo {volume} {107}},\ \bibinfo {pages} {185701} (\bibinfo {year}
  {2011})}\BibitemShut {NoStop}%
\bibitem [{\citenamefont {Morawietz}\ \emph {et~al.}(2016)\citenamefont
  {Morawietz}, \citenamefont {Singraber}, \citenamefont {Dellago},\ and\
  \citenamefont {Behler}}]{morawietz16}%
  \BibitemOpen
  \bibfield  {author} {\bibinfo {author} {\bibfnamefont {T.}~\bibnamefont
  {Morawietz}}, \bibinfo {author} {\bibfnamefont {A.}~\bibnamefont
  {Singraber}}, \bibinfo {author} {\bibfnamefont {C.}~\bibnamefont {Dellago}},\
  and\ \bibinfo {author} {\bibfnamefont {J.}~\bibnamefont {Behler}},\
  }\bibfield  {title} {\bibinfo {title} {How van der waals interactions
  determine the unique properties of water},\ }\href
  {https://doi.org/10.1073/pnas.1602375113} {\bibfield  {journal} {\bibinfo
  {journal} {Proceedings of the National Academy of Sciences}\ }\textbf
  {\bibinfo {volume} {113}},\ \bibinfo {pages} {8368} (\bibinfo {year}
  {2016})},\ \Eprint
  {https://arxiv.org/abs/https://www.pnas.org/content/113/30/8368.full.pdf}
  {https://www.pnas.org/content/113/30/8368.full.pdf} \BibitemShut {NoStop}%
\bibitem [{\citenamefont {Reinhardt}\ and\ \citenamefont
  {Cheng}(2021)}]{reinhardt21}%
  \BibitemOpen
  \bibfield  {author} {\bibinfo {author} {\bibfnamefont {A.}~\bibnamefont
  {Reinhardt}}\ and\ \bibinfo {author} {\bibfnamefont {B.}~\bibnamefont
  {Cheng}},\ }\bibfield  {title} {\bibinfo {title} {Quantum-mechanical
  exploration of the phase diagram of water},\ }\href@noop {} {\bibfield
  {journal} {\bibinfo  {journal} {Nat. Commun.}\ }\textbf {\bibinfo {volume}
  {12}},\ \bibinfo {pages} {588} (\bibinfo {year} {2021})}\BibitemShut
  {NoStop}%
\bibitem [{\citenamefont {Kresse}\ and\ \citenamefont
  {Hafner}(1994)}]{kresse94}%
  \BibitemOpen
  \bibfield  {author} {\bibinfo {author} {\bibfnamefont {G.}~\bibnamefont
  {Kresse}}\ and\ \bibinfo {author} {\bibfnamefont {J.}~\bibnamefont
  {Hafner}},\ }\bibfield  {title} {\bibinfo {title} {Norm-conserving and
  ultrasoft pseudopotentials for first-row and transition elements},\
  }\href@noop {} {\bibfield  {journal} {\bibinfo  {journal} {J. Phys.: Condens.
  Matter}\ }\textbf {\bibinfo {volume} {6}},\ \bibinfo {pages} {8245} (\bibinfo
  {year} {1994})}\BibitemShut {NoStop}%
\bibitem [{\citenamefont {Kresse}\ and\ \citenamefont
  {Joubert}(1999)}]{kresse99}%
  \BibitemOpen
  \bibfield  {author} {\bibinfo {author} {\bibfnamefont {G.}~\bibnamefont
  {Kresse}}\ and\ \bibinfo {author} {\bibfnamefont {D.}~\bibnamefont
  {Joubert}},\ }\bibfield  {title} {\bibinfo {title} {From ultrasoft
  pseudopotentials to the projector augmented-wave method},\ }\href@noop {}
  {\bibfield  {journal} {\bibinfo  {journal} {Phys. Rev. B}\ }\textbf {\bibinfo
  {volume} {59}},\ \bibinfo {pages} {1758} (\bibinfo {year}
  {1999})}\BibitemShut {NoStop}%
\bibitem [{\citenamefont {Wu}\ \emph {et~al.}(2005)\citenamefont {Wu},
  \citenamefont {Vanderbilt},\ and\ \citenamefont {Hamann}}]{wu05}%
  \BibitemOpen
  \bibfield  {author} {\bibinfo {author} {\bibfnamefont {X.}~\bibnamefont
  {Wu}}, \bibinfo {author} {\bibfnamefont {D.}~\bibnamefont {Vanderbilt}},\
  and\ \bibinfo {author} {\bibfnamefont {D.~R.}\ \bibnamefont {Hamann}},\
  }\bibfield  {title} {\bibinfo {title} {Systematic treatment of displacements,
  strains, and electric fields in density-functional perturbation theory},\
  }\href {https://doi.org/10.1103/PhysRevB.72.035105} {\bibfield  {journal}
  {\bibinfo  {journal} {Phys. Rev. B}\ }\textbf {\bibinfo {volume} {72}},\
  \bibinfo {pages} {035105} (\bibinfo {year} {2005})}\BibitemShut {NoStop}%
\end{thebibliography}%


\clearpage

\newpage

\onecolumngrid

\appendix

\begin{widetext}

\begin{Large}
	\begin{center}
\textbf{Supplementary Material}
	\end{center}
\end{Large}

\vspace*{1.5cm}

\begin{large}
   Intermolecular forces at ice and water interfaces: premelting,  surface
   freezing and regelation.
\end{large}

\vspace*{1cm}
\begin{center}
\textit{Juan Luengo-M\'arquez, Fernando Izquierdo-Ruiz,  and Luis G. MacDowell}
\end{center}

\vspace*{1cm}

\section{Modeling the dielectric functions}

The dielectric function of a material may be obtained from the refractive index or the extinction coefficient.\cite{parsegian05,tanner13}
Measurements of the refractive index are suitable in regions were no
absorption takes place, while extinction coefficients are measured \revision{more}
accurately in regions of the spectrum with high absorption. Here, we will
parametrize our model from data of extinction coefficients, available for water
and ice in all the relevant region of the spectrum, but will gauge or model
against the refractive index in the near \revision{infrared (IR)}.
A summary of current parametrizations for the spectrum of water is presented in
Table \ref{parametrizations}. In the following, we review sources of literature
data for the optical properties in order to select an optimal data set for the
purpose of modeling optical properties of water and ice in the neighborhood of
the triple point.

 \begin{table}[thb]
 	\begin{tabular}{|l|p{0.7\textwidth}|}
 		\hline
 		& \\
 		Reference & Description \\
 		& \\
 		\hline
 		& \\
 		Parsegian and Weiss\cite{parsegian81} & First accurate fit to UV band by Heller, with a
 		microwave term and five IR absorbtions. \\
 		Elbaum and Schick\cite{elbaum91b}  & Uses similar data as \cite{parsegian81}, but high energy band is underestimated. \\
 		Roth and Lenhoff\cite{roth96} & Uses the same MW and IR data than parsegian and Weiss,
 		but improves the
 		parametrization of the Heller UV band considerably.\\
 		Dagastine et al.\cite{dagastine00} & Uses improved data in the IR region and obtains the
 		dielectric
 		function at imaginary frequencies from a KK analysis. \\
 		Fernandez-Varea et al.\cite{fernandez00} & Uses the accurate representation of the Heller
 		data, enforcing the f-sum rule. \\
 		Wang and Nguyen\cite{wang17} & Full update of the UV data using results from Ref.\cite{hayashi15},
 		but does not provide results for the IR region and uses 
 		KK analysis with epsilon(iw) in tabulated form. \\
 		Fiedler et al.\cite{fiedler20} & Improved parametrization with detailed study of MW
 		and far IR
 		regions, but overestimates refractive index. \\
 		Gudarzi and Aboutalebi\cite{gudarzi21} & Uses the Hayashi band, with high energy decay
 		estimated with f-sum
 		rule and enforcement of refractive index, but parametric form
 		at low energy based on room temperature results. 
 		\\ & \\ \hline
 	\end{tabular}
 	\caption{Review of optical properties of water parametrized 
 		for use in Hamaker coefficient calculations.
 	\label{parametrizations}
 	}
 \end{table}
 

\subsection{Water}

\subsubsection{From the microwave to the visible region}

In this low energy region, ranging from $10^{-3}$ eV to about 1 eV, 
the spectrum of
water exhibits significant absorption all the way from the MW, across the IR,
and to the mid-IR, with most of the vibrational transitions ending sharply at 
ca.  0.4 eV. The remaining range of the spectrum including the \revision{visible (VIS)}, 
exhibits very
small absorption and an almost constant refractive index,
ca. $n=1.33$.\cite{lide94}
Thereof, absorption increases sharply in the Near UV, at ca. 7 eV.

There is a large number of experimental data measuring the extinction
coefficient at ambient temperature\cite{bertie96} but experiments performed
for cold or supercooled water show a significant temperature dependence in this region of the
spectrum,\cite{pinkley77,hasted87,zelsmann95,zasetsky04,wagner05} as discussed
recently by Rowe et al.\cite{rowe20}

Particularly the \revision{MW} region is well known to exhibit a strong
temperature
dependence, as it is mostly related to thermally activated libration and
rotation of water molecules. Measurements  indicate a significant
attenuation of this band as temperature decreases from ambient temperature
to 268~K. For this reason, we choose data for ice cold water measured by Zelsmann as recommended in Rowe
et al.\cite{rowe20} for the extinction coefficient in the \revision{hundredth} of
eV.\cite{zelsmann95} In practice, this choice is not
consequential for the calculation of the Hamaker function, since the
thermal Matsubara frequency at 273,15~K,  $\omega_T=0.15-$~eV
(i.e. 1200 cm$^{-1}$) is already beyond the MW region.
Whence, only three discrete Matsubara frequencies are
required to span the full IR absorption spectrum of liquid water at this temperature.
The first two, at 0.15~eV and 0.30~eV sample the decay of the libration band
and the libration+bending combination band, respectively and have extinction
coefficients well below 0.1, where experimental uncertainties preclude
detection of a clear temperature dependence.\cite{rowe20} The third frequency
falls at the high energy edge of the O-H bending and stretching overtone, which
is known to show a significant temperature dependence, with higher absorption
intensities at lower temperature.\cite{pinkley77} However, the
temperature dependence is significant only at the maximum of the band,
and the Matsubara frequency of 0.45~eV falls well on the high energy edge,
where the temperature dependence is very small and shows very small
differences with the compilation of Bertie and Lan at
293~K.\cite{bertie96,segelstein81,wieliczka89,rowe20}.

In view of this, we choose the data of Zelsmann et al.\cite{zelsmann95} for the
far-IR, but stick to the ambient temperature data of Segelstein and Wieliczka
as recommended by Bertie and Lan
to model the dielectric function of water all
the way from the MW to the near
UV.\cite{zelsmann95,segelstein81,wieliczka89,bertie96}


\subsubsection{High energy band}

Starting at about 7 eV, the extinction coefficient of water increases steeply and
spans over several decades as a result of single electron and
collective electronic excitations.\cite{heller74,seki81,hayashi15} Measurements
up to 25~eV were carried out by Heller et al. for water at 1~C.\cite{heller74} Most studies since then use this
data, or the revised analysis of Dingfelder,\cite{dingfelder98} as the
source for the parametrization of water's dielectric function
\cite{parsegian81,elbaum91b,roth96,dagastine00,fernandez00}. However,
careful analysis performed later showed that the high energy
tail of Heller's data appears to drop too fast, and should be extended to higher
energies.\cite{segelstein81,dingfelder98} Recent experiments performed
with a synchrotron source reported absorption data up to 100 eV, albeit at
ambient temperature. The results show that the principal \revision{absorption} band
extends over significantly larger energy ranges, but has a lower absorption
maximum than estimated by Heller.

Although these two sets of experiments have been performed at different
temperature, it is expected that absorptions for energies above the decade of eV
are hardly affected by thermal rearrangement. This expectation is supported by
measurements of refractive indexes in the VIS and near IR, which are known to
great accuracy and show temperature changes that are extremely small, of order
$10^{-5}$ per Kelvin.\cite{dobbins73,lide94}

Based on these observations, our complete set of absorption coefficients
for the parametrization of water at the freezing point comprises
the data of Zelsmann et al.\cite{zelsmann95} for the far-IR (2.4 meV to 70 meV),
Wieliczka et al. (0.066 to 1.01 eV),
and the synchrotron high energy band measurements of Hayashi and Hiraoke
recommended in Ref.\cite{hayashi15,wang17,fiedler20}  We call this the `Hayashi
set'. To \revision{account} for the uncertainty in the temperature effect of the high
energy band, we also consider an alternative  set with the same data for the
far-IR  to the near-UV but with Heller's high energy band instead, which
we will denote as the `Heller set'.  The choice of experimental data is
displayed in Fig.\ref{fig:kappa}-top.

In both cases
we use $\epsilon(0)=88.2$ for the static dielectric constant at
0~C.\cite{buckley58}

\subsection{Ice}

Contrary to liquid water,  \revision{modeling} of optical properties for ice
appears far less studied.\cite{elbaum91b,emfietzoglou07}
In the spectral range that is required for the
calculation of van der Waals forces, we are only aware of the
parametrization by Elbaum and Schick \revision{(ES)}.\cite{elbaum91b} However,
a large number of studies are available which update the old compilation
of data from Irvine and Pollack used in that work,\cite{irvine68} so we perform also for ice a new parametrization with modern data recommended by
Warren and Brandt.\cite{warren08}

The absorption spectrum of ice clearly resembles that of water in all
the relevant \revision{regions} from the far IR to well beyond the UV.
At lower energies the spectrum  differs significantly.   Liquid water displays
the characteristic large microwave \revision{absorption band} related to molecular rotation, while
ice displays lattice \revision{absorptions} instead.\cite{warren84}

For the region spanning MW to VIS, a compilation of several sources measured
at different temperatures  is corrected and extrapolated
to T=266~K by Warren and Brandt.\cite{warren08} This is a revised version of a previous
compilation of 1984.\cite{warren84} The updated results are qualitatively
similar, but correct the absorption in some regions by as much as a factor of
two. This serves to illustrate the extent of uncertainty that plague optical
data. It is estimated that temperature effects could change the absorption
peaks by about 1\% per Kelvin in this region,\cite{warren08} so some variations
with respect to ice at the melting point could occur. \revision{In the case of water at the triple point},
however, only three Matsubara frequencies are really affected by this problem,
since the absorption in the near IR and VIS are very small and do not
contribute significantly to the van der Waals forces.

Unfortunately, measurements of the  extinction coefficients for ice in  the
important
high energy region are scarce and do not exceed energies  of
28 eV.\cite{daniels71,seki81,kobayashi83}
We discard the results by Daniel, which are measured
in samples grown from the vapor at 78~K, and pertain therefore to either cubic or amorphous
ice. For this reason, we choose the data by Seki et al. as reviewed in
Ref.\cite{warren08}, which correspond to  monocrystal ice grown from
from pure water. Unfortunately, data acquisition was reported at a very low
temperature of 80~K, so it is difficult to rule out significant
effects in principle. In practice, we find that refractive indexes in the IR
measured recently at T=150~K differ with those reported by Warren and Brandt at
T=266~K by about 0.2\% only (c.f. 1.3080 v 1.3060 at
$\lambda=700$~nm).\cite{kofman19} As additional evidence for the small
temperature change expected in the high energy band, we note that photon
emission spectroscopy data of
ice at 90 and 250~K appear very similar, with the low temperature spectrum
exhibiting somewhat sharper bands, but no significant change in emission
energies.\cite{nordlund08}

Based on this discussion, our data set for the parametrization of ice
comprises the compilation of Warren for the spectral region between MW and UV,
and the high energy band by Warren (c.f. Fig.\ref{fig:kappa}-bottom).\cite{warren08,seki81} For the static
dielectric constant we use the value of $\epsilon(0)=91.5$ reported by Auty and
Cole.\cite{auty52}

\subsection{Fit to experimental data}

The calculation of van der Waals forces by means of DLP theory requires
evaluation of the complex dielectric function at imaginary frequencies,
$\epsilon(i\omega)$. This property, which is always real, can be obtained
in principle directly from extinction coefficients using the Kramers-Kronig
relation:
\begin{equation}
   \epsilon(i\omega') = 1 +
   \frac{2}{\pi}\int_{0}^{\infty}\frac{\omega\epsilon_{2}(\omega)}{\omega^{2}+\omega
   '^{2}}d\omega
   \label{kk_eigi}
\end{equation}
The use of this relation is somewhat inconvenient, however, as it requires
extinction coefficients from the full electromagnetic spectrum. To circumvent
this problem, it is customary to assume a parametric form for the
complex dielectric function. A simple prescription due to Parsegian and Ninham\revision{\cite{ninham70,parsegian05}}
uses a sum of Lorentz oscillators to achieve this goal, such that:
\begin{equation}
   \epsilon(\omega) = 1 +
   \sum_{k=1}^{N_{osc}}\frac{A_{k}}{1-iB_{k}\omega-C_{k}\omega^{2}}
   \label{drude_complex}
\end{equation}
This form shows readily that  evaluation of $\epsilon(\omega)$ at
a purely imaginary frequency, say $\omega=i\xi$, provides
a well behaved real valued function,
\begin{equation}
    \epsilon(i\xi) = 1 + \sum_{k=1}^{N_{osc}}\frac{A_{k}}{1 + B_{k}\xi +
    C_{k}\xi^{2}}
    \label{dielec_func_drude}
\end{equation}
which can be  used for the calculation of the Hamaker function,
Eq.2 of the main text.

In order to obtain the parameters required in \Eq{dielec_func_drude},
we notice that the experimentally available extinction coefficients
are related to the imaginary part of $\epsilon(\omega)=\epsilon_1(\omega) +
i\epsilon_2(\omega)$ as $2\kappa^2(\omega)=|\epsilon(\omega)| -
\epsilon_1(\omega)$, so that substitution of \Eq{drude_complex} into
this relation provides a parametric function for the
experimental data of $\kappa(\omega)$.

We performed fits for both water and ice using a total of 11
Lorentz oscillators. Six were used to fit the MW and IR regions down to approximately 1~eV,
and five to model the high energy band in the extreme UV region and beyond.

The parameters  obtained for the fits to Hayashi and Heller sets of water, as well as those for ice may be found in Tables \ref{table:hayashi}-\ref{table:ice}.

Fig.\ref{fig:kappa} compares our model extinction coefficients with the experimental data
set for both water (top) and ice (bottom) over the electromagnetic spectrum from
the microwave to the extreme ultra-violet. Details of the important high energy
band from the near to the extreme ultra-violet are shown in Fig.1. of the main
text.

\begin{figure}[htb!]
        \includegraphics[width=0.5\textwidth,keepaspectratio]{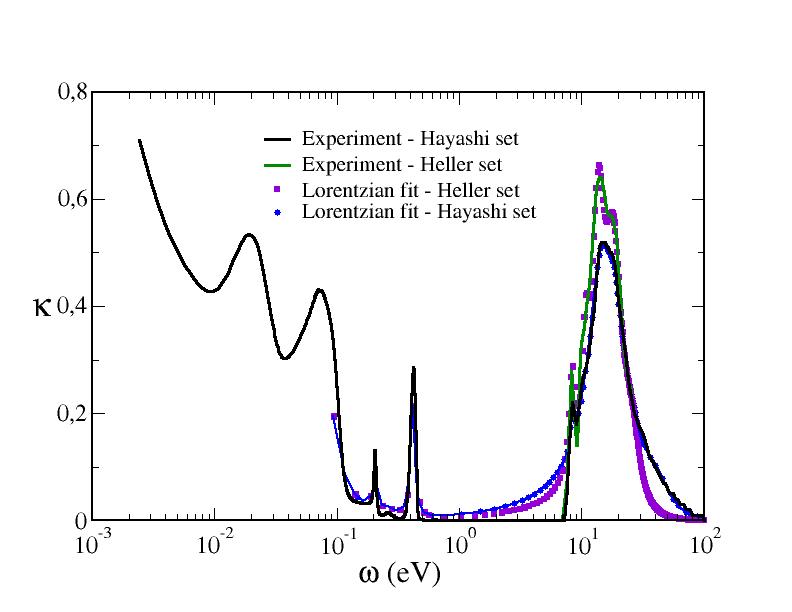}
        \includegraphics[width=0.5\textwidth,keepaspectratio]{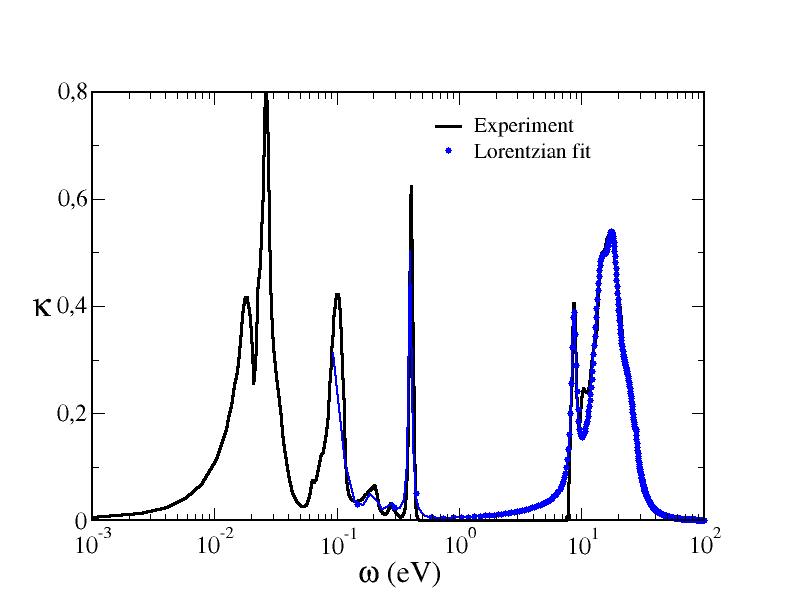}
        \caption{Experimental data of extinction index ($\kappa$), and parametric representations with Lorentzian oscillators. Experimental results for water (top), displaying the Hayashi and Heller sets  together with parametrizations using the Lorentz model from this work.
        Experimental results for ice (bottom), compared to fit with Lorentz model. Symbols correspond (roughly) to extinction indexes measured at Matsubara frequencies (some of the highly dense frequencies were removed for clarity of presentation). }
        \label{fig:kappa}
\end{figure}

For water the high energy band is very accurately reproduced in the important
region above ca. 10~eV, both for the Hayashi and Heller sets. 
The less important IR region is reproduced only qualitatively.
For ice, the high energy band is also well reproduced, except for
a small shoulder occurring at about 10~eV, which could not be reproduced
by the model without spoiling the rest of the fit. The
IR region in this case is more faithfully reproduced than it is for water.
Together with our fits, Fig.1-top of the main manuscript also displays the extinction
coefficients of liquid water and ice
predicted by the model of \revision{ES},\cite{elbaum91b} which appears
to differ significantly from our fit to the Heller data set.

\subsection{Quantum density functional theory calculations}

Since there appears to be some degree of uncertainty as regards the experimental
measurement of Dielectric functions, we have also performed quantum density
functional theory calculations (DFT). Results obtained in this way are only
approximate, but  the calculations are performed under the same footing
for both water and ice. This is not the case for experimental studies, because
it takes some approximations to resolve the signal from the instrumental
resolution, and absorption of vapor from the liquid surface hampers the data
analysis.\cite{hayashi15}

%

%

In order to simplify the calculations, we first performed classical simulations
of bulk ice and water at T=273~K using the TIP4P/Ice model.\cite{abascal05}
Because of the computational cost, simulations were performed for samples of 16
molecules under periodic boundary conditions and the  minimum image convention.
After equilibration, we performed a batch of 12 consecutive runs, and
stored the final configuration of each run for further analysis. In the
case of ice, each of the 12 configurations was obtained from an independent
hydrogen bond arrangement sampled according to the ice
rules.\cite{buch98,macdowell04b} In practice, we find the high energy band
displays a very small dependence of the selected configuration, in line with
other studies.\cite{bischoff21} The thermalized configurations are
used as input for the DFT calculations, which are used to relax the
intramolecular degrees of freedom of the H$_2$O molecules.

The Electronic structure calculations are performed  using the
Perdew-Burke-Ernzerhof (PBE) functional,\cite{perdew96} which, as other generalized gradient
functionals is widely used to study energetic properties of ice and
water.\cite{koning06,feibelman08,santos20,zhong20}
PBE does not account properly for dispersion interactions,
which are known to have an important impact in
water thermodynamic properties.\cite{santra11,morawietz16,reinhardt21} However,
here the configuration space of the nuclei is sampled from an
empirical potential anyway, and we do not expect large corrections to the
electronic properties from dispersion effects.\cite{zhong20}

To solve the Kohn-Sham equations, we employ
the Vienna Ab innitio Simulation Package
(VASP),\cite{vasp1,vasp2,vasp3} using a plane wave basis
set with cutoff at 700~eV for the valence electrons, and
core electrons treated with
pseudopotentials in the
projected augmented wave approximation (PAW).\cite{kresse94,kresse99}

In order to obtain the optical properties in the high energy region,
results from the PBE calculations are post processed under the GW/RPA
approximation,\cite{shishkin06,fuchs07,bischoff21}. For further details, the reader
is referred to references \cite{gajdos06,nunes01}. Optical properties in the
vibrational region are obtained using Density Functional Perturbation Theory
\cite{wu05}.

Figure \ref{fig:eps2dft} displays the absorption spectrum obtained from the DFT
calculations for both ice and water. Whereas it is clear that a quantitative
agreement with experiment is not achieved, the PBE functional provides
a reasonable qualitative description of the absorption spectrum.
Particularly, the calculations yield a first electronic excitation in the
range of decades of eV--\revision{admittedly}, with significantly smaller
intensity--followed
by a strong adsorption close to 20 eV. In the IR region, DFT calculations
from frozen configuration only provide the absorption frequencies, and the
spectrum shows results obtained with an assumed constant band width of 0.02~eV
for the sake of illustration. \revision{Bearing} this in mind, we find that the density
of normal modes is well reproduced and displays significantly sharper features
for ice than for water, as observed in experiments. Since no dynamics is input
into the calculations, we are unable to reproduce the MW region of liquid
water, but this fortunately is inconsequential as far as the calculation of
Hamaker functions is concerned.

For the purpose of calculating the Hamaker function, we obtain
the dielectric function at imaginary frequencies by using the Kramers-Kronig relation of \Eq{kk_eigi}. In order to ease the calculations, the resulting function is fitted to a model of Drude oscillators, \Eq{dielec_func_drude}, with zero band width (i.e. $B_i=0$). The parameters obtained from the fit may be found in Table \ref{table:DFTparam}.

\begin{figure}

  \includegraphics[width=0.48\textwidth,height=0.4\textwidth,keepaspectratio]{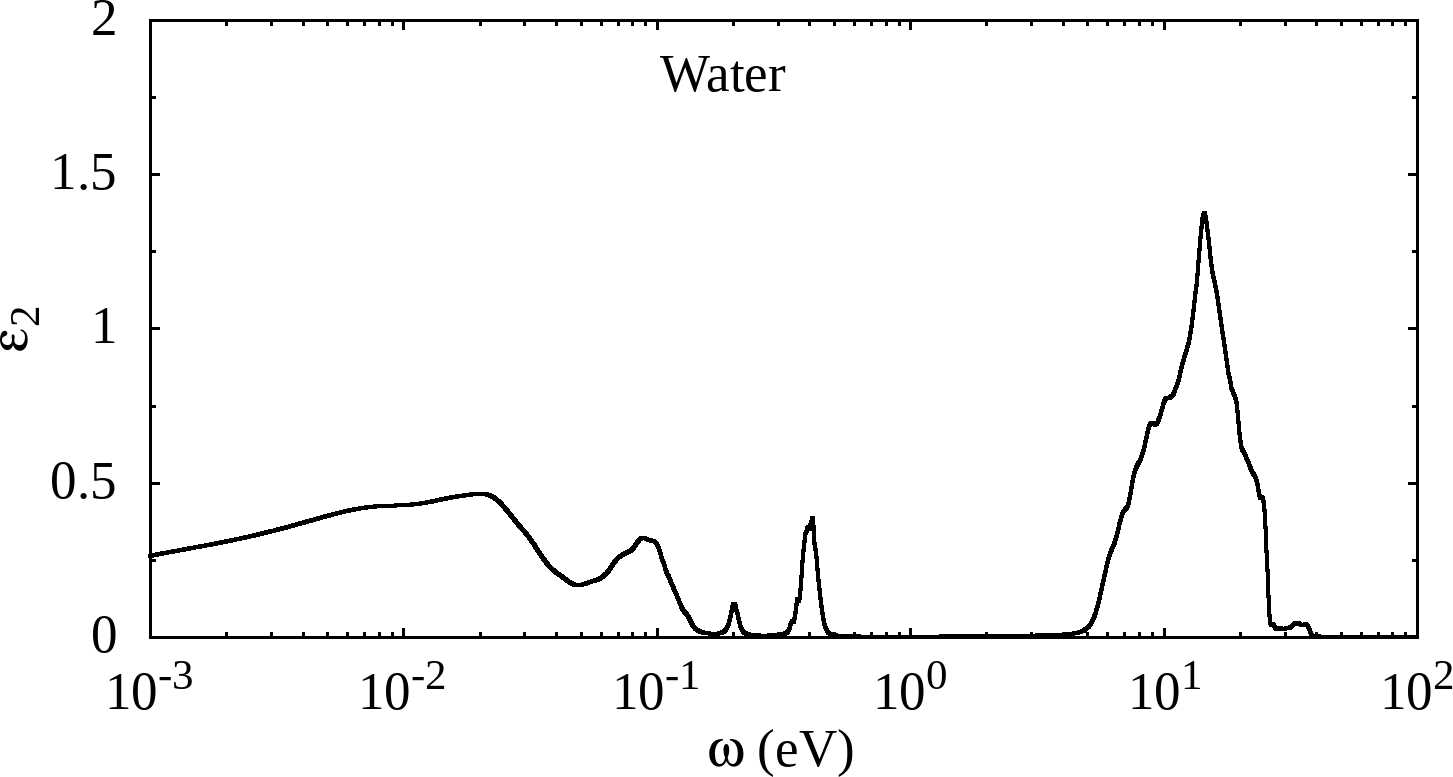} 
        \includegraphics[width=0.48\textwidth,height=0.4\textwidth,keepaspectratio]{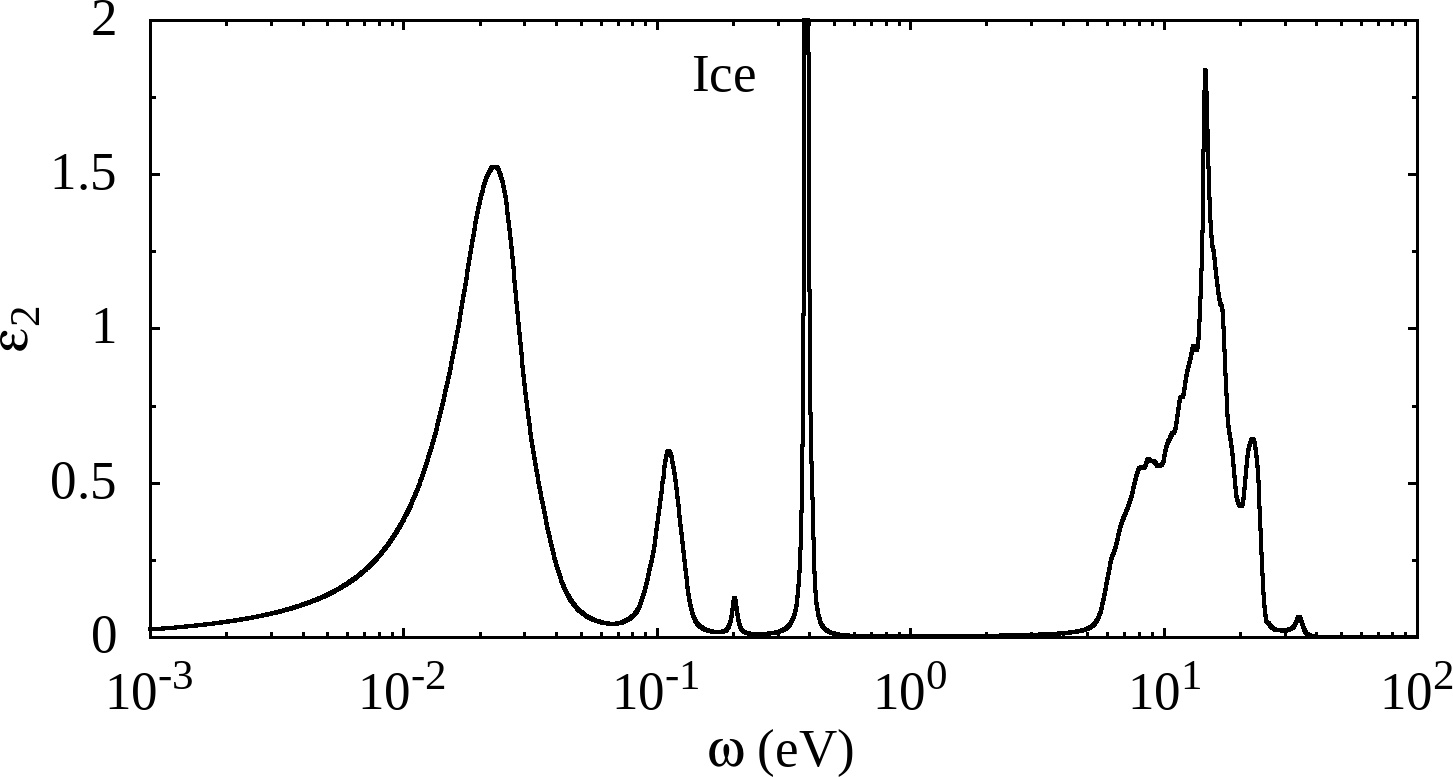}
        \caption{Real and imaginary parts of the complex dielectric function as
           obtained from DFT calculations. Compared to liquid water, the
           absorption bands appear sharper for ice due to its ordered structure.
\label{fig:eps2dft}
}
\end{figure}

\clearpage

\section{Parameters set}
\begin{table}[h]
\centering
\begin{tabular}{|c|c|c|}
\hline
$A_{k}$ & $B_{k}\cdot$eV & $C_{k}\cdot$eV$^2$ \\
\hline
4.69 & 610.70 & 2587.34 \\
\hline
2.21 & 83.09 & 2119.77 \\
\hline
0.50 & 8.13 & 186.75 \\
\hline
1.97x$10^{-2}$ & 0.40 & 24.25 \\
\hline
7.51x$10^{-3}$ & 0.94 & 14.57 \\
\hline
5.79x$10^{-2}$ & 0.22 & 5.74 \\
\hline
1.13x$10^{-2}$ & 5.22x$10^{-3}$ & 1.44x$10^{-2}$ \\
\hline
0.57 & 3.88x$10^{-2}$ & 5.02x$10^{-3}$ \\
\hline
0.17 & 2.22x$10^{-2}$ & 2.96x$10^{-3}$ \\
\hline
8.28x$10^{-2}$ & 2.04x$10^{-2}$ & 1.59x$10^{-3}$ \\
\hline
0.11 & 1.91x$10^{-2}$ & 6.05x$10^{-4}$ \\
\hline
\end{tabular}
\caption{Parameters set for Liquid Water - Hayashi. The CSF is applied to the eighth oscillator, with $\omega_{0} = 7.7 \ eV$ and $\Delta \omega = 0.1 \ eV^{4}$}
\label{table:hayashi}
\end{table}

\begin{table}[h]
\centering
\begin{tabular}{|c|c|c|}
\hline
$A_{k}$ & $B_{k}\cdot$eV & $C_{k}\cdot$eV$^2$ \\
\hline
4.69 & 610.70 & 2587.34 \\
\hline
2.21 & 83.09 & 2119.77 \\
\hline
0.50 & 8.13 & 186.75 \\
\hline
1.97x$10^{-2}$ & 0.40 & 24.25 \\
\hline
7.51x$10^{-3}$ & 0.94 & 14.57 \\
\hline
5.79x$10^{-2}$ & 0.22 & 5.74 \\
\hline
8.02x$10^{-2}$ & 1.70x$10^{-2}$ & 1.46x$10^{-2}$ \\
\hline
8.99x$10^{-2}$ & 1.38x$10^{-2}$ & 8.73x$10^{-3}$ \\
\hline
0.37 & 2.06x$10^{-2}$ & 5.58x$10^{-3}$ \\
\hline
0.31 & 2.04x$10^{-2}$ & 3.21x$10^{-3}$ \\
\hline
0.10 & 1.71x$10^{-2}$ & 1.56x$10^{-3}$ \\
\hline
\end{tabular}
\caption{Parameters set for Liquid Water - Heller. The CSF is applied to the seventh to tenth oscillators, with $\omega_{0} = 7.7 \ eV$ and $\Delta \omega = 500.0 \ eV^{4}$}
\label{table:heller}
\end{table}

\begin{table}[h]
\centering
\begin{tabular}{|c|c|c|}
\hline
$A_{k}$ & $B_{k}\cdot$eV & $C_{k}\cdot$eV$^2$ \\
\hline
0.40 & 26.22 & 3328.19 \\
\hline
0.30 & 7.66 & 1545.24 \\
\hline
0.25 & 2.72 & 105.58 \\
\hline
2.44x$10^{-2}$ & 1.10 & 25.88 \\
\hline
1.03x$10^{-2}$ & 0.62 & 12.99 \\
\hline
8.66x$10^{-2}$ & 0.16 & 6.24 \\
\hline
9.39x$10^{-2}$ & 1.23x$10^{-2}$ & 1.35x$10^{-2}$ \\
\hline
0.28 & 2.26x$10^{-2}$ & 5.03x$10^{-3}$ \\
\hline
0.24 & 1.76x$10^{-2}$ & 3.25x$10^{-3}$ \\
\hline
0.12 & 1.66x$10^{-2}$ & 1.73x$10^{-3}$ \\
\hline
1.47x$10^{-3}$ & 1.37x$10^{-3}$ & 1.29x$10^{-3}$ \\
\hline
\end{tabular}
	\caption{Parameters set for Ice. The CSF is applied to the seventh to ninth oscillators, with $\omega_{0} = 8.1 \ eV$ and $\Delta \omega = 0.1 \ eV^{4}$}
\label{table:ice}
\end{table}

\begin{table}[h]
   \centering
   \begin{tabular}{|c|c||c|c|}
	\hline
	\multicolumn{2}{|c||}{Water}      & \multicolumn{2}{c|}{Ice}          \\
	\hline
	$A_{k}$         & $C_{k}\cdot$eV$^2$         & $A_{k}$         & $C_{k}\cdot$eV$^2$         \\
	\hline
	 0.646          &  1.21x$10^{+8}$   & 0.198           &  1.23x$10^{+8}$   \\
	 \hline
	  0.281          &  4.90x$10^{+6}$   & 0.154           &  6.08x$10^{+3}$   \\
	  \hline
	   0.582          &  5.91x$10^{+4}$   & 0.836           &  1.79x$10^{+3}$   \\
	   \hline
	    0.534          &  2.84x$10^{+3}$   & 0.132           &  8.28x$10^{+1}$
	    \\ \hline
	     0.184          &  6.89x$10^{+1}$   & 0.126           &  6.11
	     \\ \hline
	      0.180          &  2.83x$10^{-2}$   & 0.279           &  1.36x$10^{-2}$
		\\ \hline
		 0.539          &  3.87x$10^{-3}$   & 0.362           &  3.13x$10^{-3}$
		 \\ \hline
	    \end{tabular}
	      \caption{Parameters sets for liquid water and ice from fitting DFT
		calculations to 7 Lorentz oscillators without width. }
		\label{table:DFTparam}
	   \end{table}

\clearpage


\end{widetext}

\end{document}